\documentclass[preprint2]{aastex}
\usepackage[utf8]{inputenc}
\usepackage[varg]{txfonts}
\usepackage[T1]{fontenc}
\usepackage{epstopdf}
\usepackage{ifpdf}
\usepackage{float} 
\usepackage{longtable}
\usepackage{multicol}
\usepackage{natbib}
\usepackage{graphicx}
\bibpunct{(}{)}{;}{a}{}{,}
\usepackage{color}
\usepackage{hyphenat}
\emergencystretch=\maxdimen
\hyphenchar\font=-1
\slugcomment{Not to appear in Nonlearned J., 45.}

\shorttitle{Stellar populations of early-type galaxies}
\shortauthors{Rosa et al.}

\begin{document}


\title{Eight luminous early-type galaxies in nearby pairs and sparse groups. \\
       I. Stellar populations spatially analysed$^{*}$}


\author{D. A. Rosa\altaffilmark{1,2}}
\affil{Divis\~ao de Astrof\'isica, Instituto Nacional de Pesquisas Espaciais, S\~ao Jos\'e dos Campos-SP, Brazil \\
Instituto de Pesquisa \& Desenvolvimento, Universidade do Vale do Para\'iba, S\~ao Jos\'e dos Campos-SP, Brazil}
\email{deiserosa@univap.br}

\author{A. C. Milone\altaffilmark{1}}
\affil{Divis\~ao de Astrof\'isica, Instituto Nacional de Pesquisas Espaciais, S\~ao Jos\'e dos Campos-SP, Brazil}
\email{andre.milone@inpe.br}

\and

\author{A. C. Krabbe\altaffilmark{2} and I. Rodrigues\altaffilmark{2}}
\affil{Instituto de Pesquisa \& Desenvolvimento, Universidade do Vale do Para\'iba, S\~ao Jos\'e dos Campos-SP, Brazil}
       \email{krabbe@univap.br; irapuan@univap.br}


\altaffiltext{1}{Instituto Nacional de Pesquisas Espaciais (INPE), Divis\~ao de Astrof\'isica, 
             Av. dos Astronautas 1758, 12227-010, S\~ao Jos\'e dos Campos-SP, Brazil\\}
\altaffiltext{2}{present address: Instituto de Pesquisa \& Desenvolvimento (IP\&D), Universidade do Vale do Para\'iba,
             Av. Shishima Hifumi 2911, 12244-000,  S\~ao Jos\'e dos Campos-SP, Brazil\\ }
             

\begin{abstract}
We present a detailed spatial analysis of stellar populations based on long-slit optical spectra in a 
sample of eight luminous early-type galaxies selected from nearby sparse groups and pairs,  
three of them may have interaction with another galaxy of similar mass. We have 
spatially measured luminosity-weighted averages of age, [M/H], [Fe/H], and [$\alpha$/Fe] in the sample 
galaxies to add empirical data relative to the influence of galaxy mass, environment, 
interaction, and AGN feedback in their formation and evolution. The stellar population of the 
individual galaxies were determined  through the well-established stellar population synthesis code  
{\scriptsize\,STARLIGHT} using semi-empirical simple stellar population models. Radial variations of luminosity-
weighted means of age, [M/H], [Fe/H], and [$\alpha$/Fe] were quantified up to half of the effective 
radius of each galaxy. We found trends between representative values of age, [M/H], 
[$\alpha$/Fe], and the nuclear stellar velocity dispersion. There are also relations between the metallicity/age
gradients and the velocity dispersion. Contributions of $1-4$ Gyr old stellar populations were
quantified in IC\,5328 and NGC\,6758 as well as $4-8$ Gyr old ones in NGC\,5812. Extended gas is present in 
IC\,5328, NGC\,1052, NGC\,1209, and NGC\,6758, and the presence of a LINER is identified in all these 
galaxies. The regions up to one effective radius of all galaxies are basically dominated by 
$\alpha$-enhanced metal-rich old stellar populations likely due 
to rapid star formation episodes that induced efficient chemical enrichment. On average, the age 
and [$\alpha$/Fe] gradients are null and the [M/H] gradients are negative, although discordant cases were 
found. We found no correlation between the stellar population properties and the 
LINER presence as well as between the stellar properties and environment or gravitational interaction, 
suggesting that the influence of progenitor mass can-not be discarded in the formation and evolution 
of early-type galaxies.
\end{abstract}


\section{Introduction}

The formation and evolution of elliptical and lenticular galaxies
(early-type galaxies, or simply ETGs)
are still open questions in Astrophysics.
The modern view of formation/evolution of massive ETGs, especially ellipticals, implies a long two-phase process \citep{2010ApJ...725.2312O}.
The initial phase at high redshift is induced by dissipative fusions
of small stellar systems formed through individual gravitational collapses.
This initial process builds up the main body of a massive elliptical with stellar populations passively evolving
such that the system rapidly turns red
\citep{2005ApJ...634..861Y,2008ApJ...680..224Z,2009ApJ...697.1493S,
2009A&A...504..751S,2011ApJ...739L..44D,2015ApJ...803...26P,2017MNRAS.470.1050F}.
In the second phase, minor and/or major dry mergers make this main body progressively bigger and redder
during many billions of years, explaining the existence of an extended stellar envelope
with smaller metallicity, bluer than the nuclear region
\citep{2013ApJ...763...26O,2016ApJ...821..114H}.

Additionally, different physical mechanisms, including internal and external ones,
play distinct roles in the formation and evolution of ETGs
under the cold dark matter paradigm for structure formation in a Universe dominated by dark energy ($\Lambda$CDM cosmology).
Some physical processes depend on the environmental conditions,
such as major merging;
minor merging;
ram pressure, which extracts the interstellar medium;
and strangulation, which strips the galactic halo 
\citep{1972ApJ...178..159C,2003ApJ...597..893N,2004cgpc.symp..305V,2009MNRAS.394.1213W}.
Other physical processes are intrinsic to the galaxies themselves such as
secular evolution, which modifies the internal structure;
active galactic nuclei (AGN) feedback;
and stellar feedback, produced by supernovae explosions
\citep{2008gady.book.....B,2006MNRAS.365...11C,2015MNRAS.449..528H}.
Galactic winds may interrupt the star formation by removing the gas reservoir.
They can be mainly induced by the AGN feedback compared with the stellar feedback.
In addition, the efficiency of galactic winds is, in some sense, related to the galaxy gravitational potential.

While around half the galaxies in the Universe belong to groups and clusters, ETGs are less abundant in the nearby groups.
Galaxy groups are a transition between clusters and field from the hierarchical point of view.
They are physical entities disconnected from the cosmic expansion by mutual gravity attraction among their members,
which are packed into a volume of about a few Mpc wide.
Groups either hold galaxies before they are gravitationally engulfed by a cluster,
or may never be captured by the gravitational potential of a cluster.
Sparse groups can be gravitationally relaxed entities, especially when they hold an x-ray emitting intra-group gas.
Velocity dispersion of galaxies in a group is comparable with the velocity dispersion of stars inside a galaxy
(they span from $100$ to $500$ km\,s$^{-1}$).
Consequently, encounters between two or more galaxies in a group as well as the galaxy movement inside it
can play an important role for internal perturbations and perhaps galactic morphological mutation.

{\bf
For the radial gradients in stellar metallicity and age classic models by 
\citet{1974MNRAS.166..585L,1975MNRAS.173..671L} and \citet{1984ApJ...286..403C,1984ApJ...286..416C}
show that dissipative gravitational collapse of protogalaxies with no or slow rotation produce very deep negative metallicity gradients.
On the other hand, the two-phase formation process, studied by \citet{2010ApJ...725.2312O},
induces more realistic negative gradients in both metallicity and age.
}
Mergers and strong mutual interactions are supposed
to shallow these gradients by mixing the stellar populations and gas over the system
\citep{1996ApJ...464..641M,2006MNRAS.366..499D,2006A&A...457..809S,2006A&A...457..787S,2014A&A...563A..49S,
2014MNRAS.444.2005R}.
In fact, the gradients are strongly determined by the assembly history of the galaxy
\citep{2004MNRAS.347..740K},
and they are spatially dependent on the Star Formation History (SFH), which can also be affected by the environment.

An open question about galaxy formation is the roles played
by the progenitor mass and environment for the formation/evolution of massive ETGs.
This issue is the debate known as ``nature versus nurture'',  
\citep[e.g.][]{2007MNRAS.376.1445C, 2009A&A...503..379T, 2012MNRAS.427.1530M}.
Detailed characterization of stellar populations associated or not to the ionized gas
within the central region of ETGs from different nearby low-density regions
provides a lot of information about the internal SFH and 
radial gradients of simple stellar populations.
In this case, nearby massive ETGs ($z\leq0.05$) are excellent physical laboratories as they
can be observationally analysed in detail even with small aperture telescopes
that have good instrumentation.

The current work focuses on a sample of 8 nearby luminous ETGs, in which  
one is an isolated galaxy and 7 are in sparse groups. Basic data on each galaxy in the 
sample is presented in Table \ref{tabwww}. Six out of eight have companion galaxies that may characterize them as 
galaxies in pairs, according to our simple criterion 
(see next section), three of which are possibly in interaction. 
The sample also spans slow and fast rotators based on stellar 
kinematics measurements from the 
literature.

To homogeneously recover stellar population properties (age, [M/H], [Fe/H], and [$\alpha$/Fe]) over the sample ETGs,
we have applied the well-established spectral synthesis code {\scriptsize\,STARLIGHT} 
\citep{2004MNRAS.355..273C,2005MNRAS.358..363C,2006MNRAS.370..721M,2007MNRAS.375L..16C,2007MNRAS.381..263A}.
No previous work has analysed this set of ETGs based on a stellar population synthesis like we have done. 
Another work goal is obviously to add knowledge about formation and evolution
of massive $z=0$ ETGs from pairs and sparse groups, including interacting galaxies and LINERs.

This paper is organized as follows.
Section~\ref{amostra} has details about the sample selection.
In Section~\ref{obs_redu}, we present the spectroscopic data and data reduction.
The stellar population synthesis and the resulting properties are explored in Section~\ref{stellar}, 
including a detailed galaxy-by-galaxy analysis (stellar content spatially 
resolved, radial gradients of stellar population properties and chemical enrichment at observed regions).
In Section~\ref{discute}, all results are discussed for the ETGs sample as a whole. 
Finally, in Section~\ref{conclude}, we present our conclusions.

\begin{table*}
 \centering
  \caption{Main characteristics and parameters of sample galaxies. ENV means environment classification.
  NED means NASA/IPAC Extragalactic Database. $E(B-V)_{\rm G}$ is the
foreground $(B-V)$ colour excess due to our Galaxy.}
  \begin{footnotesize}
  \resizebox{\textwidth}{!}{%
  \scalebox{0.8}{
  \begin{tabular}{@{}lrrccccccccc@{}}
  \hline
Galaxy ID & Morph\,$^{[1]}$ &ENV\,$^{[NED]}$ & $cz_{\rm helio}$\,$^{[NED]}$ & $m_{B}$\,$^{[3]}$&$(B-V)_{\rm e}$\,$^{[1]}$ & $M_{B}$\,$^{[2]}$ &$E(B-V)_{\rm G}$\,$^{[4]}$  
&$r_{\rm e}$\,$^{[1]}$ &$r_{\rm 25}$\,$^{[1]}$&$\epsilon$\,$^{[1]}$&$\mu_{\rm e}$\,$^{[1]}$
\\
  &           &       & $(\rm km/s)$ &$(\rm mag)$&(\rm mag)& $(\rm mag)$&(arcsec)&(arcsec)& &($\rm mag$\,arcmin$^{-2}$)&($\rm mag$)\\
   \hline
  \noalign{\smallskip}
 IC\,5328 & E4        &P, G(4)    &3137 &12.27&0.97    &-21.00 &0.013    &22.2 &75.4 &0.3834&12.18\\
 NGC\,1052& E4        &wP, G(12)  &1510 &11.45&1.00    &-20.20 &0.023    &33.7 &90.6 &0.3082&12.15\\
 NGC\,1209& E6?       &wP, G(22)  &2600 &12.35&1.00    &-20.65 &0.033    &18.5 &72.0 &0.5214&11.83\\
 NGC\,5812& E0        &P, G(6)    &1970 &12.19&1.03    &-20.51 &0.076    &25.5 &64.1 &0.1290&12.33\\
 NGC\,6758& E+/cD?    &G(7)       &3404 &12.61&1.05    &-21.14 &0.058    &20.3 &67.2 &0.2238&12.29\\
 NGC\,6861& SA0$^{-s}$ &wP, G(17) &2829 &12.08&1.03    &-21.14 &0.048    &17.7 &84.6 &0.3543&11.46\\
 NGC\,7507& E0        &P, G(7)    &1566 &11.38&1.00    &-20.56 &0.044    &30.7 &82.6 &0.0228&11.90\\
 NGC\,7796& E+/cD     &Field      &3364 &12.44&1.00    &-20.93 &0.009    &21.2 &65.6 &0.1290&12.20\\
\hline
\noalign{\smallskip}
\end{tabular}
}
}
\begin{minipage}[l]{18cm}
{\it References:} [1] \citet{1991S&T....82Q.621D}; [2] \citet{2014A&A...570A..13M} ; [3]\citet{2000A&AS..146...19P};
[4] \citet{2011ApJ...737..103S};\\
{\it Notes:} In ENV column, we used P for pair with possible interaction, wP for weak pair, G for group (with 
the number of galaxies in parenthesis)\\ 
and Field for isolated galaxies. \\
\end{minipage}
\label{tabwww}
\end{footnotesize}
\end{table*}

\section {The early-type galaxies sample}
\label{amostra}

The observed sample of eight nearby luminous ETGs is composed of
IC\,5328, NGC\,1052, NGC\,1209, NGC\,5812, NGC\,6758, NGC\,6861, NGC\,7507, and NGC\,7796.
All of them are brighter than the absolute blue magnitude of the \lq\lq knee'' ($M_{B}^{\star}$)
of the global galaxy luminosity function at zero redshift
\citep{2003MNRAS.344..307L, 2003Ap&SS.285..175D}, and their $M_{B}$ covers almost 1\,mag.
Table~\ref{tabwww} shows the global properties of the sample galaxies.

These luminous ETGs were selected to produce a sample of galaxies under possible interaction with any other galaxy.
The criteria we adopted to verify pair and group membership of a target are:
{\bf (i) sparse group:} composed by galaxies with projected linear separation of $\leq400$\,kpc, and relative 
heliocentric radial velocity difference of $\leq500$\,km\,s$^{-1}$;
{\bf (ii) galaxy pair:} composed of galaxies with projected linear separation of $\leq100$\,kpc and heliocentric 
radial velocity difference of $\leq500$\,km\,s$^{-1}$;
{\bf (iii) interacting pair:} if besides the previous criterium, it shows some signs of tidal distortion.

To apply the above criteria, we used the Physical Companions tool of
NED\footnote{NASA/IPAC Extragalactic Database (NED)
operated by the Jet Propulsion Laboratory, California Institute of Technology,
under contract with the National Aeronautics and Space Administration; http://ned.ipac.caltech.edu/.}
to find the group members and interacting pairs upon the standard $\Lambda$CDM cosmology
(h$_{0}=0.73$, $\Omega_{\rm matter}=0.27$, and $\Omega_{\rm dark-energy}=0.73$).
We also take into account the spatial scale corrected by the cosmology.
The Nearby Optical Galaxy (NOG) database of \citet{2000ApJ...543..178G}
was also adopted as a cross check of the environment classification if possible,
because specifically IC\,5328, NGC\,7507, and NGC\,7796 are not catalogued by NOG.

IC\,5328, NGC\,5812, and NGC\,7507 are identified as galaxies in possible 
interacting pairs. NGC\,1052, NGC\,1209, and NGC\,6861 are galaxies in weak pairs, but for all of them, 
signs of interactions have been reported in the past \citep{1986AJ.....91..791V, 2000AJ....120.1946S, 
2009AJ....138.1417T, 2012MNRAS.419..687R, 2009AJ....138.1417T,2012MNRAS.419..687R, 
2015MNRAS.449..612E, 2010ApJ...711.1316M}, although they currently do not have a close companion.
Except for NGC\,7796 (an isolated field galaxy), all the other galaxies in the sample 
are members of sparse groups.

Effective surface brightness $\mu_{\rm e}$, integrated visual magnitude,
and structural details, such as effective radius $r_{\rm e}$, diameter at 25\,mag\,arcmin$^{-2}$,
and apparent ellipticity $\epsilon$ (listed in Table~\ref{tabwww}),
were also taken into account to fine tune selection of the targets 
to make the long slit spectroscopic observations possible with the chosen instrumentation (see next section).
Short description of each galaxy is presented in Appendix A, showing their main relevant characteristics.

Three sample ellipticals (NGC\,1052, NGC\,1209, and NGC\,6758) are classified as LINER 
hosts \citep{1997ApJS..112..315H, 2010A&A...519A..40A, 2011A&A...528A..10P}. Stellar kinematics 
have been measured in all sample ETGs by a myriad of previous works
\citep{1985MNRAS.216..429L, 1989Msngr..56...31L, 1998A&AS..130..267L, 1998AJ....116....1D, 2005ApJ...621..673T, 2007MNRAS.381.1711I,
2007A&A...464..853L, 2013MNRAS.431..440F, 2013ApJS..207...19B}. 
The kinematical measurements in the literature enabled us to split the sample ETGs into two categories:
(i) slow rotators (IC\,5328, NGC\,7507, and NGC\,7796), and (ii) fast rotators (NGC\,1052, NGC\,1209, NGC\,5812, NGC\,6758 and NGC\,6861). 
We dynamically classified them based on the normalized stellar kinematical anisotropy parameter
{\scriptsize\,$(V_{rot}^{max}/\sigma_{\rm v}^{0})^{\ast}$},
adopting the transition between the two categories at {\scriptsize\,$(V_{rot}^{max}/\sigma_{\rm v}^{0})^{\ast}=0.50$},
such that {\scriptsize\,$(V_{rot}^{max}/\sigma_{\rm v}^{0})^{\ast}<0.50$} for slow rotators
and {\scriptsize\,$(V_{rot}^{max}/\sigma_{\rm v}^{0})^{\ast}\geq0.50$} for fast rotators.

Radial gradients of stellar populations have been empirically quantified in 6 out of 8 ETGs in our sample
\citep{2007A&A...469...89M,2007A&A...463..455A,2005MNRAS.358..419P,2001MNRAS.324.1087R,2005A&A...433..497R}.
Four of them were analysed by a single work \citep{2007A&A...463..455A}.
IC\,5328 and NGC\,7507 do not have any spatial analysis of stellar population properties 
(age, metallicity and $\alpha$-iron ratio).
\citet{2007A&A...463..455A} measured luminosity-weighted averages of age, $Z$, and [$\alpha$/Fe]
over 7 radial apertures in NGC\,1052, NGC\,1209, NGC\,5812, and NGC\,6758
by adopting long slit spectra mainly across photometric major axis
and performing a comparison of Lick indexes against simple stellar population (SSP) model predictions.
They reached radial distances up to half of effective radius corrected by the ellipticity.
However, no previous work had applied a detailed stellar population synthesis
based on a homogeneous approach that we have done for this whole set of ETGs.

\section {Observation and data reduction}
\label{obs_redu}

Long slit spectroscopy observations of the sample galaxies were carried out
with a Boller \& Chivens spectrograph at the Cassegrain focus of the 1.60-m telescope
of Observat\'orio do Pico dos Dias (OPD)
\footnote{$^{*}$Based on observations carried out at the Observat\'orio do Pico dos Dias,
which is operated by the Laborat\'orio Nacional de Astrof\'{\i}sica (LNA, MCTIC, Brazil).},
which is operated by Laborat\'orio Nacional de Astrof\'{\i}sica (LNA).

\begin{figure*}
\centering
\includegraphics[angle=0,width=0.331\textwidth]{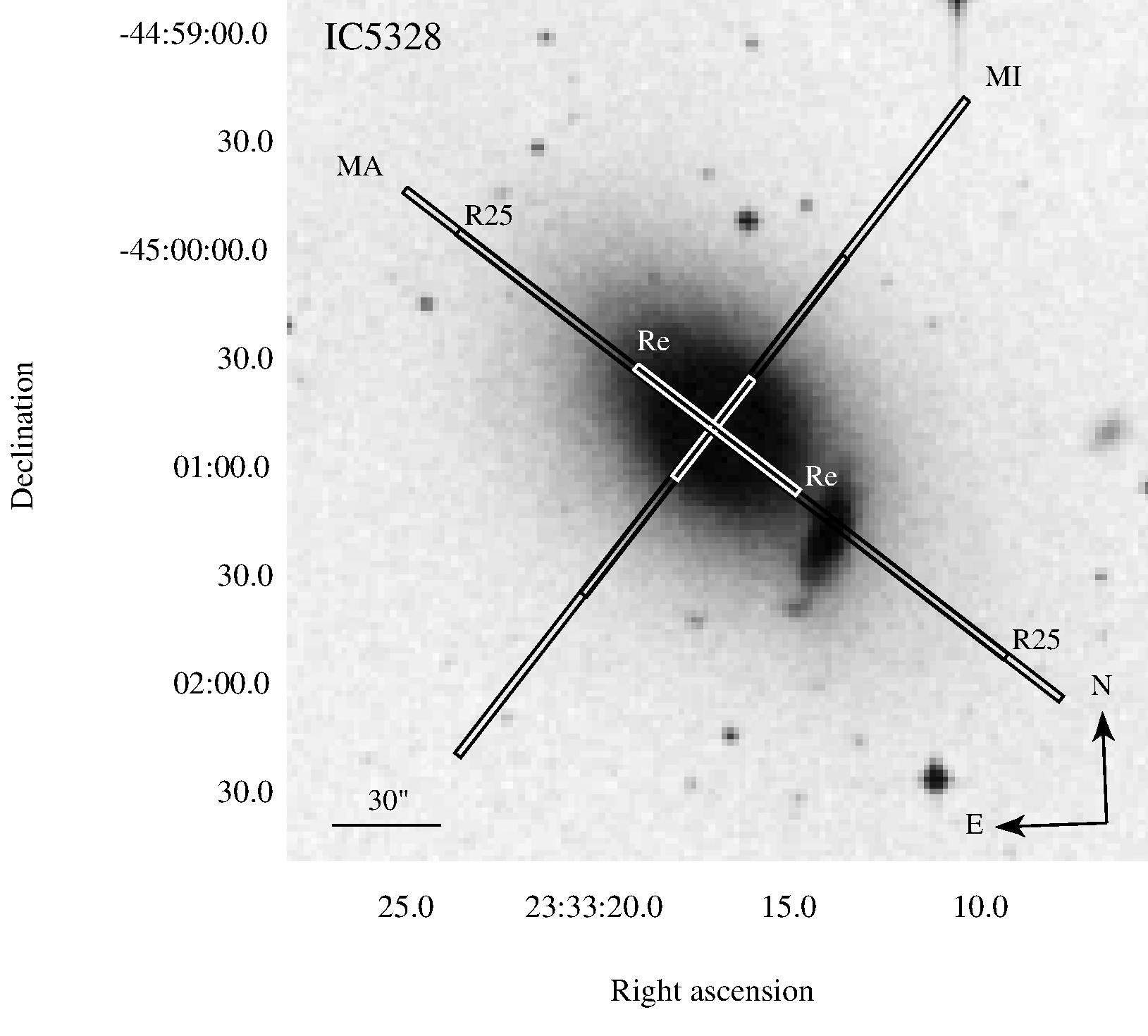}
\includegraphics[angle=0,width=0.331\textwidth]{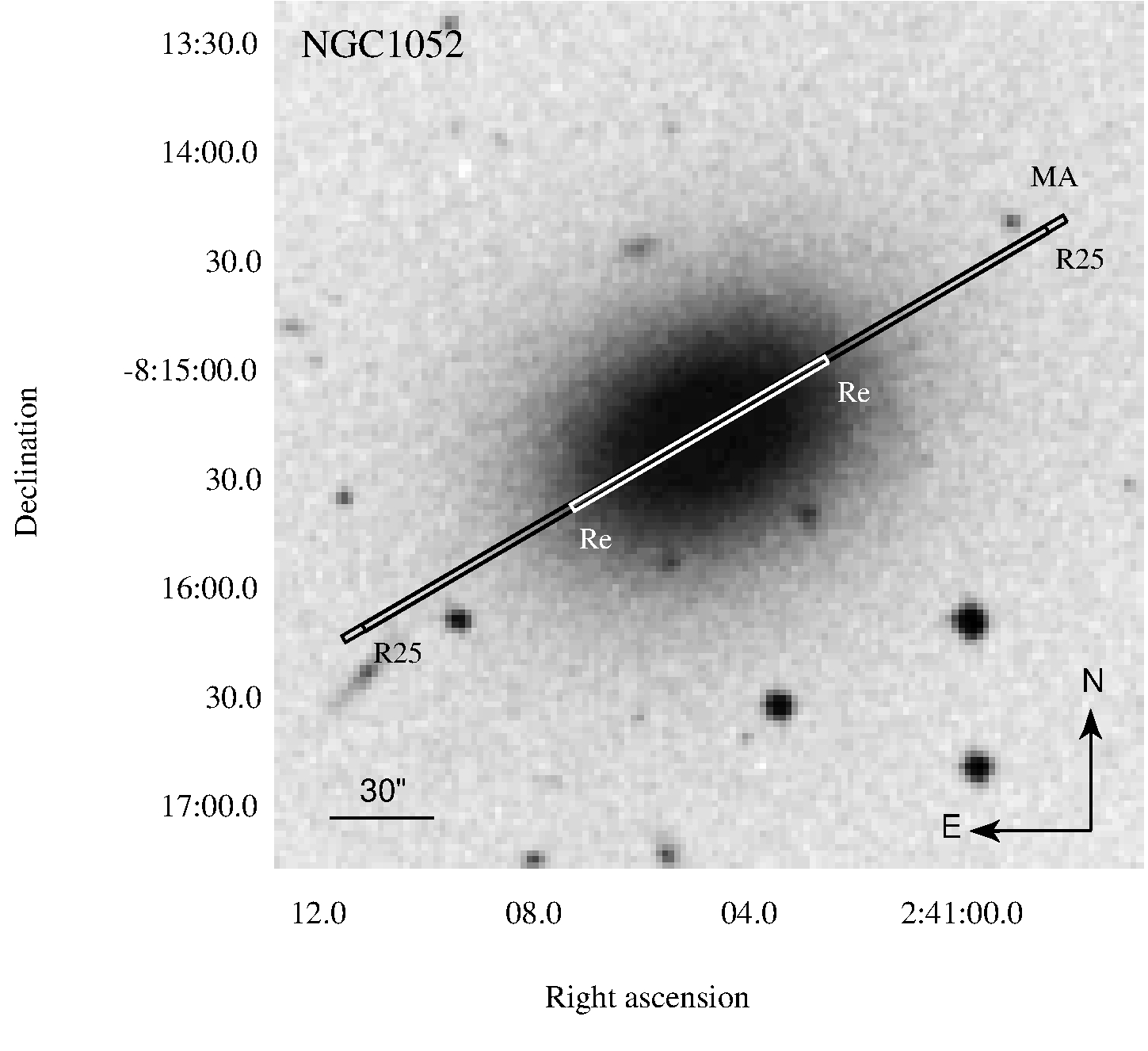}
\includegraphics[angle=0,width=0.331\textwidth]{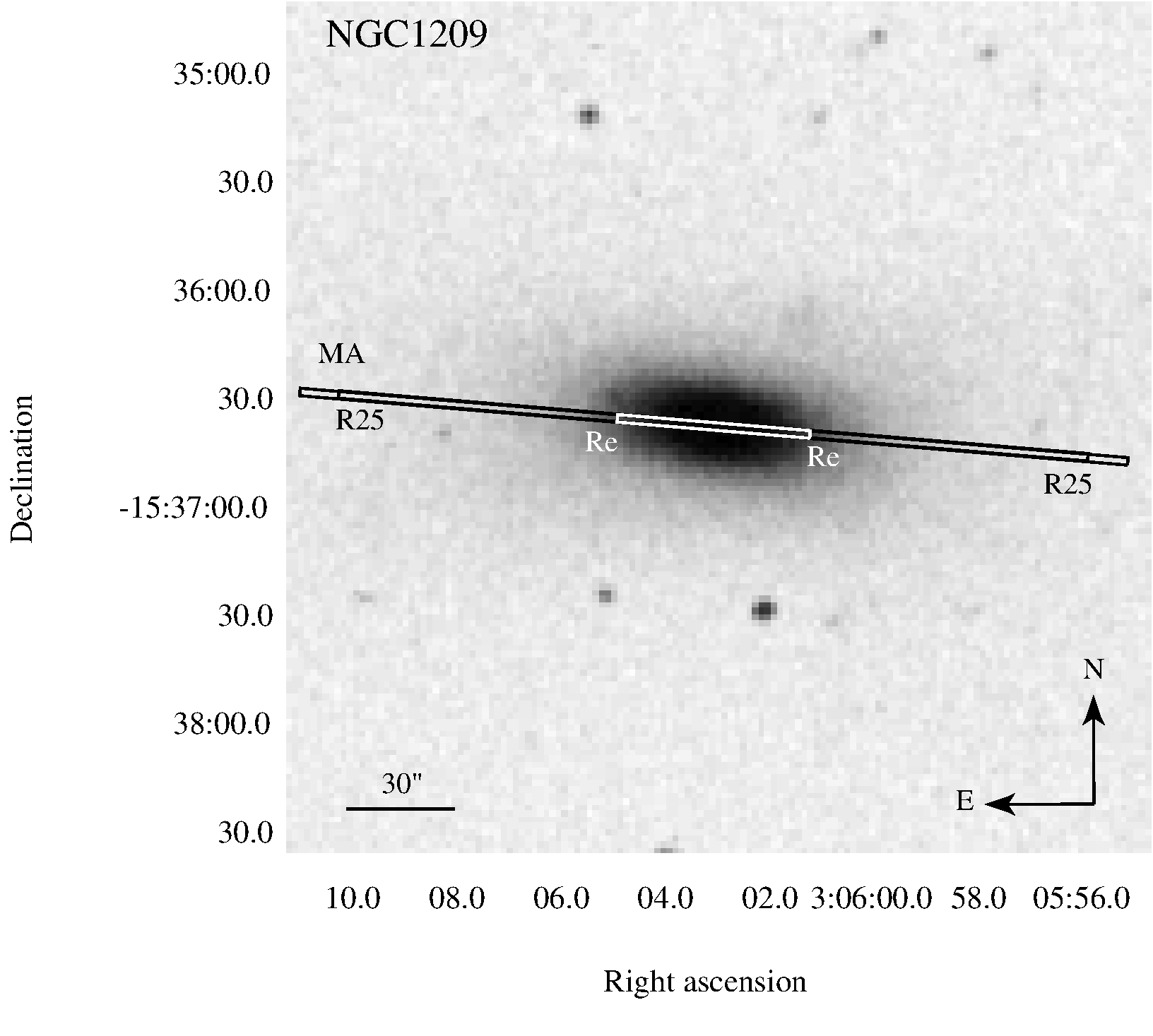}
\includegraphics[angle=0,width=0.331\textwidth]{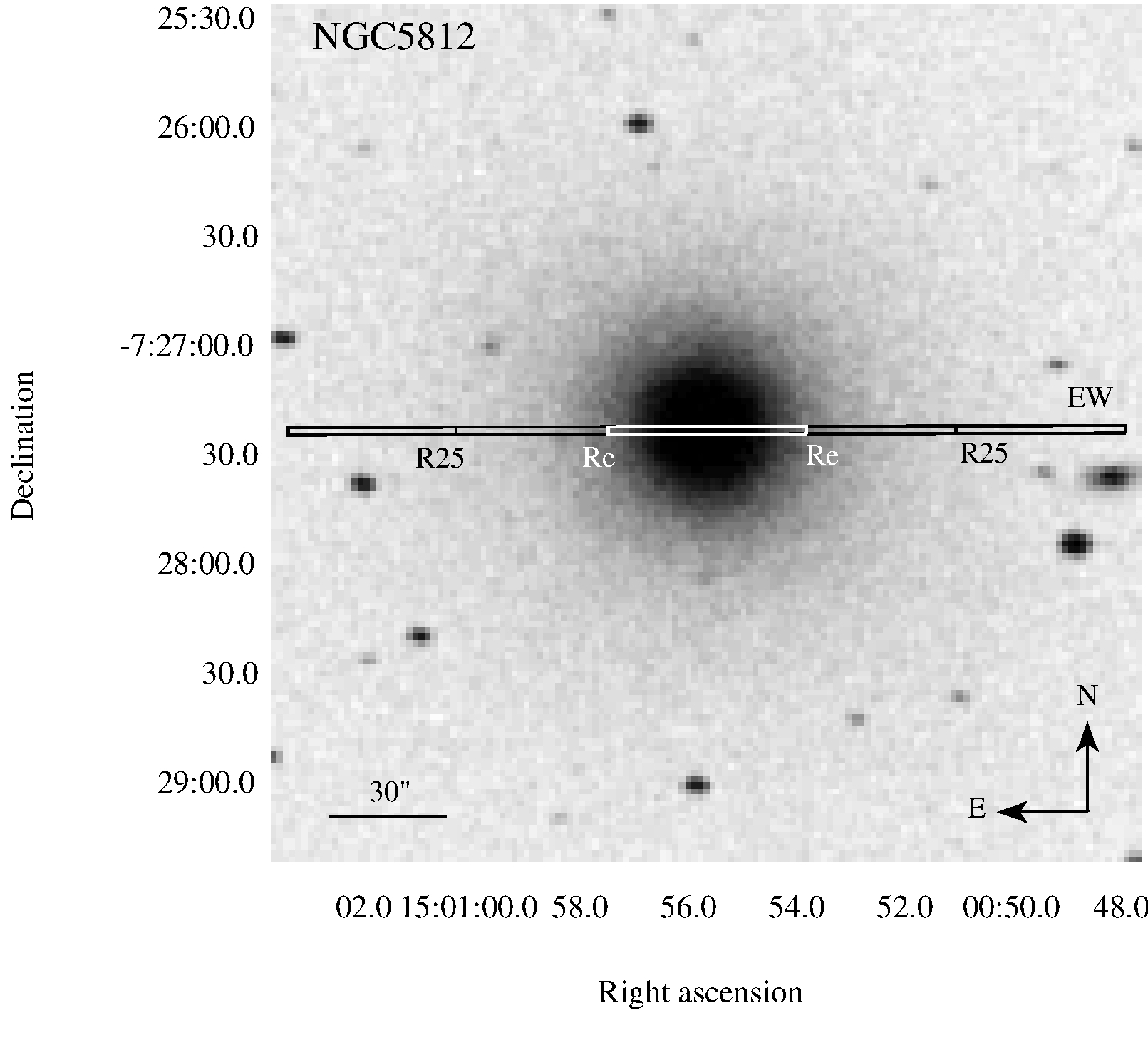}
\includegraphics[angle=0,width=0.331\textwidth]{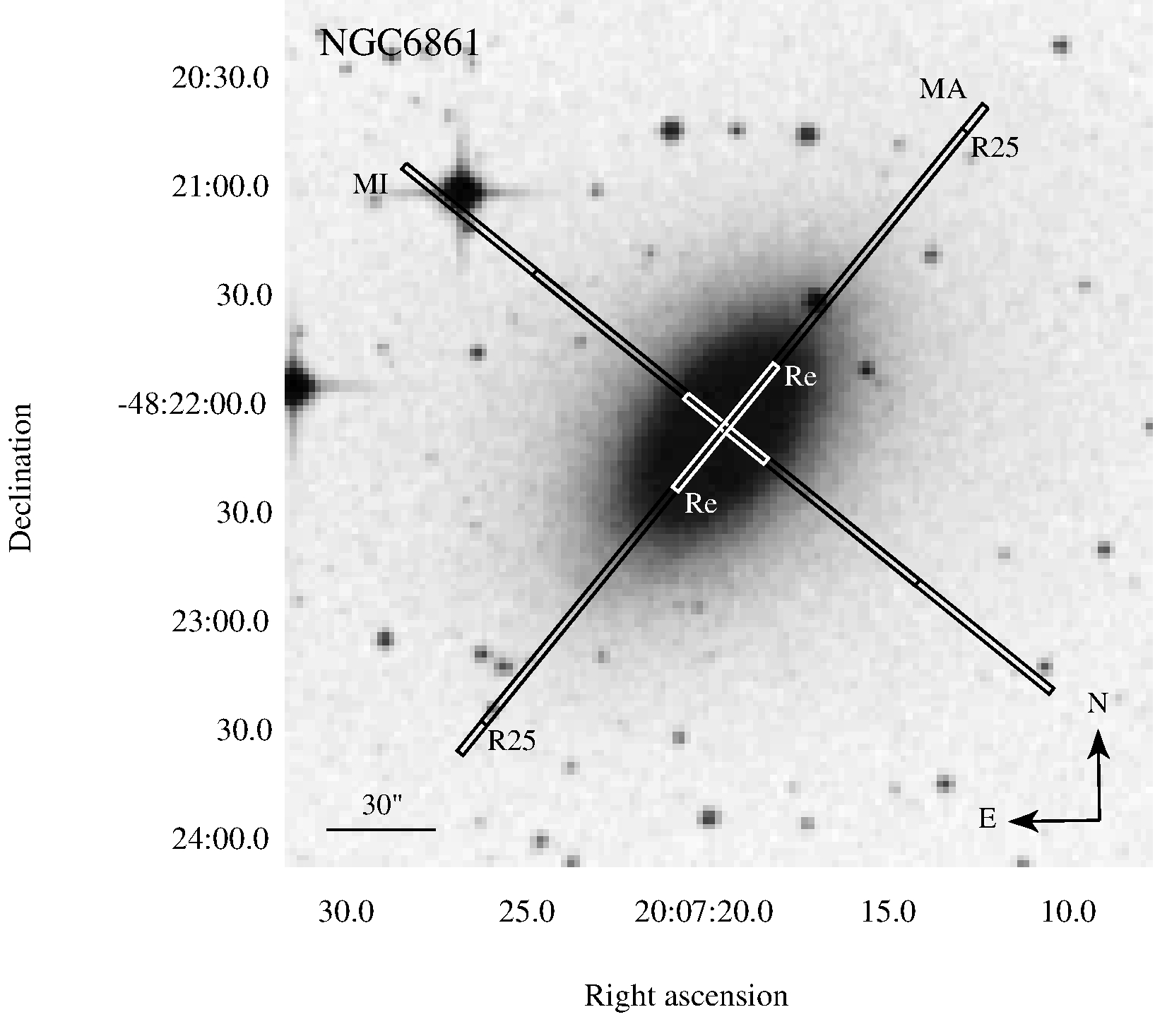}
\includegraphics[angle=0,width=0.331\textwidth]{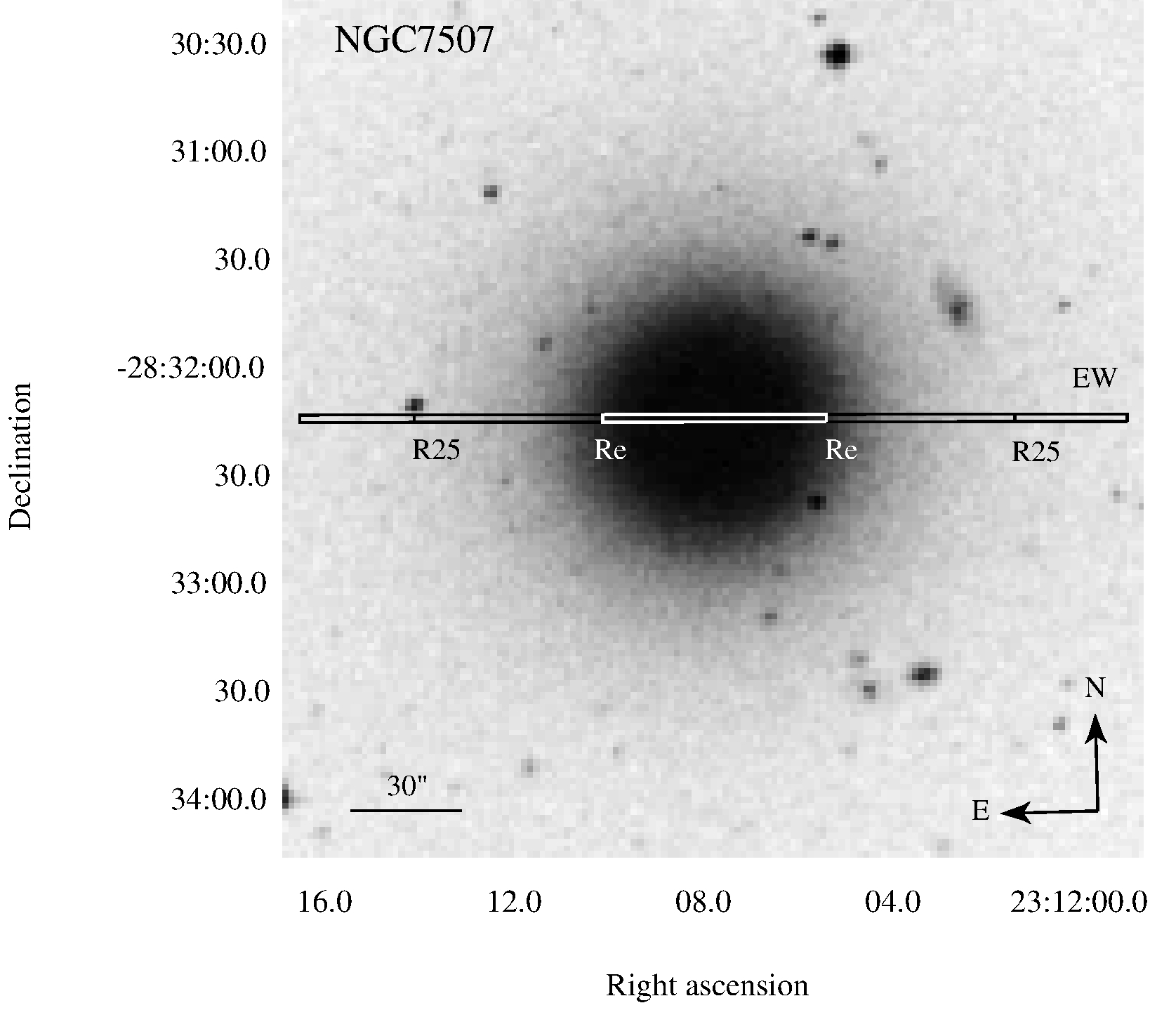}
\includegraphics[angle=0,width=0.331\textwidth]{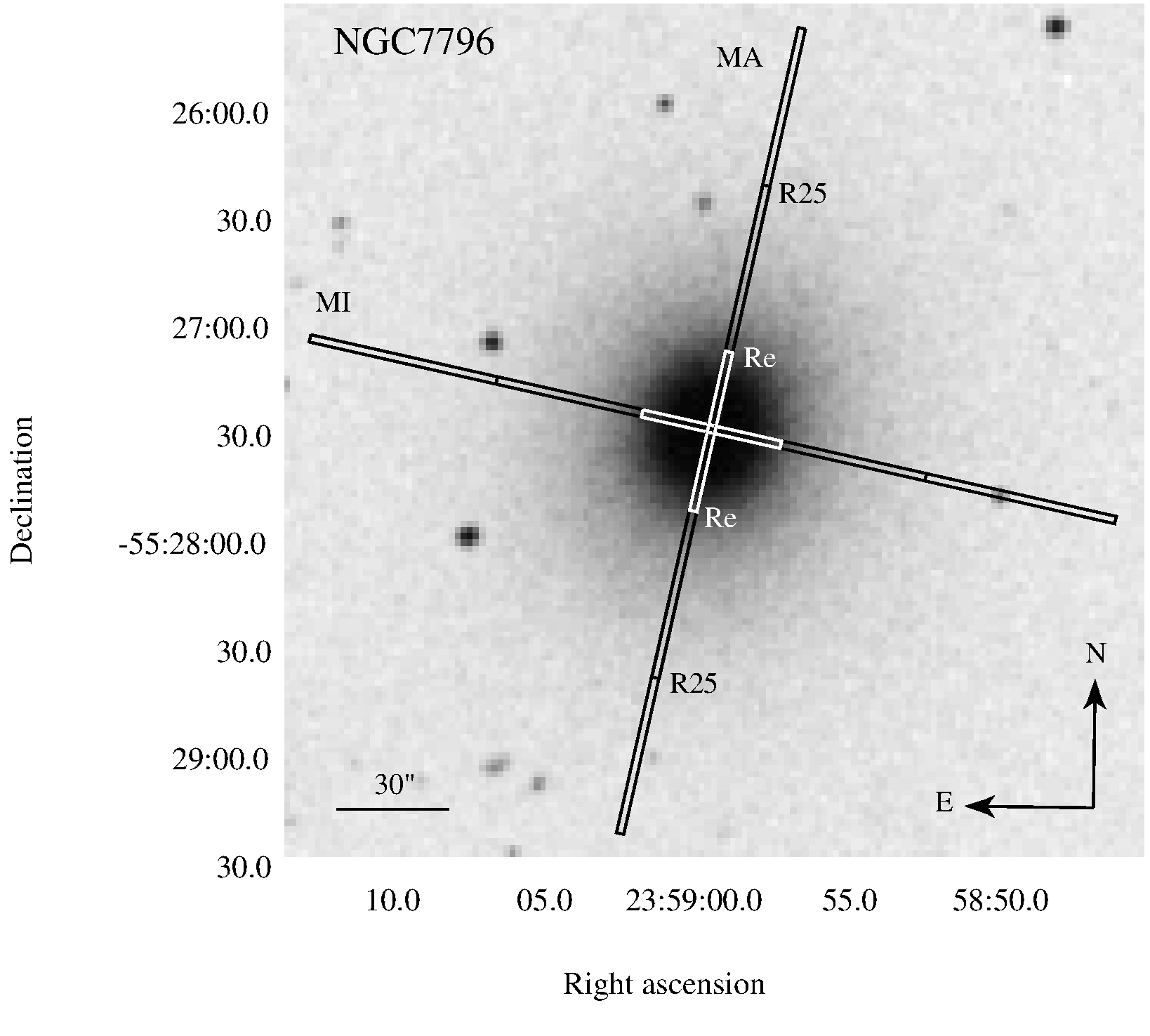}
\caption{Images of our sample of galaxies with the observed slit positions. 
The extensions of $R_{\rm e}$ and $R_{\rm 25}$ (both corrected by galaxy apparent ellipticity) 
are denoted in each optical image (IIIaJ photographic band centred at $\lambda$4680{\AA}. 
The Digitized Sky Survey from The Space Telescope Science Institute as provided by NED, NASA/IPAC Extragalactic Database).
MA means major axis direction, MI minor axis direction, and EW east-west direction.
}
\label{sample}
\end{figure*}

Stellar and galaxy spectra in the range $\lambda\lambda4320-6360$\,{\AA}
were collected with two observing runs (one night in 1998 Oct and two nights in 1999 Aug)
with a sampling of 2.01\,{\AA}\,pixel$^{-1}$ by using the CCD\,\#106 with $1024\times1024$\,pixels
(square pixel: $24\times24$\,$\mu$m) and a grating of $600$\,lines\,mm$^{-1}$ (spectrograph resolving power $R=1800$).
The spectrograph slit length is 230\,arcsec and its width was fixed to 2.08\,arcsec for all exposures in both observing runs,
i.e. slightly above the maximum value of atmospheric seeing of all observing nights
(seeing denoted by FWHM$_{\rm seeing}$),
which in fact varied from 1.76 up to 2.00\,arcsec.
FWHM$_{\rm seeing}$ is estimated across the spatial axis of $2-\rm D$ raw spectroscopic images of standard stars
(average spectroscopic flux profile at $\lambda\lambda4320-6360$\,{\AA}),
whose exposures were taken across each night.
The exposure times for sample galaxies were divided in 2 or 3 integrations of 30\,min each
to remove cosmic ray effects and to enhance the spectrum signal-to-noise ratio
(except for NGC\,7507, for which we took 4 integrations of 20\,min each).
The spatial scale is 1.092\,arcsec\,pixel$^{-1}$.
The instrumental spectrum resolution, estimated
from the broadening of isolated emission lines in the wavelength calibration spectra,
was FWHM$_{\rm inst}=4.96\pm0.08$\,{\AA} on the 1998 Oct night,
$5.07\pm0.20$\,{\AA} on the first night of 1999 Aug run, and $5.35\pm0.11$\,{\AA} on the second night of 1999 Aug run.
The average instrumental broadening of the three nights corresponds
to $\sigma_{inst}=126.5\pm5.0$\,km\,s$^{-1}$ at 5155\,{\AA}.
Figure~\ref{sample} shows the slit position(s) for each galaxy.
Table~\ref{tabw} gives the journal of observations.
Up to three spectrophotometric standard stars were observed each night at different airmasses
to provide acceptable flux calibration for the stellar population synthesis.

\begin{table*}[!tb]
 \centering
  \caption{Journal of observations (convention for the slit orientation according to Figure~\ref{sample}).} 
\label{tabw}
\scalebox{0.7}{
\begin{tabular}{@{}llllcclc@{}}
\hline
\noalign{\smallskip}
Galaxy   & Date   & \multicolumn{1}{c}{PA$_{\rm slit}$} & \multicolumn{1}{c}{Orientation }  & Exposure   & FWHM$_{\rm seeing}$  & Spatial        & $R_{\rm e}$ \\
   &    &  & \multicolumn{1}{c}{ Slit}  &  Time  &   &   Scale      &  \\
  
\noalign{\smallskip}
            &        & \multicolumn{1}{c}{($\degr$)}&          & \multicolumn{1}{c}{(s)}      & \multicolumn{1}{c}{($\arcsec$)}          & \multicolumn{1}{c}{(pc\,pix$^{-1}$)} & \multicolumn{1}{c}{(kpc)} \\
\noalign{\smallskip}
\hline
\noalign{\smallskip}
IC\,5328 & 1999 August 11 &  40&MA & 3$\times$1800 & 1.91 & 209 & 5.40 \\
         & 1999 August 11 & 130&MI & 2$\times$1800 & 1.79 & --- & 3.33 \\
NGC\,1052& 1999 August 12 & 120&MA & 3$\times$1800 & 2.00 &  93 & 3.44 \\
NGC\,1209& 1999 August 11 &  85&MA & 2$\times$1800 & 1.91 & 175 & 4.28 \\
NGC\,5812& 1999 August 12 &  90&EW & 2$\times$1800 & 1.79 & 156 & 3.65 \\
NGC\,6758& 1999 August 11 & 121&MA & 3$\times$1800 & 1.76 & 239 & 5.05 \\
         & 1999 August 11 & 211&MI & 2$\times$1800 & 1.91 & --- & 3.92 \\
NGC\,6861& 1999 August 12 & 140&MA & 3$\times$1800 & 1.79 & 193 & 3.56 \\
         & 1999 August 12 & 230&MI & 2$\times$1800 & 1.70 & --- & 2.52 \\
NGC\,7507& 1998 August 23 &  90&EW & 4$\times$1200 & 1.91 &  92 & 2.58 \\
NGC\,7796& 1999 August 12 & 168&MA & 3$\times$1800 & 1.91 & 229 & 5.77 \\
         & 1999 August 12 & 258&MI & 2$\times$1800 & 2.00 & --- & 4.15 \\
\hline
\noalign{\smallskip}
\end{tabular}
}
\end{table*}

The spectroscopic data reduction was processed
using the {\scriptsize\,NOAO} (National Optical Astronomical Observatory)
package of {\scriptsize\,IRAF}\footnote{Image Reduction and Analysis Facility,
http://iraf.noao.edu} following the usual procedures:
trimming of the $2-\rm D$ raw images,
bias subtraction,
dark subtraction for long integrations,
flat field correction,
extraction of the stellar spectra,
extraction of aperture spectra for each galaxy integration across the flux profile
with an adequate sky level subtraction,
wavelength calibration of the $1-\rm D$ spectra,
flux calibration of each galaxy aperture spectrum,
and averaging of each galaxy aperture spectrum with elimination of spurious noises.

The galaxy aperture spectra across the slit direction for each integration
was extracted from each side of the brightness profile centre at different distances with different widths
to keep decreasing the spectral signal-to-noise ratio in a low rate
over the extracted spectra across the light profile (or radial distance towards the galaxy centre).
In fact, the radial distances and widths of aperture spectra
depend on the surface brightness distribution across the slit direction
that is essentially determined by the galaxy morphology.
We made a set of aperture spectrum extractions
to proceed with the stellar population synthesis (see Sec.~\ref{stellar}).
We assigned an equal width for the nuclear and its two adjacent apertures
(extraction width equal to 2.0\,arcsec, which is the highest value of FWHM$_{\rm seeing}$ of all nights),
such that the other apertures have progressively greater widths
(see Table~\ref{tabw} and Table~\ref{tcorr1} for IC\,5328 as an example).
Besides the extraction of non-concentric aperture spectra at each side of the light profile,
we also extracted spectrum inside one effective radius distance,
{\bf $R_{\rm e}$\footnote{{\scriptsize\,$R_{\rm e}$ = $r_{\rm e}$(1-$\epsilon$)$^{-1/2}$} if major axis
and {\scriptsize\,$R_{\rm e}$ = $r_{\rm e}$(1-$\epsilon$)$^{1/2}$} if minor axis,
where $r_{\rm e}$ is the standard effective radius
and $\epsilon$ is the apparent ellipticity,
which has been corrected by the galaxy apparent ellipticity $\epsilon$ over the slit direction}}
across the major axis direction
(excepts for NGC\,5812 and NGC\,7507, which were done in the EW direction).
The sky windows of spectral extractions
were defined at radial distances greater than $R_{\rm 25}$.
The relative error on flux calibration is about a few percent in the flux scale
(the maximum value was 6\,per\,cent on the night of Aug 11) 
making the aperture spectra suitable for stellar population synthesis.

To proceed with the stellar population synthesis,
we had to prepare each aperture spectrum by following two more steps:
(i) de-reddening taking into account the respective Galactic line of sight extinction denoted by $E(B-V)_{\rm G}$ (foreground extinction),
and (ii) close transformation onto the wavelength-of-rest.
Galactic extinctions and heliocentric velocities were those provided
by the NED database. 
The second step adopted the catalogue heliocentric velocity as a first approximation
associated to the heliocentric velocity correction (estimated inside  {\scriptsize\,IRAF}),
since the stellar population synthesis approach also provides a radial velocity correction,
fine tuning the wavelength-of-rest calibration.

\section{Stellar population synthesis}
\label{stellar}

\begin{figure*}
\centering
\includegraphics*[angle=-90,width=1.65\columnwidth]{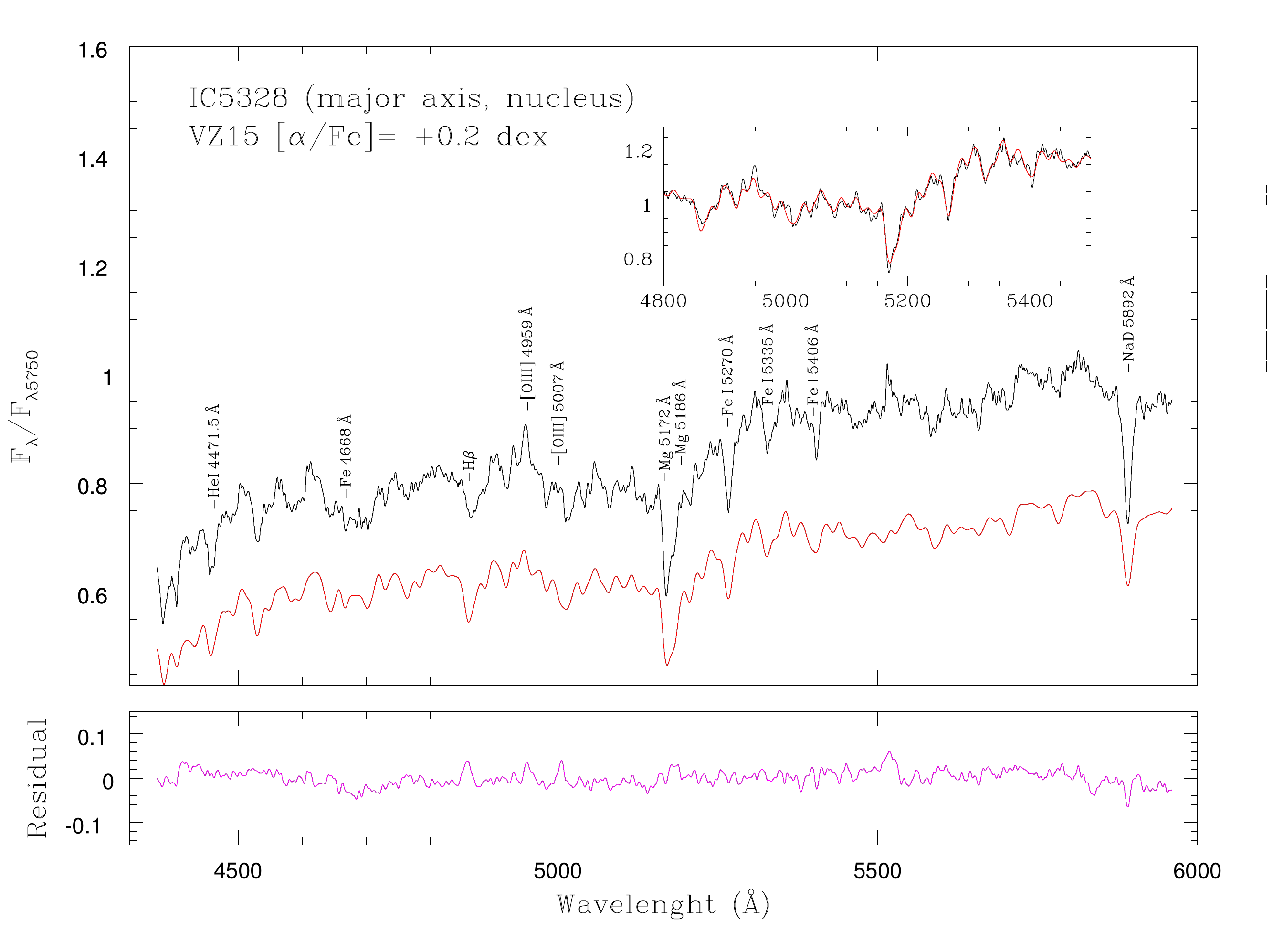}
\includegraphics*[angle=-90,width=1.65\columnwidth]{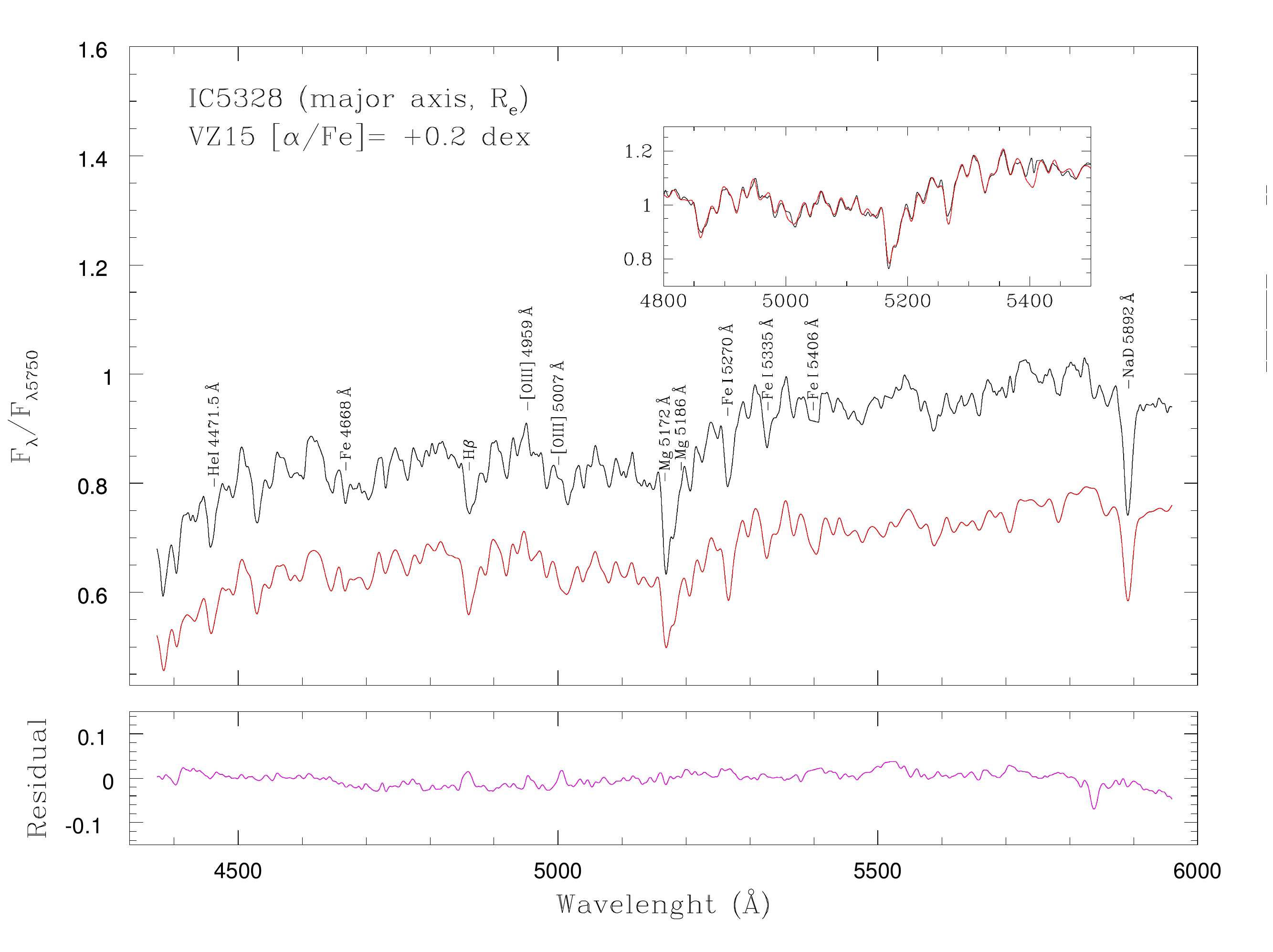}
\caption{Stellar population synthesis for the nuclear and one $R_{\rm e}$ regions across the major axis of IC\,5328 
(left and right plots, respectively).
In the top panel of each plot, the observed spectrum (in black) and the synthesized spectrum (in red) are shown.
In the bottom panel, we present the flux difference between observed and synthesized spectra.
The main absorption and emission features are identified in both top panels.
The sub-panels illustrate the spectral fit of stellar population synthesis
carried out over the spectral range used to derive [$\alpha$/Fe] ($\lambda\lambda4800-5500$\,{\AA}).}
\label{sintese_002}
\end{figure*}

The study of stellar populations in galaxies must enable quantification or estimation of important parameters
such as age, metallicity (global $Z$ and [M/H], or [Fe/H] instead), and [$\alpha$/Fe] abundance ratio
that can be represented by mean values or their distributions over specific key regions of those stellar systems.

We applied the stellar population synthesis approach
to obtain the spatial distribution of stellar populations in every sample galaxy.
The {\scriptsize\,STARLIGHT} code 
\citep{2004MNRAS.355..273C,2005MNRAS.358..363C,2006MNRAS.370..721M,2007MNRAS.375L..16C,2007MNRAS.381..263A} was adopted for this purpose.
In short, {\scriptsize\,STARLIGHT} tries to fit the observed integrated spectrum ($O_{\lambda}$) of a composite stellar system
through a linear combination of simple stellar population spectra.

We adopted the set of semi-empirical SSP models from \citet{2015MNRAS.449.1177V}, 
with the Kroupa Universal initial mass function ($\Gamma_{b}=1.30$), 
ranging 18 ages ($t=0.1$, $0.2$, $0.3$, $0.5$, $0.7$, $1$, $2$, $3$, $4$, $5$, $6$, $7$, $8$, $9$, $10$, $11$, $12$, and $13$\,Gyr),
six metallicities ([M/H]$=-0.96$, $-0.66$, $-0.35$, $0.06$, $0.26$, and $0.40$)
and two $\alpha$/Fe abundance ratios ($0.0$ and $+0.4$\,dex).
The spectral energy distributions (SEDs) of these SSPs models have reliable flux-calibrated response
($F_{\lambda}/L_{\odot}$\,{\AA}$^{-1}$\,M$_{\odot}^{-1}$ unity, where $L_{\odot}=3.826\times10^{33}$\,erg\,s$^{-1}$).
Their mid-resolution spectra (constant FWHM$\,=\,2.51$\,{\AA} or $\sigma=64$\,km\,s$^{-1}$ at $\lambda5000$\,{\AA})
cover the wavelength range $\lambda\lambda3540.5-7409.6$\,{\AA} under a sampling of 0.9\,{\AA}.

The synthetic spectrum \textit{M}$_{\lambda}$ is solved by the {\scriptsize\,STARLIGHT} code according to the following equation:
\begin{equation}
\label{my}
\centering
M_\lambda = M_{\lambda_{0}} \left[\sum_{j=1}^{N_\star} \vec{x}_j b_j,_{\lambda} r_{\lambda} \right] \otimes G(\nu_{\star},\sigma_{\star}),
\end{equation}
where b$_{j}$,$_{\lambda}$ is the reddened spectrum of the $j^{\rm th}$ SSP model that is flux normalized at $\lambda_{0}$\,=\,5750\,{\AA};
$r_\lambda\equiv10^{-0.4(A_{\lambda}-A_{\lambda_{0}})}$ is the extinction term;
$\otimes$ represents the convolution operator;
G$(\nu_{\star},\sigma_{\star})$ is the Gaussian line of sight velocity distribution (LOSVD) centred at velocity $\nu_{\star}$
and with a dispersion $\sigma_{\star}$;
\textit{M}$_{\lambda_{0}}$ is the synthetic flux at that reference wavelength;
and {\bf \textit{x}} is the stellar population vector.
The stellar population vector {\bf \textit{x}} represents the flux fractional contribution of SSPs models at $\lambda_{0}$
that are distributed in terms of age and metallicity ($t_{j}$, $z_{j}$).
The population vector can be also expressed as a function of the SSP mass fractional contribution and is designed by the
vector {\bf \textit{m}}.

Every observed galaxy aperture spectrum $\textit{O}_{\lambda}$ is foreground de-reddened and in wavelength-of-rest.
The resulting model spectrum \textit{M}$_{\lambda}$
and residual spectrum ($O_{\lambda}$ - $M_{\lambda}$) are also in wavelength-of-rest.
All of them are flux normalized at $\lambda5750$\,{\AA}.
\textit{M}$_{\lambda}$ and ($O_{\lambda}$ - $M_{\lambda}$)
are not corrected by the galaxy intrinsic extinction,
even though $A_{V}$ is estimated by the code for each galaxy aperture.
The synthetic spectrum is broadened by a Gaussian function
taking into account a galaxy line-of-sight velocity dispersion
as estimated by the code for every aperture spectrum.

The intrinsic galactic reddening of every extracted aperture is estimated by the {\scriptsize\,STARLIGHT} code itself.
The extinction law by \citet{1989ApJ...345..245C} is adopted
(taking into account $A_{V}=R_{V}\,E(B-V)$ where $R_{V}=3.1$),
because we are dealing with early-type galaxies.
See results in Table~\ref{tcorr1} specifically for IC\,5328 (major axis long slit spectra only),
and in the online table for the minor axis apertures of IC\,5328 and other galaxies.

The {\scriptsize\,STARLIGHT} code provides a vector of stellar population contributions over the whole adopted SSP model grid.
However, according to \citet{2005MNRAS.358..363C} the individual components  {\bf \textit{x$_{j}$}} are very uncertain.
To troubleshoot this limitation, \citet{2005MNRAS.358..363C} proposed that the resulting SSP vector should be re-binned in age.
In the current work, every population vector in flux contribution is represented by the following age bins or components:
Young Population, $f_{y}$ ($0.1\times10^{9}<\,t\,\leq1\times10^{9}$ years);
Intermediate-young Population, $f_{iy}$ ($1\times10^{9}<\,t\,\leq4\times10^{9}$ years);
Intermediate-old Population, $f_{io}$ ($4\times10^{9}<\,t\,\leq8\times10^{9}$ years);
and Old Population, $f_{o}$ ($8\times10^{9}<\,t\,\leq13\times10^{9}$ years), which have uncertainties lower for signal-to-noise $\rm S/N\geq10$
(spectral signal-to-noise ratio S/N was estimated in $\lambda\lambda5800-5850$\,{\AA}).
The same age components are employed for the SSP mass fractional contribution {\bf \textit{m$_{j}$}} ($m_{y}$, $m_{iy}$, $m_{io}$, $m_{o}$).
The quality of the resulting fit is quantified by the parameters $\chi^{2}$ and $adev$ (average deviation in flux).
The latter gives the relative mean deviation
$|(\textit{O}_{\lambda}-\textit{M}_\lambda)|/\textit{O}_{\lambda}$ over all 
considered wavelengths (spectral dispersion scale binned to 1 {\AA} sampling as {\scriptsize\,STARLIGHT} works), 
between the observed and model spectra.

Another important issue is that a stellar population synthesis code (including {\scriptsize\,STARLIGHT})
does not usually provide any estimate of uncertainties for the resulting standard solution
such as the luminosity-weighted average age, [M/H], [Fe/H], and [$\alpha$/Fe] and their distributions (SSP output vectors).
To solve this, the internal uncertainties of the parameter averages are estimated as a function of the spectrum quality denoted by
the spectral signal-to-noise ratio S/N.
We performed 20 levels of random flux perturbations over every single galaxy in one $R_{\rm e}$ aperture spectrum,
ran the stellar population synthesis for each perturbed spectrum,
and calculated differences between the new resulting parameters and standard ones.

\subsection{Spectral synthesis results}
\label{stellar2}

\begin{table*}[!tb]
\caption{Stellar population synthesis results for all aperture spectra across the major axis of IC\,5328.
In the last row, we present the resulting SSP flux and mass fractions for the one $R_{\rm e}$ aperture (age re-binned as well).
The results of the other seven sample galaxies and minor axis analysis of IC\,5328
are available in the electronic version only.}
\scalebox{0.77}{
\begin{tabular}{lllllllllllclllll}
\hline
$R/R_{\rm e}$&Area&S/N&$f_{y}$\,(\%)&$f_{iy}$\,(\%)&$f_{io}$\,(\%)&$f_{o}$\,(\%)&$m_{y}$\,(\%)&$m_{iy}$\,(\%)&$m_{io}$\,(\%)&$m_{o}$\,(\%)
&log[$\rm age$]&
${\rm [M/H]}$ &$\chi^{2}$&$adev$&$A_{V}$\,\\
&(arcsec$^{2}$)&&&&&&&&&&(yr)&(dex)&&&(mag)\\
 \hline
                            &     \multicolumn{12}{c}{IC\,5328\,MA} \\
\noalign{\smallskip}
-0.41SW  &23.81 &29 &9.1    &0.0   &0.0   &   89.5  &   0.4  &   0.0  &   0.0  &   99.6  &10.06& 0.19 &   1.1  &  2.32  &  0.28 \\
-0.25SW  &14.73 &48 &6.5    &0.0   &5.9   &   87.6  &   0.3  &   0.0  &   2.1  &   97.6  &10.04& 0.20 &   1.2  &  1.95  &  0.26 \\
-0.15SW  &8.74  &73 &6.9    &0.0   &0.0   &   92.9  &   0.4  &   0.0  &   0.0  &   99.6  &10.04& 0.26 &   1.4  &  1.65  &  0.28 \\
-0.07SW  &4.16  &79 &6.4    &2.2   &0.0   &   90.3  &   0.3  &   0.3  &   0.0  &   99.4  &10.06& 0.28 &   1.9  &  1.42  &  0.36 \\
 0.00    &4.16  &92 &6.3    &0.0   &0.0   &   92.3  &   0.3  &   0.0  &   0.0  &   99.7  &10.07& 0.37 &   2.3  &  1.34  &  0.49 \\
 0.07NE  &4.16  &76 &0.0    &1.8   &0.0   &   97.3  &   0.0  &   0.4  &   0.0  &   99.6  &10.09& 0.22 &   2.0  &  1.41  &  0.25 \\
 0.15NE  &8.74  &78 &0.0    &33.1  &3.4   &   62.9  &   0.0  &   20.3 &   1.4  &   78.4  &09.98& 0.30 &   1.5  &  1.69  &  0.08 \\
 0.25NE  &14.73 &66 &9.3    &0.0   &0.0   &   90.8  &   0.3  &   0.0  &   0.0  &   99.7  &10.07& 0.22 &   1.1  &  1.84  &  0.22 \\
 0.41NE  &23.81 &26 &7.5    &0.0   &0.0   &   94.1  &   0.6  &   0.0  &   0.0  &   99.4  &10.09& 0.16 &   1.1  &  2.20  &  0.18 \\
 1.00    &117.73&189&10.6   &0.0   &0.0   &   88.3  &   0.5  &   0.0  &   0.0  &   99.5  &10.04& 0.19 &   2.6  &  0.81  &  0.35 \\
\hline
\end{tabular}
}
\begin{minipage}[l]{13.0cm}
\end{minipage}
\label{tcorr1}
\end{table*}

We performed the stellar population synthesis in two steps for each aperture spectrum:
(i) the first one by choosing the spectral region as $\lambda\lambda$4800\,--\,5500\,{\AA},
from which we just measured a unique solution of [$\alpha$/Fe],
and (ii) the second one by expanding the spectral region to $\lambda\lambda$4320\,--\,5990\,{\AA},
from which we derived the definitive output SSP distributions in age and [M/H]
by choosing a single grid of SSPs based on the previous solution in [$\alpha$/Fe]
(the scaled-solar SSP models with [$\alpha$/Fe]$=0.0$ dex
or the $\alpha$-enhanced SSP models with [$\alpha$/Fe]$=+0.4$ dex).
\citet{2015MNRAS.449.1177V} recommended
the first spectral region to estimate [Mg/Fe] 
(Mg is a kind of proxy for $\alpha$-elements), due to the 
many relevant Mg- and Fe-sensitive features in there
such as Mg\,$b$ triplet, Fe\,I\,$\lambda$5015\,{\AA}, Fe\,I\,$\lambda$5270\,{\AA},
Fe\,I\,$\lambda$5335\,{\AA}, and Fe\,I\,$\lambda$5406\,{\AA}.
The second region is adequately wide enough to extract information about stellar age 
from the spectral continuum and about stellar chemical composition from absorption features.
The adopted averages of age, [M/H], and [$\alpha$/Fe] were weighted in luminosity (or flux),
although we also provided the mass fractions of the SSP resulting distribution
(associated to the mass fraction SSP components {\bf \textit{m$_{j}$}}).
The flux fraction SSP components {\bf \textit{x$_{j}$}},
or simply the flux fraction SSP vector {\bf \textit{x}},
were directly provided by the {\scriptsize\,STARLIGHT} code.
The stellar population synthesis solutions in age, [M/H], and [$\alpha$/Fe]
derived from both {\scriptsize\,STARLIGHT} steps over two different spectral regions
were cross-checked between each other at every aperture spectrum.
We can conclude that both solutions were in good agreement within the observational errors
in the averages of age, [M/H], and [$\alpha$/Fe].

The abundance ratio [$\alpha$/Fe]
was specifically derived from three stellar population synthesis runs as proposed by
\citet{2015MNRAS.449.1177V}:
the first run adopting the scaled-solar SSP models only ([$\alpha$/Fe]$=0.0$ dex),
the second one with the $\alpha$-enhanced SSP models only ([$\alpha$/Fe]$=+0.4$ dex),
and the third one with all SSP models.
The best solution for [$\alpha$/Fe] is given as function of both $\chi^{2}$ and $adev$,
looking for the minimum values of these merit figures through a
second order polynomial fit (e.g. $\chi^{2}$ versus [$\alpha$/Fe])
assuming [$\alpha$/Fe]$=+0.2$\,dex for the third run.
For instance, the minimum in a $\chi^{2}$ fitting curve is very well located across the abscissas axis [$\alpha$/Fe]
such that the typical error that can be estimated for [$\alpha$/Fe] is around 0.05 dex.

Afterwards we were able to estimate [Fe/H] for each aperture, which was computed by taking
into account a homogeneous abundance ratio relative to iron over all $\alpha$-elements based on the following equation:
[Fe/H]\,=\,[M/H]\,-\,0.75\,[Mg/Fe], whose validity covers the ETG chemical composition 
\citep{2013MNRAS.435..952S,2015MNRAS.449.1177V}.
We emphasize that magnesium is assumed as a proxy of $\alpha$-elements.

Figure~\ref{sintese_002} illustrates, as an example, the spectral fit of the best stellar population synthesis
for both adopted spectral ranges ($\lambda\lambda$4800\,--\,5500\,{\AA} and $\lambda\lambda$4320\,--\,5990\,{\AA}).
This figure shows the stellar population synthesis for the nuclear and one $R_{\rm e}$ regions of IC\,5328,
whose spectra were extracted across its major photometric axis.
Table~\ref{tcorr1} also contains as an example
the resulting SSP flux ($f_{y}$, $f_{iy}$, $f_{io}$, and $f_{o}$)
and mass re-binned fractions ($m_{y}$, $m_{iy}$, $m_{io}$, and $m_{o}$),
age and [M/H] luminosity-weighted averages, $\chi^{2}$, $adev$, and $A_{V}$
derived from the best stellar population synthesis solution for every major axis aperture spectrum of IC\,5328.
The spatial variation in the contribution of the stellar population components is shown in Fig.~\ref{ic5328_A} for IC\,5328. 
Respective figures and stellar population synthesis results of other sample galaxies are in Appendix \ref{ap_synth1}.

The results from stellar population synthesis are separately presented galaxy-by-galaxy as follows.

\subsubsection{IC\,5328}

\begin{figure*}
\centering
\includegraphics*[angle=0,width=\columnwidth]{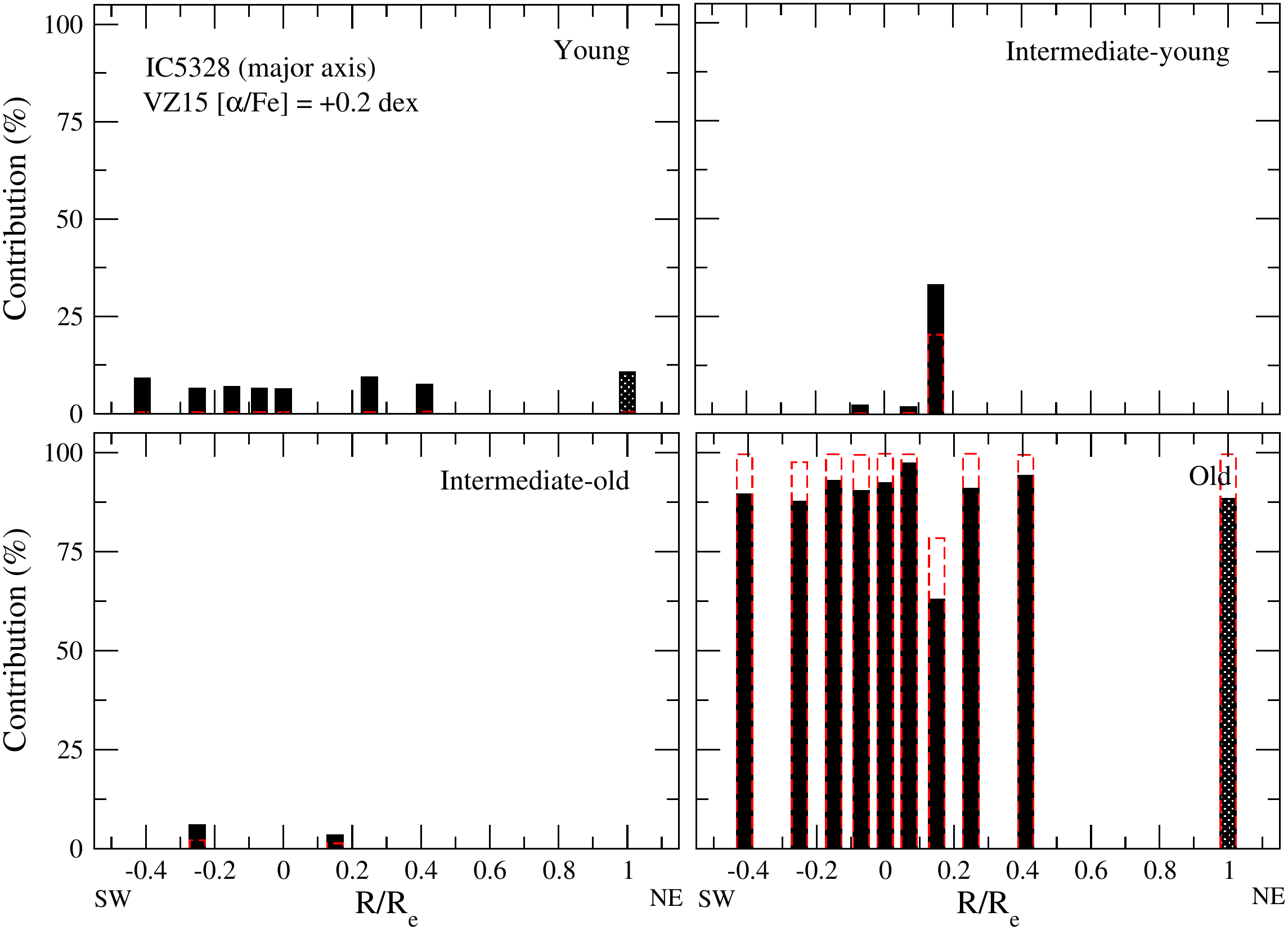}{(a)}
\includegraphics*[angle=0,width=\columnwidth]{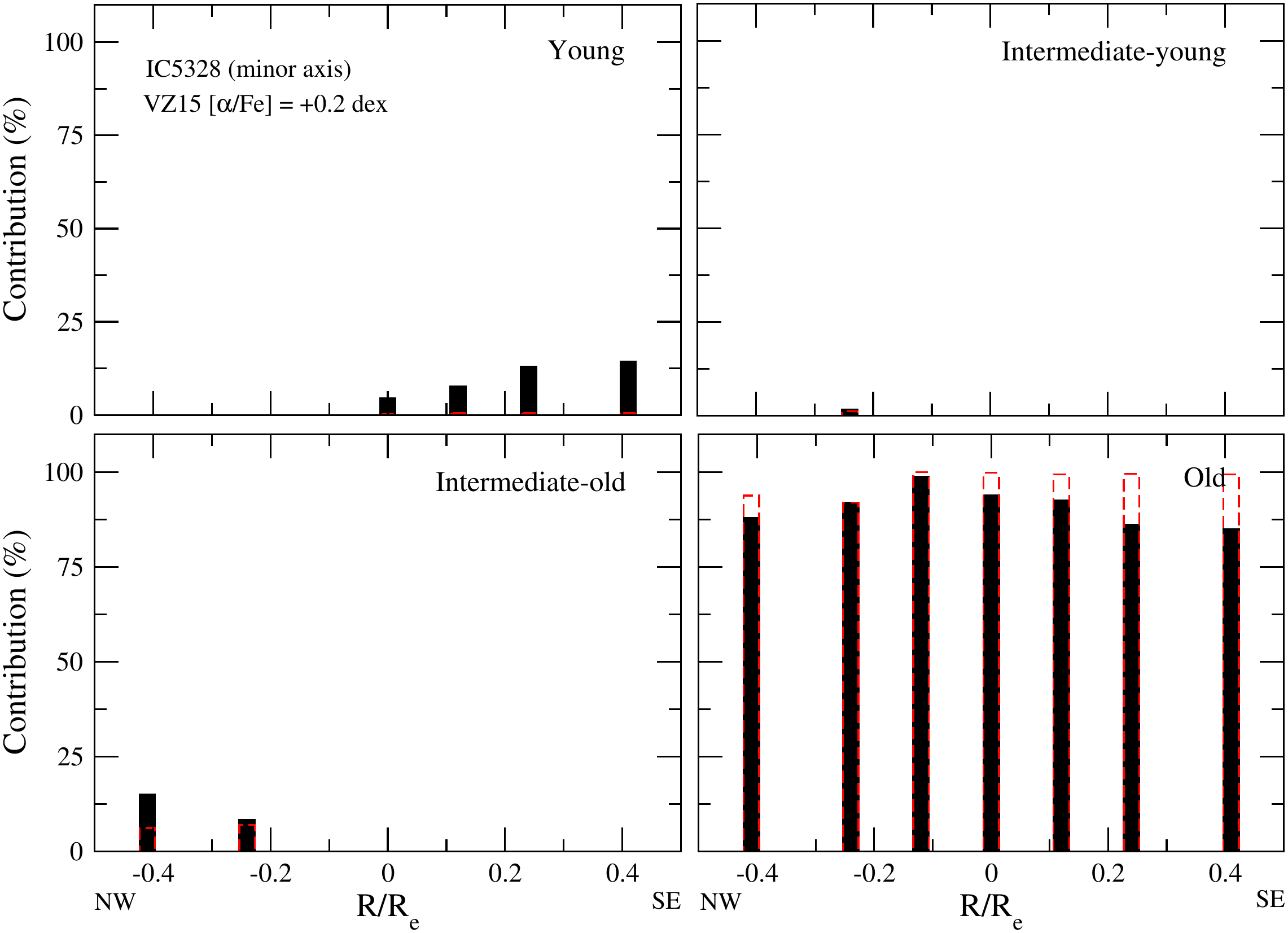}{(b)}
\caption{Flux and mass percental fractions (respectively black solid bars and red dashed open bars) of SSP contributions
from the best solution of {\scriptsize\,STARLIGHT} population synthesis
as a function of the radial distance (in $R_{\rm e}$ unity)
from one side of brightness centre to other side of IC\,5328 across the major (a) and minor (b) photometric axes.
Each sub-panel shows the SSP flux and mass fractions in each age bin.
The SSP contributions are also presented for the region inside one $R_{\rm e}$ 
(only derived from the major axis data analysis as shown in the top panel (a) at the farthest right bin at one $R_{\rm e}$).
}
\label{ic5328_A}
\end{figure*}

The bulk of stellar populations of IC\,5328 are old (8$\times10^{9}<\,t\,\leq13\times10^{9}$ years),
reaching an average value of 88.5\,per\,cent in light fraction
and 97\,per\,cent in mass fraction over the observed regions across its major photometric axis (see Table~\ref{tcorr1}).
The light and mass contributions from the intermediate-old population are negligible
over all radial distances (fractions around and less than 5\,per\,cent).
There is a contribution of about 30\,per\,cent in flux and 20\,per\,cent in mass from the intermediate-young population at the
bin $R/R_{\rm e}=0.15$ in the NE direction,
as can be seen in Fig.~\ref{ic5328_A}.
The light contribution of the young population is always less than 10\,per\,cent and is negligible in mass.
Across the minor axis, the population synthesis found light fractions around 14\,per\,cent by the young population
at $R/R_{\rm e}=0.24$ and $0.41$ (SE direction, 0.5\,per\,cent in mass fraction) and intermediate-old at $R/R_{\rm e}=0.41$
(NW direction, 6\,per\,cent in mass fraction).
The contributions of integrated stellar populations inside one  R$_{\rm e}$ 
indicate a predominant contribution of the old population in light fraction (around 90\,per\,cent) as well in mass (almost 100\,per\,cent).
However, the a flux contribution is about 10\,per\,cent for the young population.
Curiously, there is no contribution from the intermediate-young population at all.
In terms of chemical composition, the stellar populations are metal rich with a moderate $\alpha$-enhancement.
The best solutions of {\scriptsize\,STARLIGHT} stellar population synthesis suggest [$\alpha$/Fe]$=+0.14$\,dex
(assumed as +0.20\,dex to run the code)
for all aperture spectra.
The bulk of stellar populations has [Fe/H]$=+0.25$\,dex and [M/H]$=+0.40$\,dex over all observed regions,
reaching an average light fraction around $\sim$75\,per\,cent.
The stellar population synthesis from the integrated light inside one R$_{\rm e}$ 
points to a global contribution of about 90\,per\,cent for the bin [Fe/H]$=+0.25$\,dex ([M/H]$=+0.40$\,dex),
but with also a 10\,per\,cent of contribution for [Fe/H]$=-1.11$\,dex ([M/H]$=-0.96$\,dex).

\subsubsection{NGC\,1052}

\begin{figure*}
\centering
\includegraphics*[angle=270,width=\columnwidth]{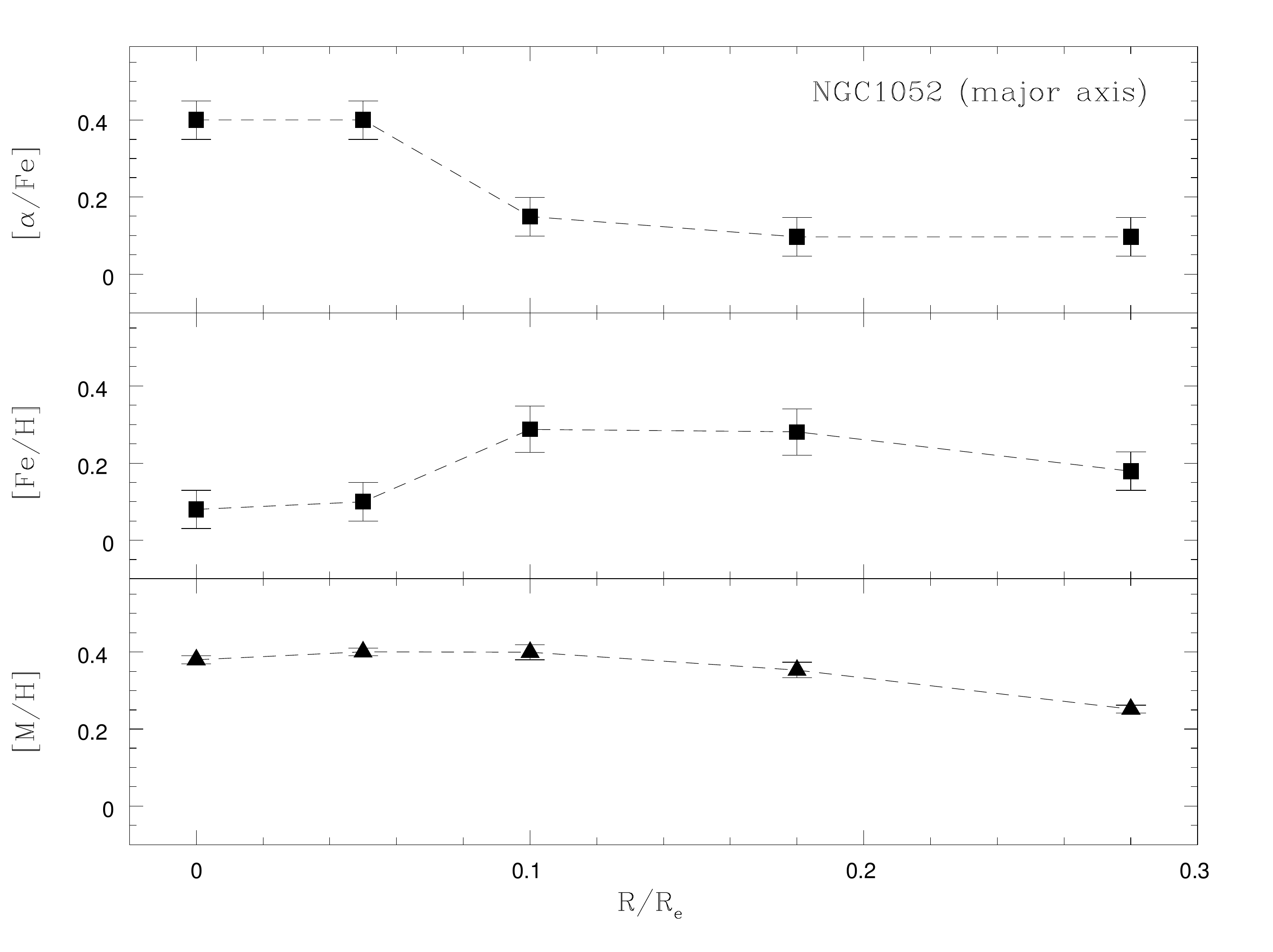}{(a)}
\includegraphics*[angle=270,width=\columnwidth]{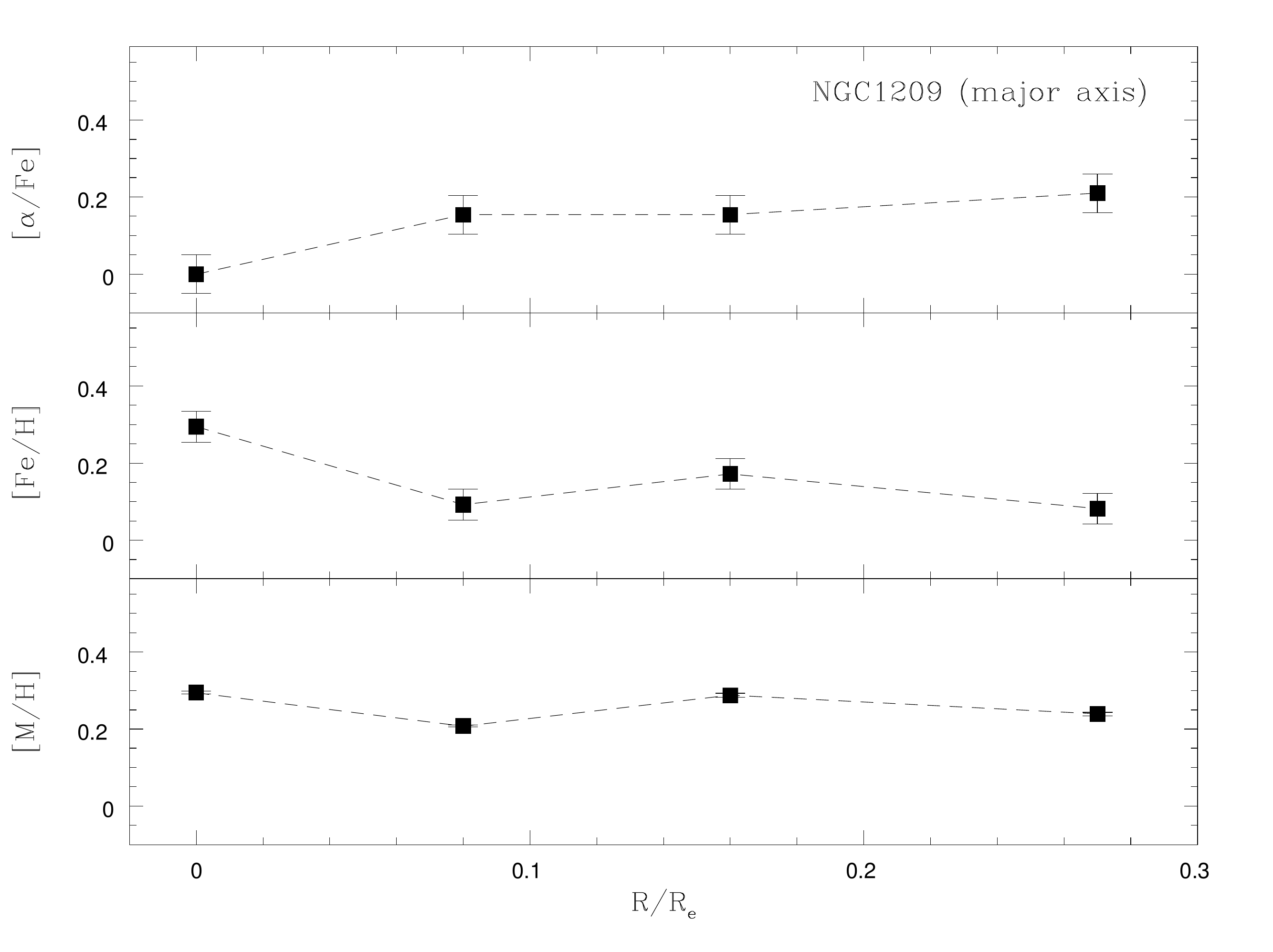}{(b)}
\caption{[M/H], [Fe/H], and [$\alpha$/Fe] derived from the stellar population synthesis
as a function of the normalized radial distance ($R_{\rm e}$ unity)
for just two galaxies, NGC\,1052 (panel (a)) and NGC\,1209 (panel (b)),
in which [M/H], [Fe/H], and [$\alpha$/Fe] are correlatively variable across the major photometric axis.
}
\label{gra_1052_1209}
\end{figure*}

The population synthesis results in Fig.~\ref{n_bins_nova}
indicate that the stellar light and mass across the disc of NGC\,1052 (major axis direction)
are mainly dominated by old stellar populations,
with null or negligible contributions of the young, intermediate-young, and intermediate-old populations
over all radial distances (fractions less than $\sim$5\,percent).
The contributions of integrated stellar populations inside one R$_{\rm e}$  also indicate a predominant contribution
of the old population in light (fraction around 89\,percent) as well in mass (fraction around 96\,percent).
The residual fractions in both flux and mass contributions are due to the intermediate-old population 
(see specifically Fig.~\ref{n_bins_nova}).
Regarding the chemical composition, the stellar populations are metal rich with a variable $\alpha$-enhancement that decreases outwards
([M/H]\,$\geq$\,+0.2\,dex, 0.0\,$\leq$\,[$\alpha$/Fe]\,$\leq$\,+0.4\,dex).
The best solutions of population synthesis suggest [$\alpha$/Fe]$=0.0$\,dex for the external observed apertures,
[$\alpha$/Fe]$=+0.2$ dex at $R/R_{\rm e}=0.10$ in both sides of the brightness profile,
and [$\alpha$/Fe]$=+0.4$ dex for the nuclear region and two nearest apertures at $R/R_{\rm e}=0.05$ 
(see Fig. \ref{n_bins_nova}).
While [M/H] radially decreases outwards, [Fe/H] increases from the nucleus up to $R/R_{\rm e}=0.18$ 
decreasing a bit from this point outwards
(also see panel (a) of Fig.~\ref{gra_1052_1209}, 
which shows the correlated radial variations of [M/H], [Fe/H], and [alpha/Fe] for NGC\,1052 and NGC\,1209).
Most stellar populations have [Fe/H]\,$\geq$\,$+0.10$, $+0.25$, and $+0.40$\,dex (and [M/H]\,$\geq$\,+0.40\,dex) 
over all observed regions,
reaching an average light fraction around $\sim$85\,percent.
The stellar population synthesis from the integrated light inside one R$_{\rm e}$ points to a global contribution of about 77\,percent
for the bin [Fe/H]$=+0.40$ dex ([M/H]$=+0.40$ dex),
but with also a 12\,percent contribution for [Fe/H]$=-0.35$ dex ([M/H]$=-0.35$ dex),
such that [$\alpha$/Fe] curiously reaches the solar value from the best solution of population synthesis.

\subsubsection{NGC\,1209}

The results of the stellar population synthesis for NGC\,1209 are presented in
Fig.~\ref{n_bins_nova}.
NGC\,1209 is light-dominated by old stellar populations, with a contribution of about $\sim$86\,percent in the nuclear region
and higher than 91\,percent in the external region towards the SW direction.
The light contribution of the young population is always less than 15\,percent and is negligible in mass.
The contributions of integrated stellar populations inside one R$_{\rm e}$  also indicate a predominant contribution
of the old population in light (fraction around 86\,percent) as well in mass (fraction around 99\,percent).
However, there is a flux contribution of about 12\,percent for the young population.
The light and mass contributions from the intermediate-young and intermediate-old populations are negligible
over all radial distances (fractions around and less than 5\,per\,cent).
Regarding the chemical composition, the stellar populations are metal rich with a variable $\alpha$-enhancement
that increases outwards ([M/H]\,$\geq$\,+0.2\,dex, 0.0\,$\leq$\,[$\alpha$/Fe]\,$\leq$\,$+0.2$\,dex).
The best solutions of population synthesis suggest [$\alpha$/Fe]$=0.0$\,dex for the nuclear region and
[$\alpha$/Fe]$=+0.2$ dex from $R/R_{\rm e}=0.08$ outwards in both sides of the brightness profile 
(see Fig. \ref{n_bins_nova}).
While [M/H] is radially constant, [Fe/H] decreases outwards up to $R/R_{\rm e}=0.27$ 
(also see panel (b) of Fig. \ref{gra_1052_1209}).
The bulk of stellar populations has [Fe/H]\,$\geq$\,$+0.31$\,dex (and [M/H]\,$\geq$\,$+0.40$\,dex) over all observed regions,
reaching an average light fraction around $\sim$74\,percent.
The stellar population synthesis from the integrated light inside one  R$_{\rm e}$  points to a global
contribution of about 78\,percent for the bin [Fe/H]$=+0.25$ dex ([M/H]$=+0.40$ dex), but with also a 10\,percent of
contribution for [Fe/H]$=+0.11$ dex ([M/H]$=+0.26$ dex), such that [$\alpha$/Fe] reaches over-solar value
from the best solution of population synthesis.

\subsubsection{NGC\,5812}

The majority of stellar populations of NGC\,5812 is old,
reaching an average value of ~95\,percent
in percentage of light fraction and $\sim$97\,per\,cent in mass fraction over the observed regions
(see Fig.~\ref{n_bins_nova}).
The intermediate-old population at the nuclear region contributes to about 24\,per\,cent in flux and 21\,percent in mass.
The light and mass contributions from the young and intermediate-young population are negligible over
all radial distances (fractions around and less than 8\,percent).
The contributions of integrated stellar populations inside one R$_{\rm e}$ indicate a predominant contribution
of the old population in light fraction and mass (100\,per\,cent).
Curiously, intermediate-old population does not contribute in the nuclear region.
In terms of chemical composition, the stellar populations are metal rich with a moderate $\alpha$-enhancement.
The best solutions of {\scriptsize\,STARLIGHT} stellar population synthesis suggest [$\alpha$/Fe]$=+0.2$ dex for all aperture spectra.
Most of stellar populations have [Fe/H]$=+0.25$\,dex and [M/H]$=+0.40$\,dex over all observed regions,
reaching an average light fraction around $\sim$94\,percent.
The stellar population synthesis from the integrated light inside one R$_{\rm e}$ 
points to a global contribution of about 81\,percent for the bin [Fe/H]$=+0.25$ dex ([M/H]$=+0.40$ dex),
but with also a $\sim$19\,percent contribution for [Fe/H]$=+0.11$ dex ([M/H]$=+0.26$ dex).

\subsubsection{NGC\,6758}

The population synthesis results in Fig.~\ref{n_bins_nova}
indicate that the stellar light and mass across the disc of NGC\,6758 are
mainly dominated by old stellar populations, with null or negligible contributions of the
intermediate-old population over all radial distances (fractions less than $\sim$5\,per\,cent) across both major and minor axes.
The light contribution of the young population is always less than 13\,per\,cent (major axis direction) and is negligible in mass.
The intermediate-young population contributes $11-20$\,per\,cent in flux and $6-9$\,per\,cent in mass
at the  bin $R/R_{\rm e}=0.11$ in the SW direction and nuclear region across the minor axis, respectively.
The contributions of integrated stellar populations inside one R$_{\rm e}$ across in the major axis direction also
indicate a predominant contribution of the old population in light (fraction around $\sim$94\,per\,cent) as well in mass 
(reaching 100\,per\,cent).
In terms of chemical composition, the stellar populations are metal rich with a moderate $\alpha$-enhancement.
The best solutions of {\scriptsize\,STARLIGHT} stellar population synthesis suggest [$\alpha$/Fe]$=+0.2$\,dex
or all aperture spectra.
The majority of stellar populations has ([Fe/H]$=+0.25$\,dex and [M/H]$=+0.40$\,dex) over all observed regions
(major and minor directions), reaching an average light fraction around $\sim$95\,per\,cent 
(see Fig.~\ref{n_bins_nova}).
The stellar population synthesis from the integrated light inside one  R$_{\rm e}$ 
points to a global contribution of about 94\,per\,cent for the bin [Fe/H]$=+0.25$\,dex ([M/H]$=+0.40$\,dex),
but with also 5\,per\,cent (almost negligible indeed) contribution for [Fe/H]$=-1.11$\,dex ([M/H]$=-0.96$\,dex).

\subsubsection{NGC\,6861}

The results of the stellar population synthesis for NGC\,6861 are presented in
Fig.~\ref{n_bins_nova}. NGC\,6861 is completely dominated by old stellar populations, marked by the contribution of both light and mass,
with a contribution of nearly 100\,percent over all observed regions (in both sides of the brightness profile
and inside one R$_{\rm e}$) across both major and minor axes. The light and mass contributions
from the young, intermediate-young and  intermediate-old population are negligible over all radial distances
(fractions around and less than 5\,percent).
In relation to chemical composition, the stellar populations are metal rich with a high $\alpha$-enhancement.
The best solutions of {\scriptsize\,STARLIGHT} stellar population synthesis suggest [$\alpha$/Fe]$=+0.40$ dex for all aperture spectra.
The bulk of stellar populations has [Fe/H]$=+0.10$\,dex (and [M/H]$=+0.40$\,dex) over all observed
regions (in both sides of the brightness profile and inside one  R$_{\rm e}$), reaching an average light fraction
around $\sim99$\,per\,cent.

\subsubsection{NGC\,7507}

Most of the stellar populations of NGC\,6861 are old, reaching 
an average value of almost 100\,percent 
in percentage of light and mass fractions over the observed regions across the slit position.
(see Fig.~\ref{n_bins_nova}).
The contributions of integrated stellar populations inside one R$_{\rm e}$ also indicate a predominant
contribution of the old population in light (fraction around 90\,percent)
as well as in mass (fraction around 97 percent), but also including a contribution of 9\,per\,cent in light fraction by an young population.
The light and mass contributions from the young, intermediate-young, and intermediate-old populations are negligible
over all radial distances (fractions around and less than 8\,percent).
Regarding the chemical composition, the stellar populations are metal rich with a moderate $\alpha$-enhancement.
The best solutions of {\scriptsize\,STARLIGHT} stellar population synthesis suggest [$\alpha$/Fe]$=+0.2$ dex for all aperture spectra.
The bulk of stellar populations has [Fe/H]$=+0.25$\,dex (and [M/H]$=+0.40$\,dex) over all observed regions,
reaching an average light fraction around $\sim$84\,percent.
The stellar population synthesis from the integrated light inside one  R$_{\rm e}$ points to a global contribution
of about 86\,percent for the bin [Fe/H]$=+0.25$ dex ([M/H]$=+0.40$ dex), but with small contributions
at the bins [M/H]\,=\,$-0.96$, $-0.66$, and $-0.35$ dex (almost negligible indeed).

\subsubsection{NGC\,7796}

In Fig.~\ref{n_bins_nova}, 
the population synthesis results indicate that the stellar light and mass across the disc of NGC\,7796 are
mainly dominated by old stellar populations, with null or negligible contributions of the
young, intermediate-young, and intermediate-old populations (fractions less than $\sim$5\,percent) over all
radial distances of the observed regions across both major and minor axes.
The stellar populations are metal rich with a moderate $\alpha$-enhancement
for all aperture spectra (best solutions give [$\alpha$/Fe]$=+0.2$ dex).
The bulk of stellar populations has [Fe/H]$=+0.25$\,dex (and [M/H]$=+0.40$\,dex) over all observed regions
(in both sides of the brightness profile and inside one R$_{\rm e}$), reaching an average light fraction around
$\sim97$\,percent.

\subsection{Radial gradients of the stellar population properties}
\label{gradi_2}

\begin{table*}
\centering
\caption{Radial gradients of two stellar population syntheses resulting in parameters for sample galaxies:
age ($\Delta$(log(age))/$\Delta$($R/R_{\rm e}$))
and global metallicity ($\Delta$([M/H])/$\Delta$($R/R_{\rm e}$))
with their errors. The nuclear values of age and [M/H] are also showed.
The values of both parameters are light-weighted means.
Means and standard deviations of gradients 
and nuclear values of age and metallicity are shown in the last two rows (data taken along the EW 
direction are considered in the mean computations of both major and minor axis directions).
}
\scalebox{0.81}{
\begin{tabular}{lllllllllll}
\hline
galaxy ID\,\,\,\,\,\,\,&      slit direction
                 &                    \multicolumn{2}{c}{age}                & &           \multicolumn{2}{c}{[M/H]}        \\
                 &&   gradient [dex/($R/R_{\rm e}$)] &  age ($R/R_{\rm e}=0$) [Gyr] & &  gradient [dex/($R/R_{\rm e}$)]    &  [M/H] ($R/R_{\rm e}=0$) [dex] \\
\cline{3-4}
\cline{6-7}
\noalign{\smallskip}
IC\,5328~~~~&MA   &$+0.01\pm0.05$               &$12\pm1$   &       &$-0.36\pm0.10$    &$0.32\pm0.02$   \\
IC\,5328~~~~& MI  & $-0.13\pm0.07$              &$12\pm1$   &       &$-0.52\pm0.19$    &$0.33\pm0.05$   \\
                  &                             &           &       &                  &                \\
NGC\,1052 &MA     &$-0.11\pm0.03$               &$13\pm1$   &       &$-0.79\pm0.22$    &$0.42\pm0.04$   \\
                  &                             &                   &       &                  &                \\
NGC\,1209& MA     &$+0.11\pm0.11$               &$11\pm1$   &       &$-0.14\pm0.18$    &$0.27\pm0.03$   \\
                  &                             &                   &       &                  &                \\
NGC\,5812& EW     &$-0.18\pm0.07$               &$14\pm1$   &       &$-0.25\pm0.15$    &$0.41\pm0.03$   \\
                  &                             &                   &       &                  &                \\
NGC\,6758 &MA     &$-0.04\pm0.10$               &$12\pm1$   &       &$-0.25\pm0.14$    &$0.40\pm0.03$   \\
NGC\,6758 &MI     &$+0.06\pm0.14$               &$12\pm1$   &       &$-0.15\pm0.11$    &$0.39\pm0.02$   \\
                  &                             &                   &       &                  &                \\
NGC\,6861 &MA     &$+0.02\pm0.02$               &$13\pm1$   &       &$-0.01\pm0.01$    &$0.40\pm0.00$   \\
NGC\,6861 &MI     &$+0.05\pm0.03$               &$13\pm1$   &       &$+0.16\pm0.10$    &$0.36\pm0.02$   \\
                  &                             &                   &       &                  &                \\
NGC\,7507 &EW     &$+0.08\pm0.13$               &$12\pm1$   &       &$-0.46\pm0.39$    &$0.41\pm0.04$   \\
                  &                             &                   &       &                  &                \\
NGC\,7796 &MA     &$-0.02\pm0.02$               &$13\pm1$   &       &$-0.08\pm0.05$    &$0.40\pm0.01$   \\
NGC\,7796 &MI     &$+0.01\pm0.01$               &$13\pm1$   &       &$+0.01\pm0.01$    &$0.40\pm0.00$   \\                  
          &       &                             &           &       &                  &         \\
Means	  &MA/EW  &$-0.02\pm0.10$               &$13\pm1$   &       &$-0.29\pm0.25$    &$0.39\pm0.05$\\	
Means	  &MI/EW  &$-0.02\pm0.11$               &$13\pm1$   &       &$-0.20\pm0.26$    &$0.38\pm0.03$\\	
\hline
\end{tabular}
}
\label{tab33}
\end{table*}

\begin{figure*}
\centering
\includegraphics*[angle=-90,width=0.8\columnwidth]{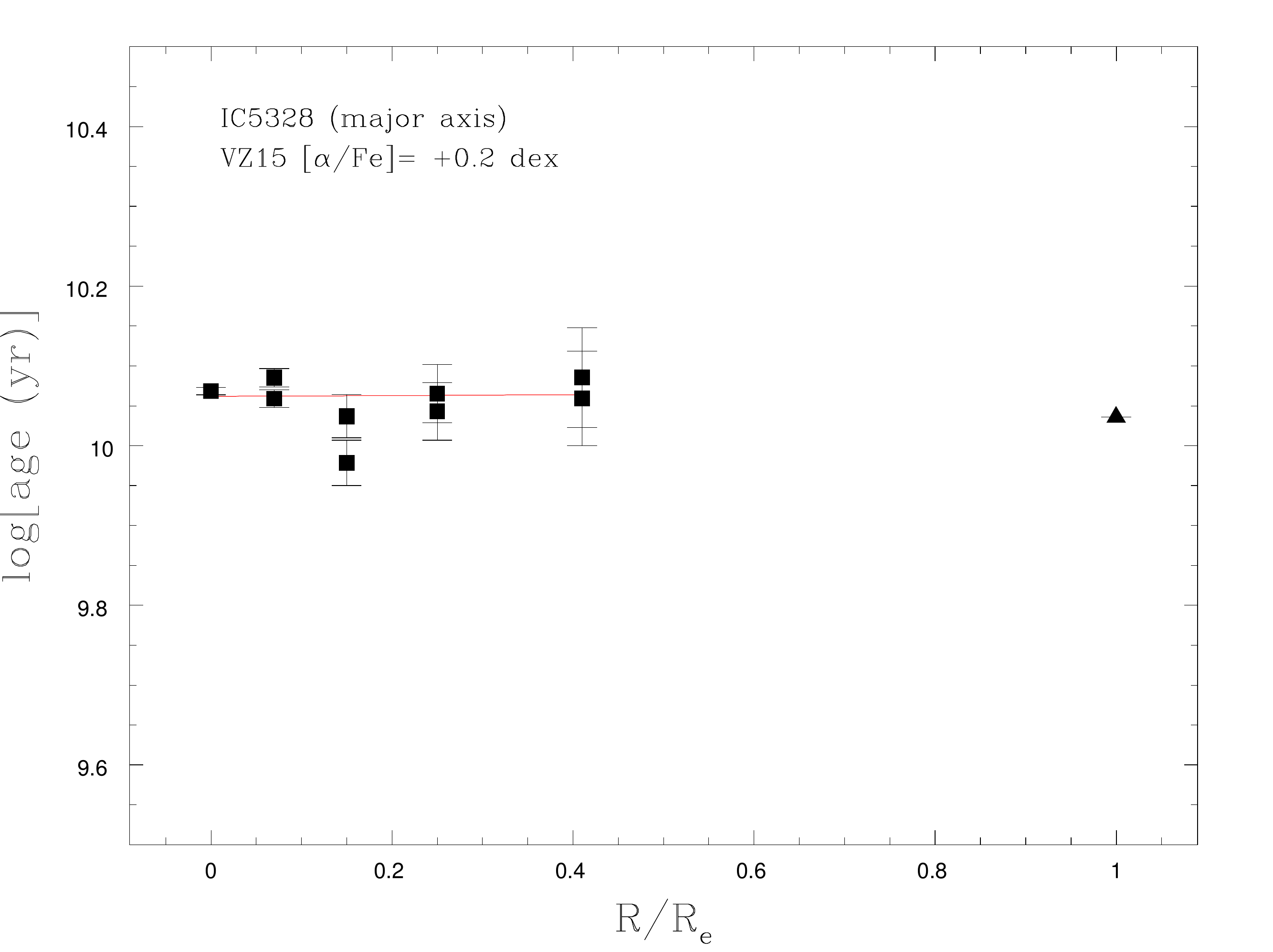}
\includegraphics*[angle=-90,width=0.8\columnwidth]{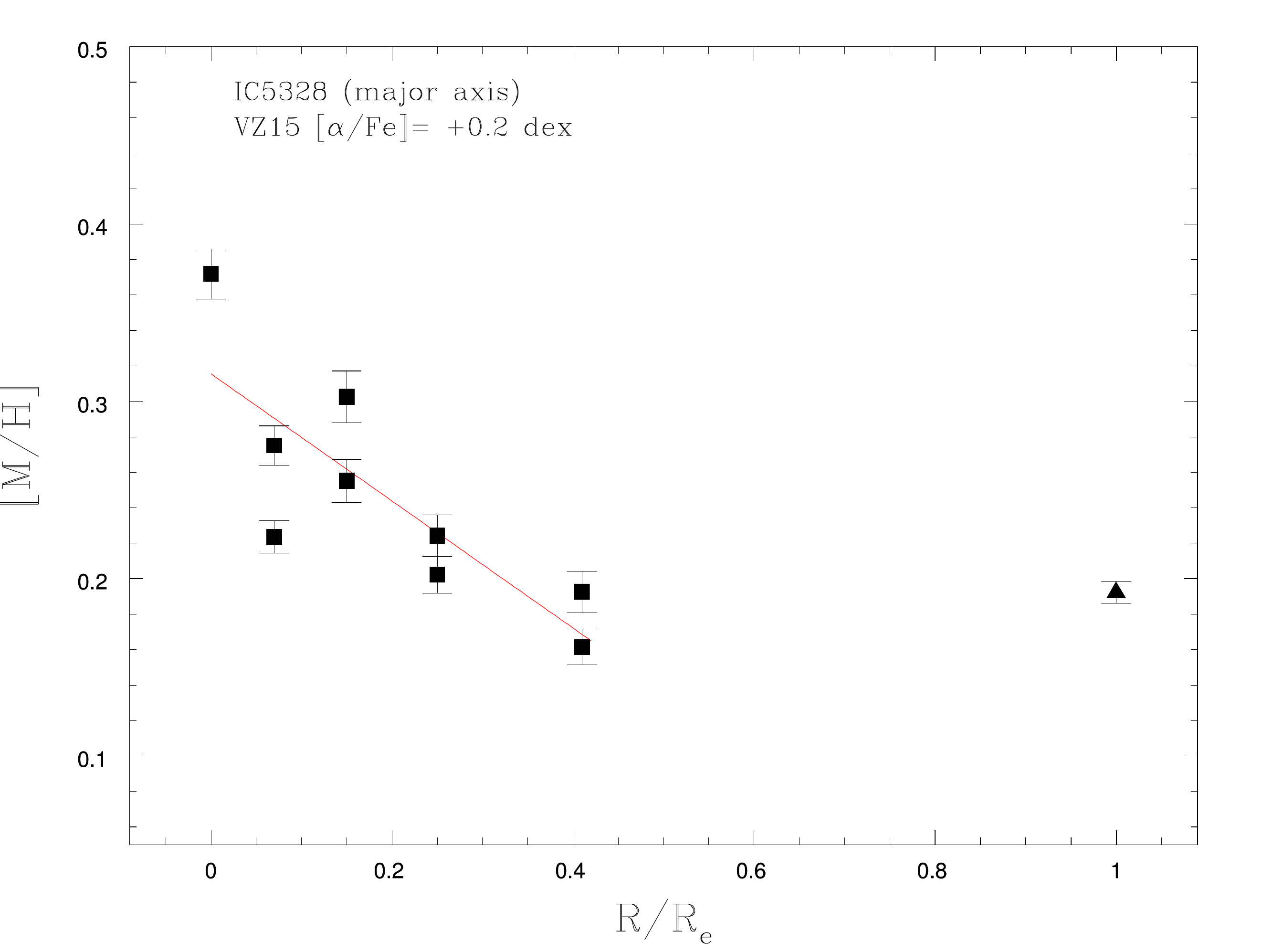}
\includegraphics*[angle=-90,width=0.8\columnwidth]{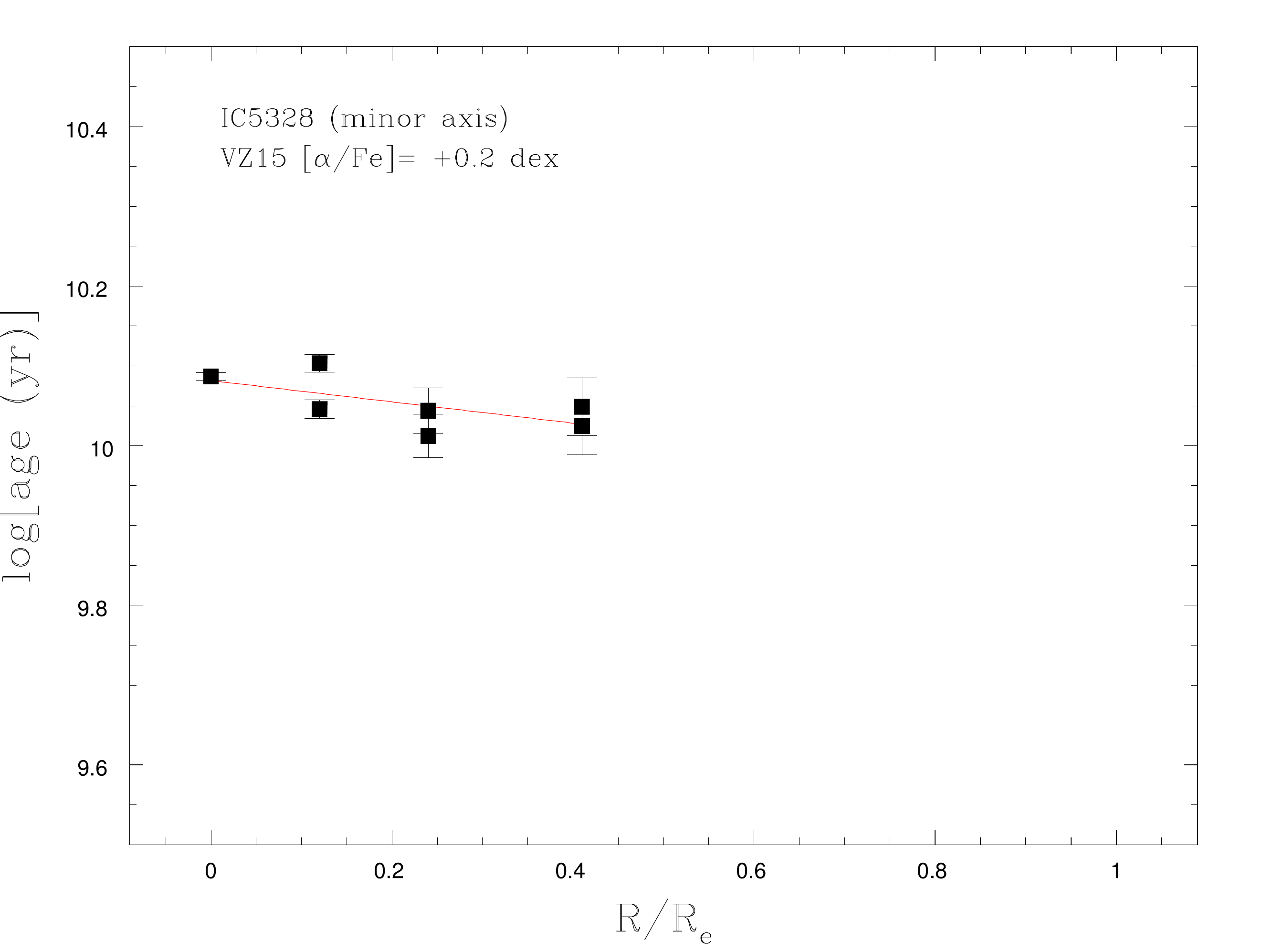}
\includegraphics*[angle=-90,width=0.8\columnwidth]{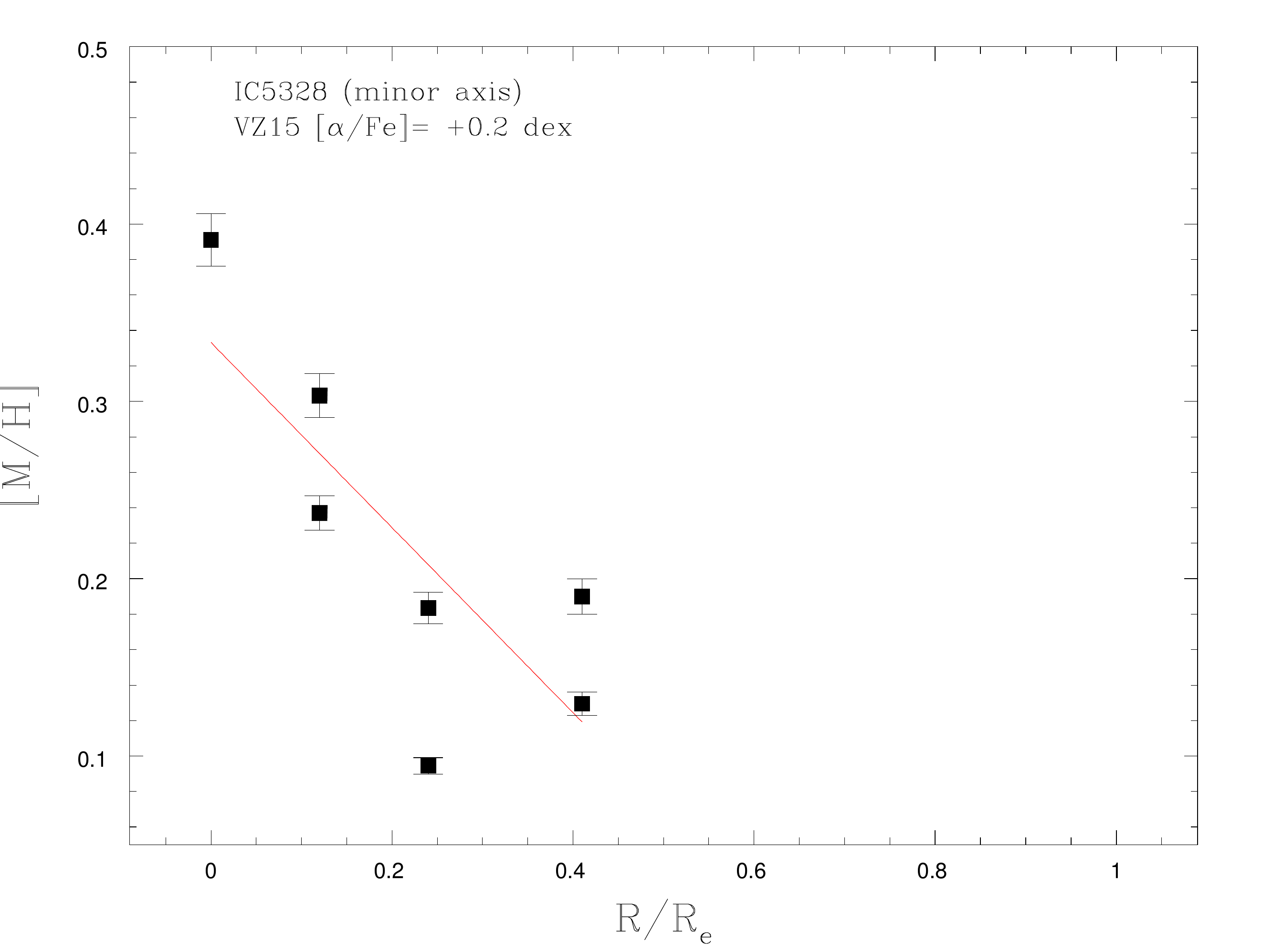}
\includegraphics*[angle=-90,width=0.8\columnwidth]{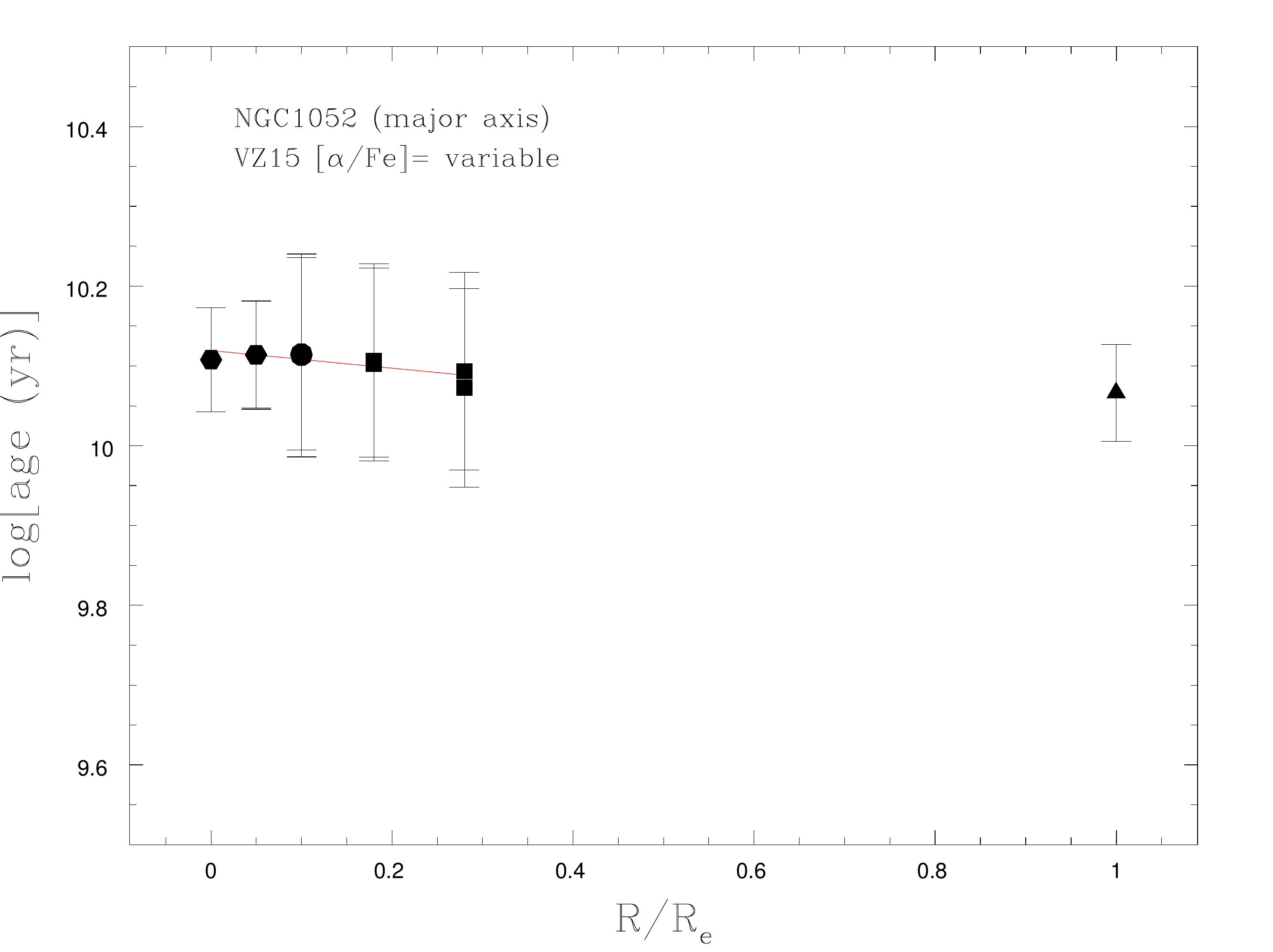}
\includegraphics*[angle=-90,width=0.8\columnwidth]{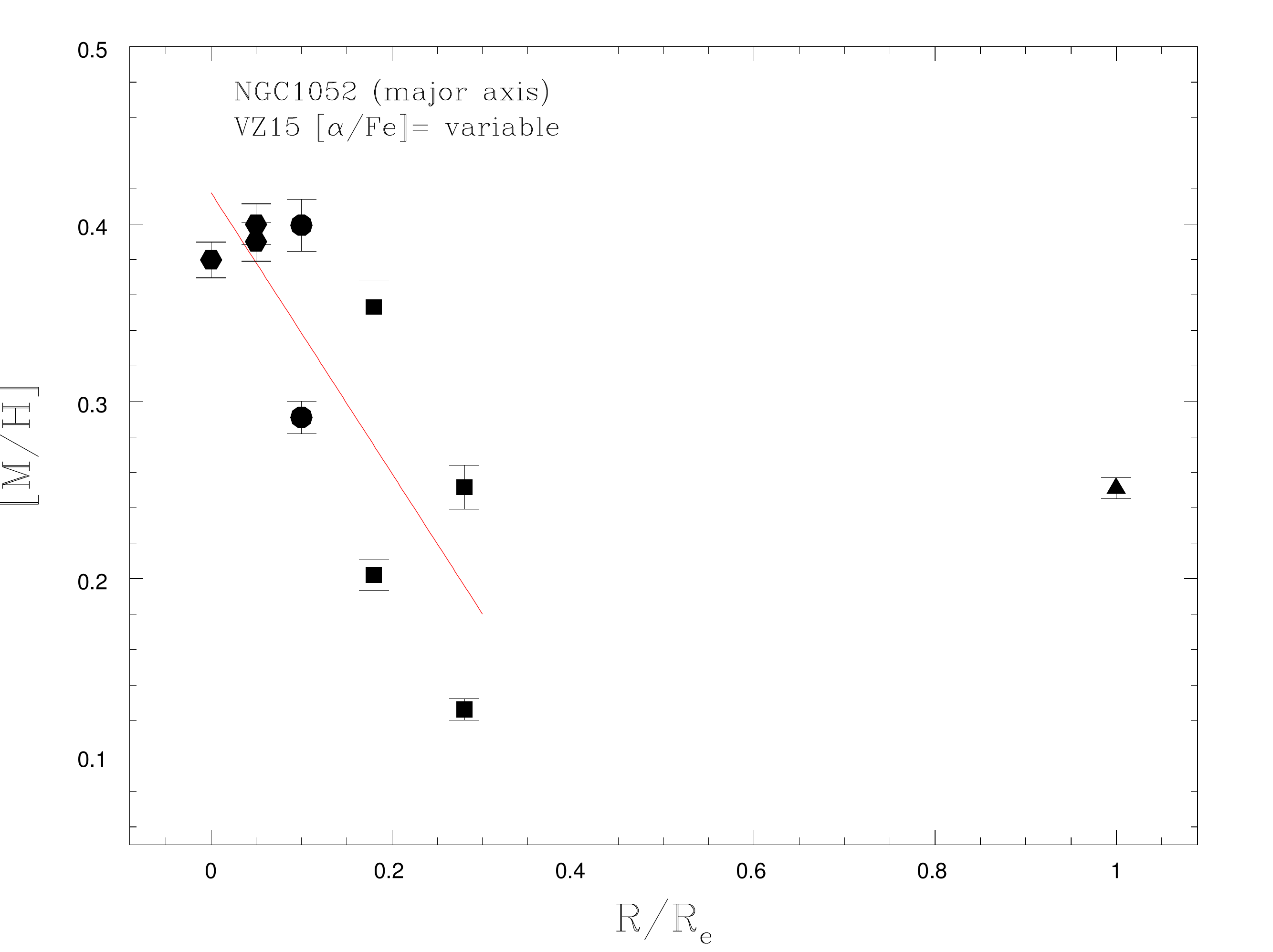}
\includegraphics*[angle=-90,width=0.8\columnwidth]{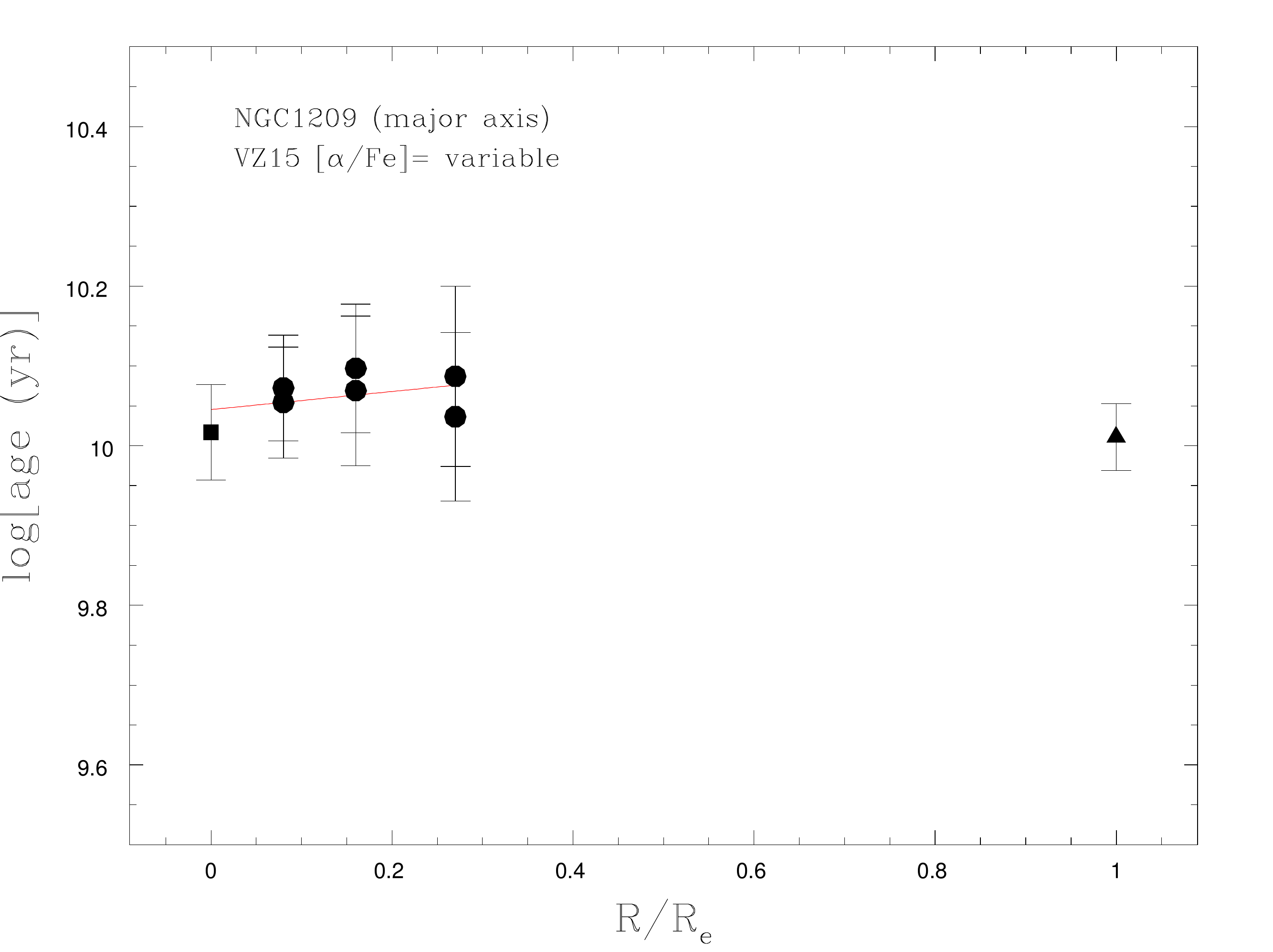}
\includegraphics*[angle=-90,width=0.8\columnwidth]{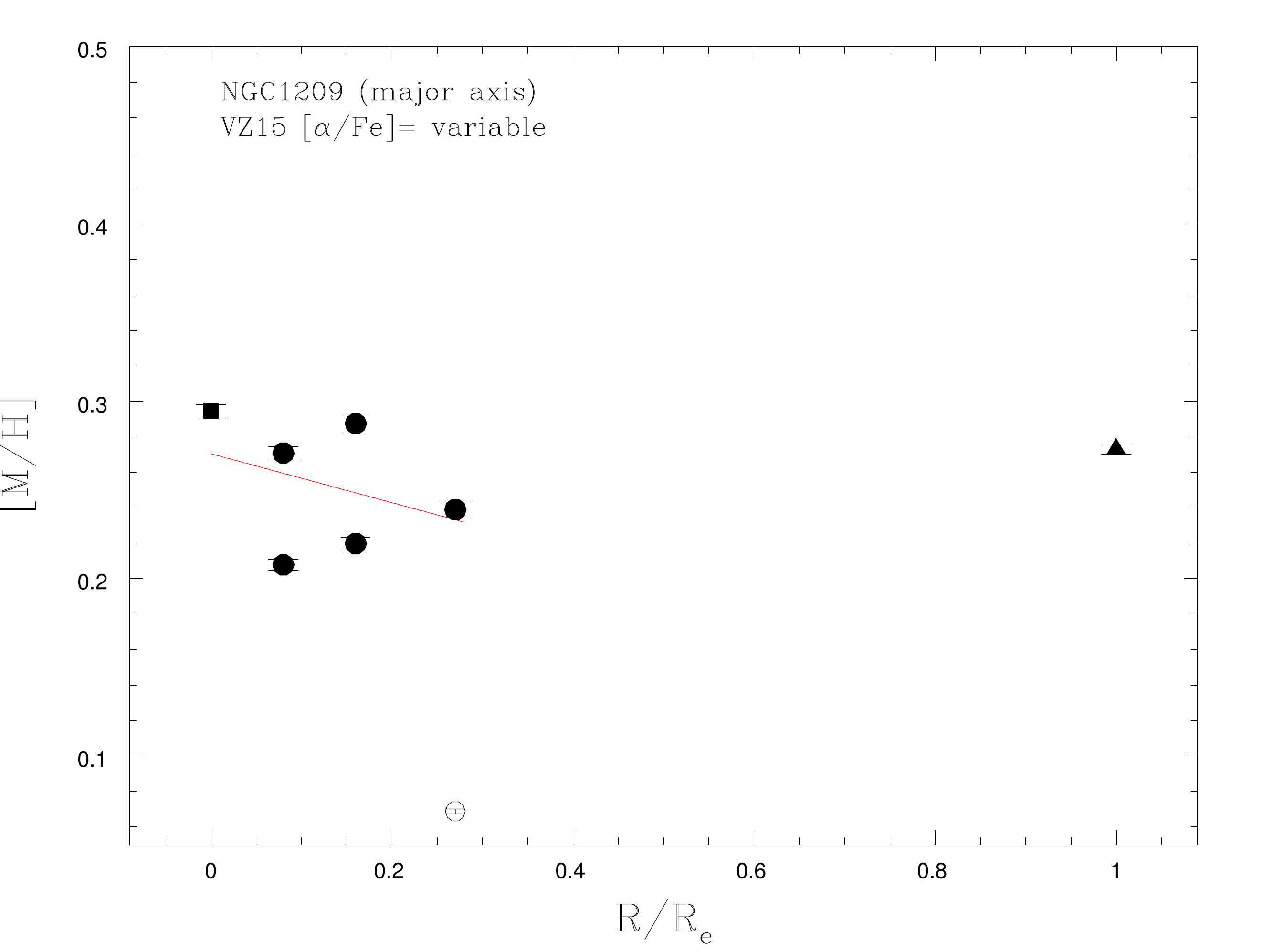}
\caption{Luminosity-weighted means of age and [M/H] derived from the stellar population synthesis as a 
function of the normalized radial distance ($R_{\rm e}$ unity), for IC\,5328, NGC\,1052, and  NGC\,1209. The direction, across which these 
resulting parameters were derived, is also indicated after the galaxy name (top left corner).
The applied {\it lsq} linear fit with two sigma clipping is represented by a red line 
to derive the respective radial gradients of age and [M/H]. The triangles represent the regions of one  R$_{\rm e}$.
The outlier points are assigned by open symbols. 
The results for NGC\,5812 and NGC\,7507 have been derived across the east-west direction only.
}
\label{age_all_a}
\end{figure*}

\begin{figure*}
\centering
\includegraphics*[angle=-90,width=0.8\columnwidth]{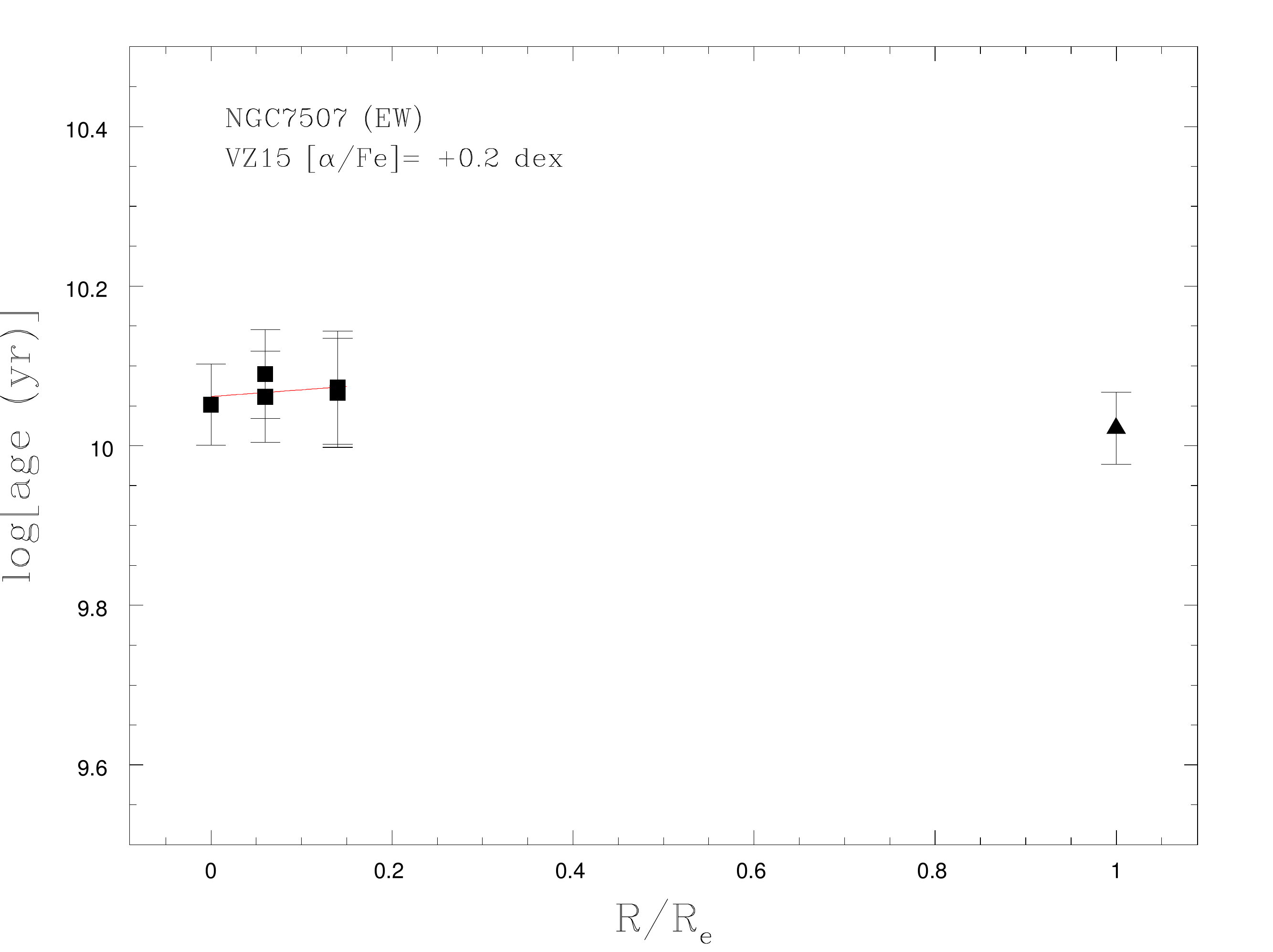}
\includegraphics*[angle=-90,width=0.8\columnwidth]{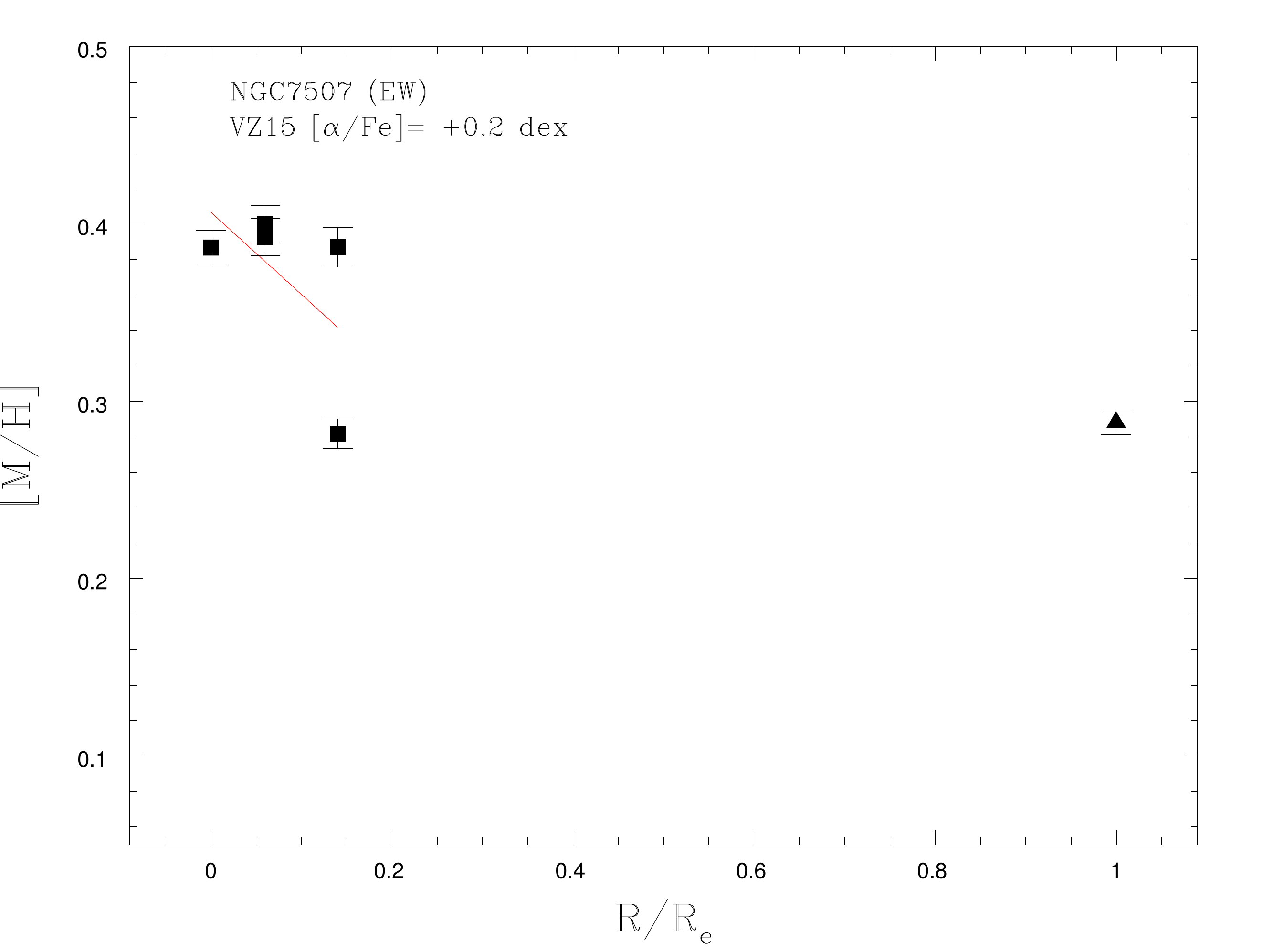}
\includegraphics*[angle=-90,width=0.8\columnwidth]{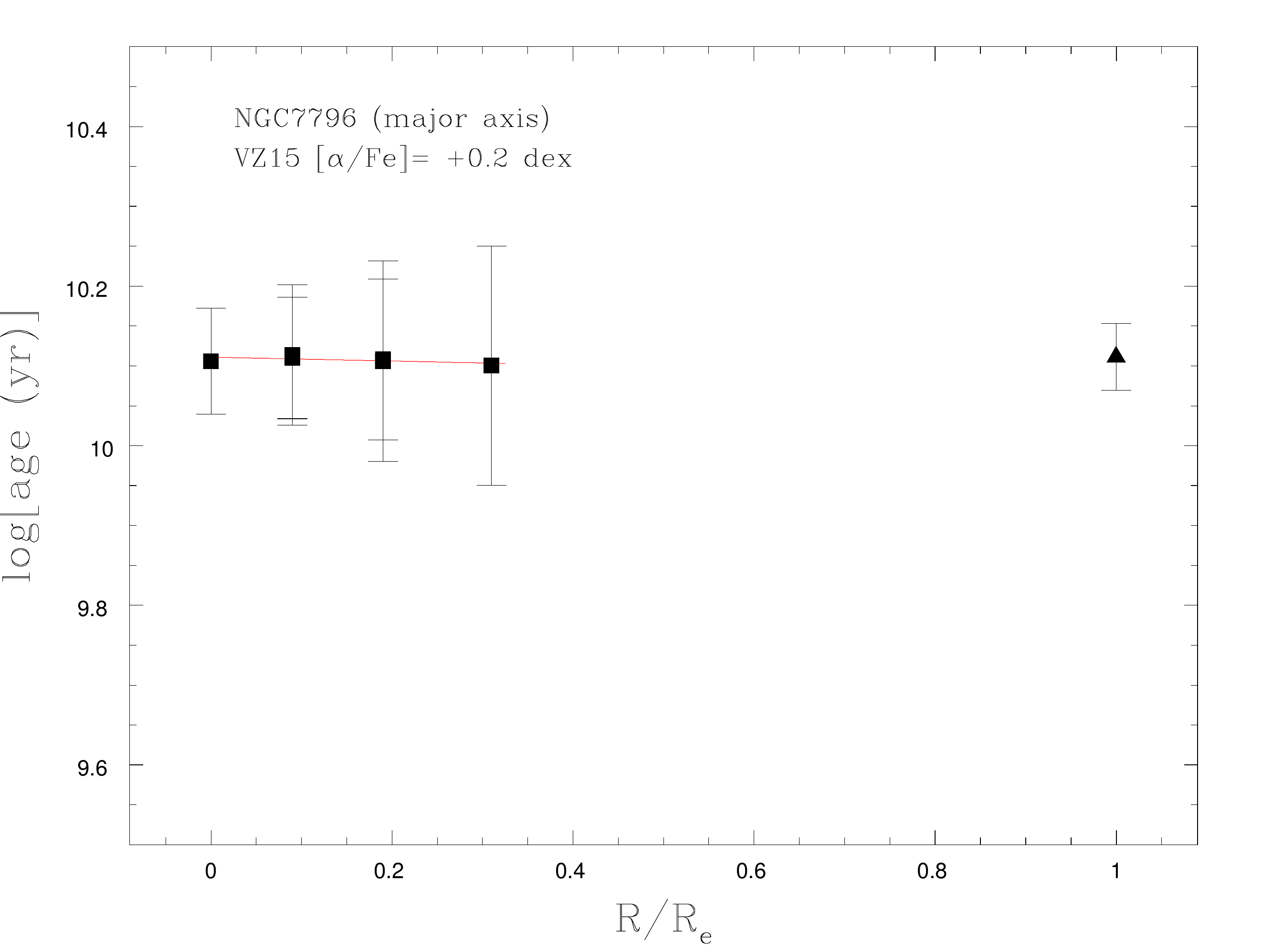}
\includegraphics*[angle=-90,width=0.8\columnwidth]{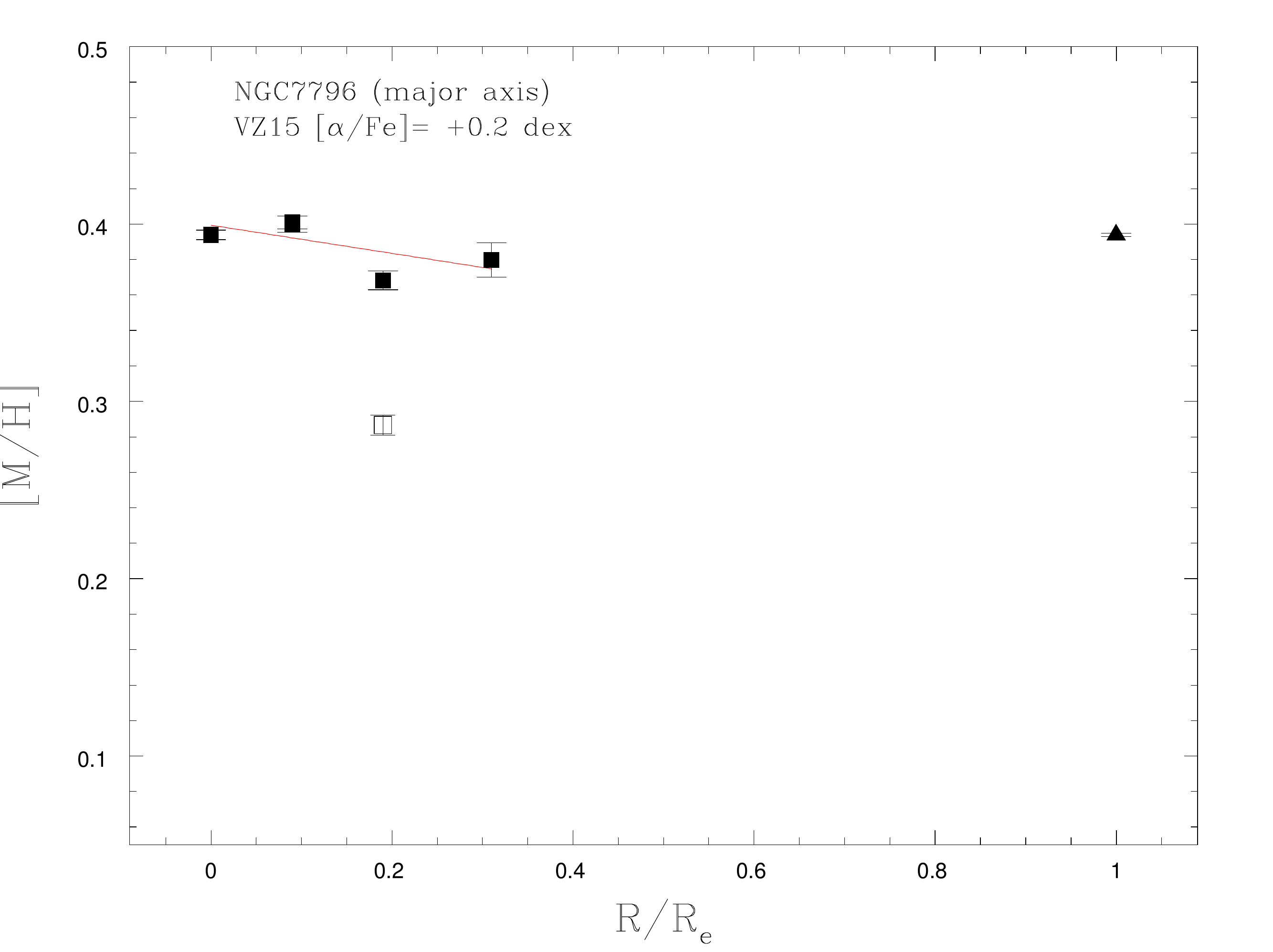}
\includegraphics*[angle=-90,width=0.8\columnwidth]{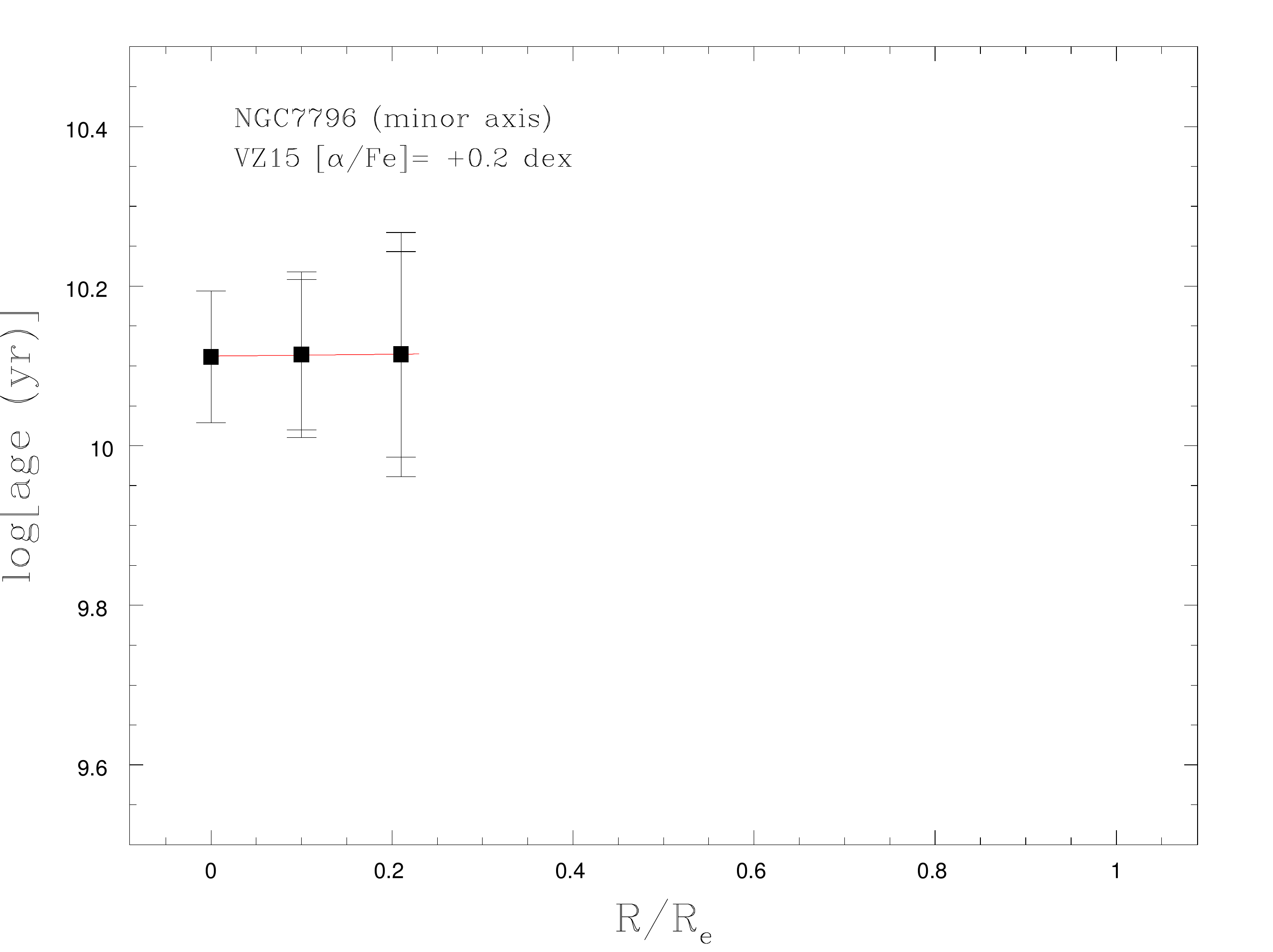}
\includegraphics*[angle=-90,width=0.8\columnwidth]{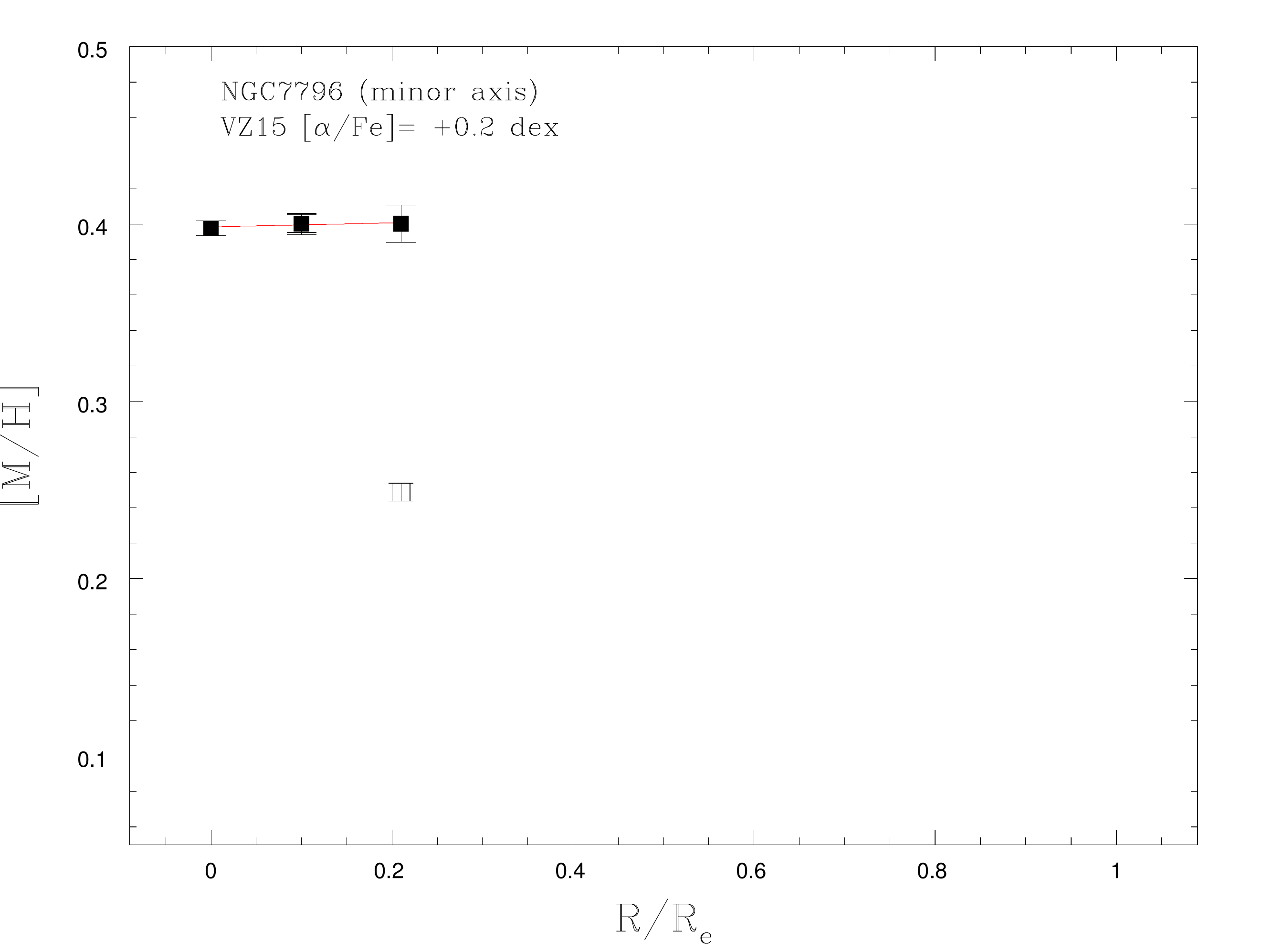}
\caption{Same as Fig.~\ref{age_all_a}, but for NGC\,7507 and  NGC\,7796.}
\label{age_all_h}
\end{figure*}

By compiling the results of {\scriptsize\,STARLIGHT} stellar population synthesis for all
aperture spectra of every sample galaxy,
we obtained the spatial distributions of luminosity-weighted average age, [M/H], and 
[$\alpha$/Fe] across the observed slit directions.
For this, we computed radial gradients of age and [M/H], expressed by
($\Delta$(log(age))/$\Delta$($R/R_{\rm e}$)) and ($\Delta$([M/H])/$\Delta$($R/R_{\rm e}$)), respectively.
These gradients are also used to estimate the values at $R=0$ of age and global metallicity. 
Table \ref{tab33} lists these results. Figures~\ref{age_all_a}\,-\,\ref{age_all_h} show age and [M/H] as a
function of the normalized radial distance (in $R_{\rm e}$ unity) for all sample galaxies, in which we
also present the linear $lsq$ fit. A two-sigma clipping is applied on these fits, in which the outliers are designed as open symbols.
Table \ref{result_compare_sint} compiles the stellar parameters derived from the stellar population synthesis
(nuclear and R$_{\rm e}$ integrated values of age, [M/H], [$\alpha$/Fe], [Fe/H]).
In Appendix~\ref{ap_radial_grad}, the results about the 
radial gradients of stellar population properties are presented galaxy-by-galaxy.

\begin{table*}
\centering
\begin{footnotesize}
\caption{Luminosity-weighted means of age, [M/H], [Fe/H], and [$\alpha$/Fe] as directly estimated by the 
stellar population synthesis for the apertures of nucleus ($r$\,$\leq$\,1 arcsec)
and one R$_{\rm e}$ corrected by ellipticity ($r$\,$\leq$\,$R_{\rm e}$).}
\vspace{0.3cm}
\scalebox{0.9}{
\begin{tabular}{lcc c cc c cc c cc c } 
\hline 
galaxy ID  
&  \multicolumn{2}{c}{age$_{mean}$}   & &  \multicolumn{2}{c}{[M/H]$_{mean}$}  & &   \multicolumn{2}{c}{[Fe/H]$_{mean}$}  
&&   \multicolumn{2}{c}{[$\alpha$/Fe]$_{mean}$}   \\
 &   nucleus &  $r$\,$\leq$\,$R_{\rm e}$  & &  nucleus &  $r$\,$\leq$\,$R_{\rm e}$ & & nucleus &  $r$\,$\leq$\,$R_{\rm e}$ && nucleus &  $r$\,$\leq$\,$R_{\rm e}$   \\
\cline{2-3}
\cline{5-6}
\cline{8-9}
\cline{11-12}
\noalign{\smallskip}             
IC\,5328  &$12\pm1$    &$11\pm1$  & &$+0.37\pm0.01$     &$+0.19\pm0.01$    &     &$+0.27\pm 0.05$    &$+0.09\pm 0.04$  &       
          &$+0.14\pm0.05$     &$+0.14\pm0.05$  \\         
          &                   &                 &     &                  &                \\
NGC\,1052 &$13\pm1$     &$12\pm1$   &     &$+0.38 \pm0.01$    &$+0.25\pm0.01$  &    & $+0.08\pm0.05$    &$+0.25\pm0.04$  &
          &$+0.40\pm0.05$     &$+0.00\pm0.05$   \\             
          &                   &                 &     &                  &                \\
         
NGC\,1209 &$10\pm1$     &$10\pm1$   &     &$+0.295 \pm0.004$    &$+0.273\pm0.003$  &    & $+0.29\pm0.04$    &$+0.12\pm0.04$  &
          &$+0.00\pm0.05$     &$+0.21\pm0.05$   \\
          &                   &                 &     &                  &                \\
         
NGC\,5812 &$10\pm1$      &$13\pm1$   &     &$+0.40\pm0.01$    &$+0.37\pm0.01$  &    & $+0.30\pm0.05$    &$+0.20\pm0.05$  &
          &$+0.13\pm0.05$     &$+0.23\pm0.05$   \\
          &                   &                 &     &                  &                \\
         
NGC\,6758 &$10\pm1$     &$11\pm1$   &     &$+0.40\pm0.01$    &$+0.32\pm0.004$   &    & $+0.30\pm0.05$     &$+0.22\pm0.04$  &
          &$+0.14\pm0.05$     &$+0.14\pm0.05$   \\
          &                   &                 &     &                  &                \\
         
NGC\,6861 &$12\pm1$     &$13\pm1$   &     &$+0.40\pm0.01$    &$+0.38\pm0.01$  &    & $+0.10\pm0.05$    &$+0.08\pm0.04$  &
          &$+0.40\pm0.05$     &$+0.40\pm0.05$   \\
          &                   &                 &     &                  &                \\
NGC\,7507 &$11\pm1$     &$10\pm1$   &     &$+0.39\pm0.01$    &$+0.29\pm0.01$  &    & $+0.29\pm0.05$    &$+0.19\pm0.05$  &
          &$+0.13\pm0.05$     &$+0.13\pm0.05$   \\
          &                   &                 &     &                  &                \\
          
NGC\,7796 &$13\pm1$     &$13\pm1$   &     &$+0.394\pm0.003$    &$+0.394\pm0.001$  &    & $+0.31\pm0.04 $   &$+0.27\pm0.04$  &
          &$+0.11\pm0.05$     &$+0.17\pm0.05$   \\
          &                   &                 &     &                  &                \\
\hline
\end{tabular}
}
\label{result_compare_sint}
\end{footnotesize}
\end{table*}

\begin{figure*}
\centering
\includegraphics*[angle=-90,width=0.8\columnwidth]{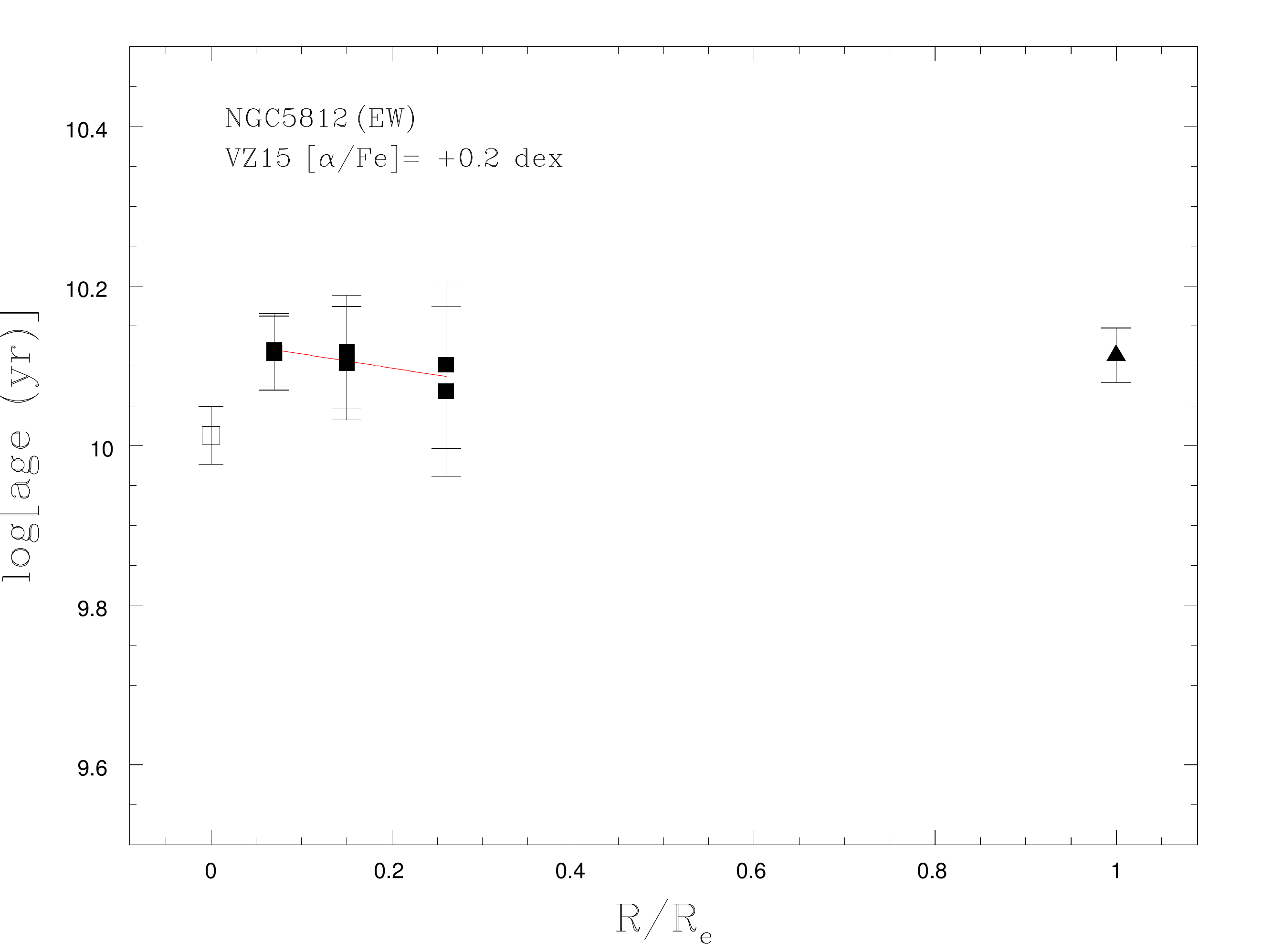}
\includegraphics*[angle=-90,width=0.8\columnwidth]{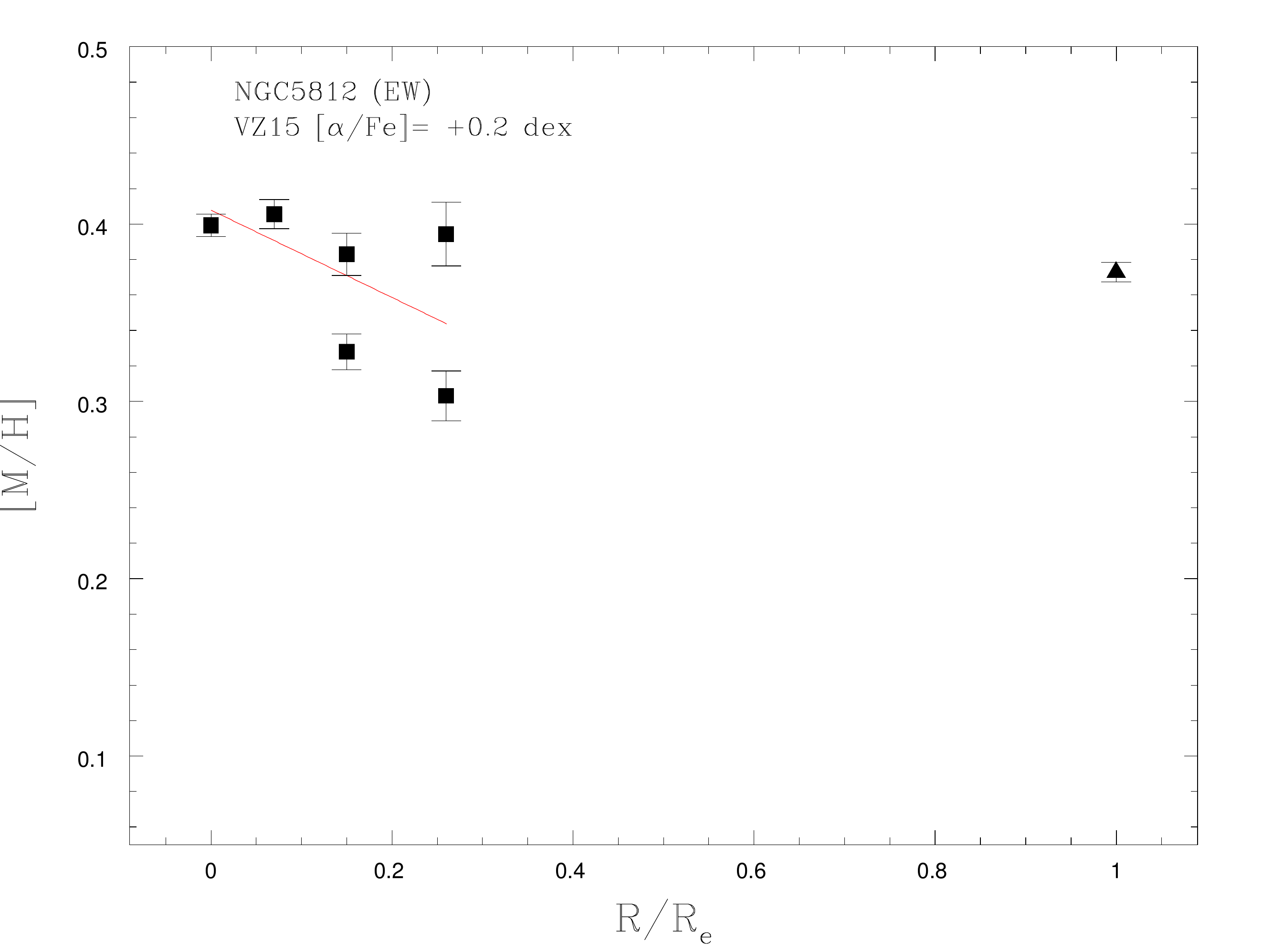}
\includegraphics*[angle=-90,width=0.8\columnwidth]{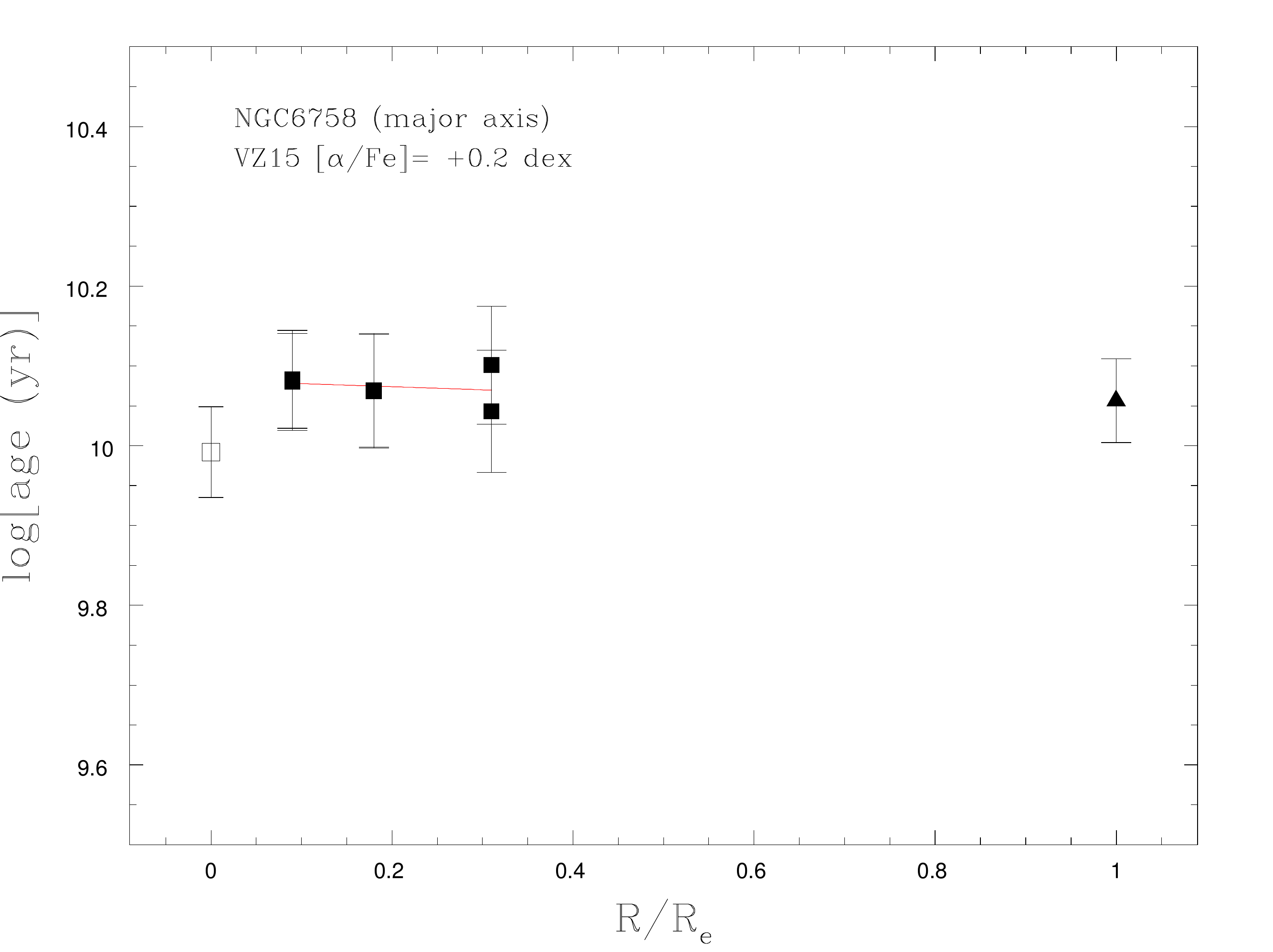}
\includegraphics*[angle=-90,width=0.8\columnwidth]{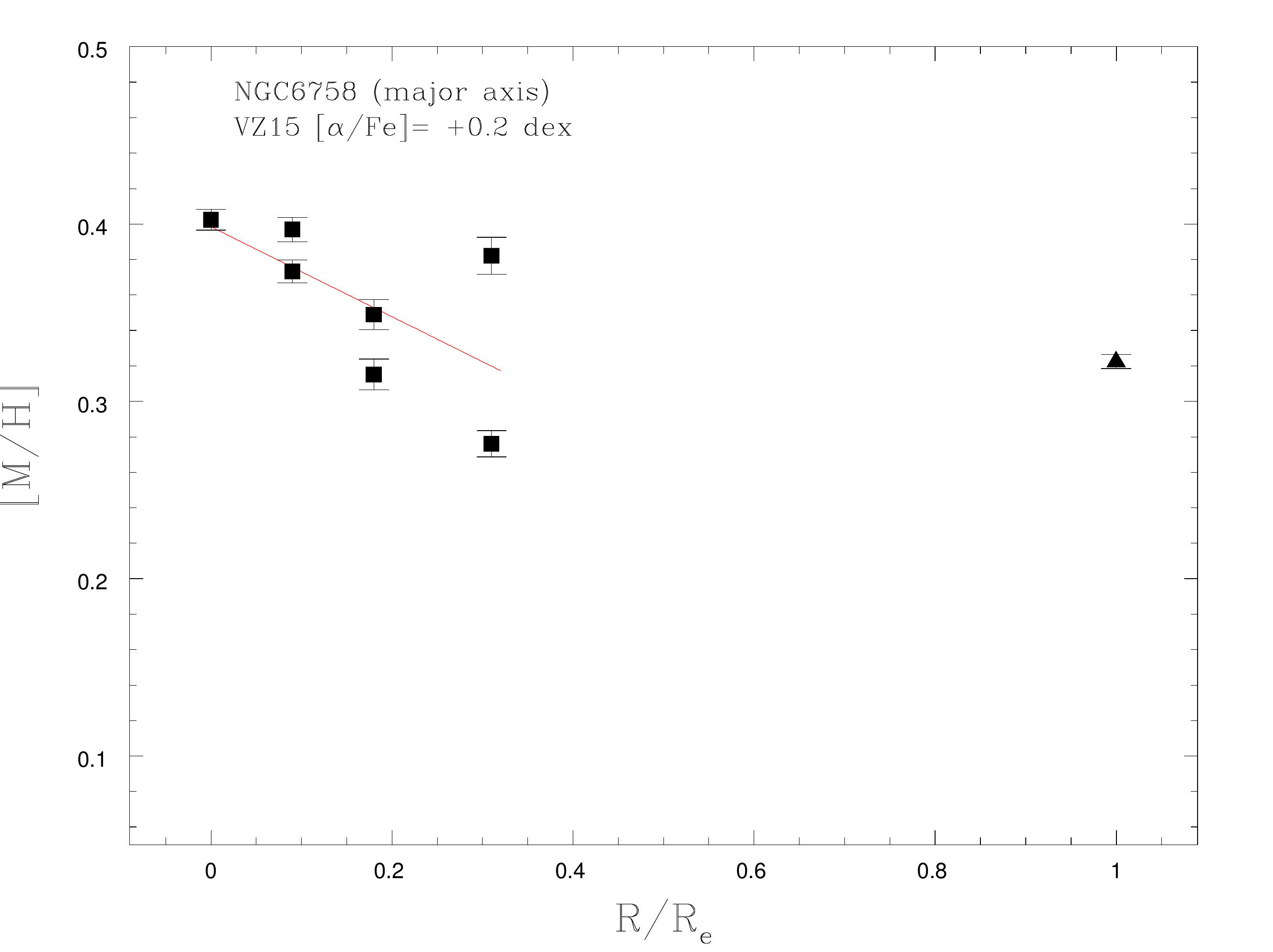}
\includegraphics*[angle=-90,width=0.8\columnwidth]{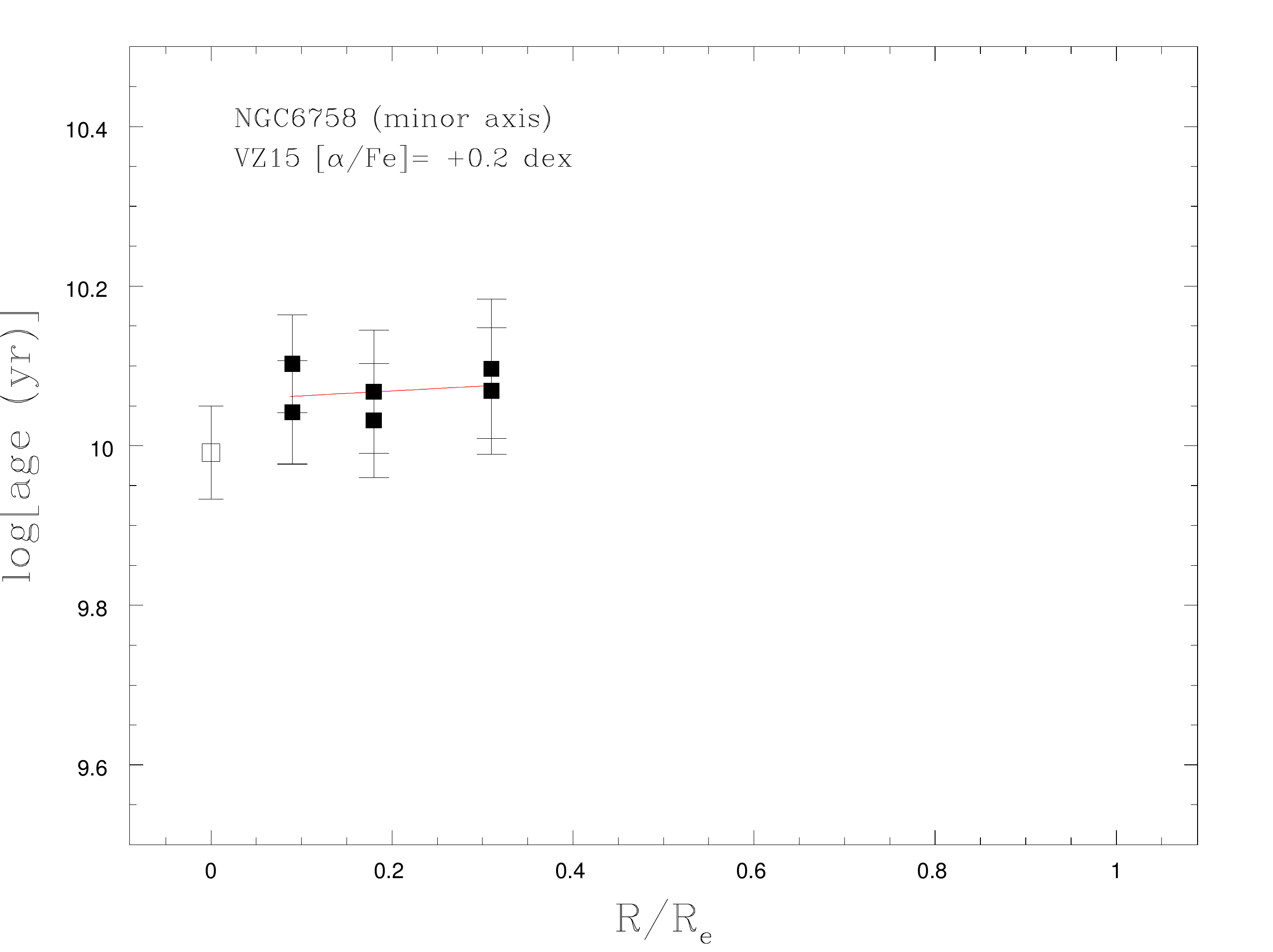}
\includegraphics*[angle=-90,width=0.8\columnwidth]{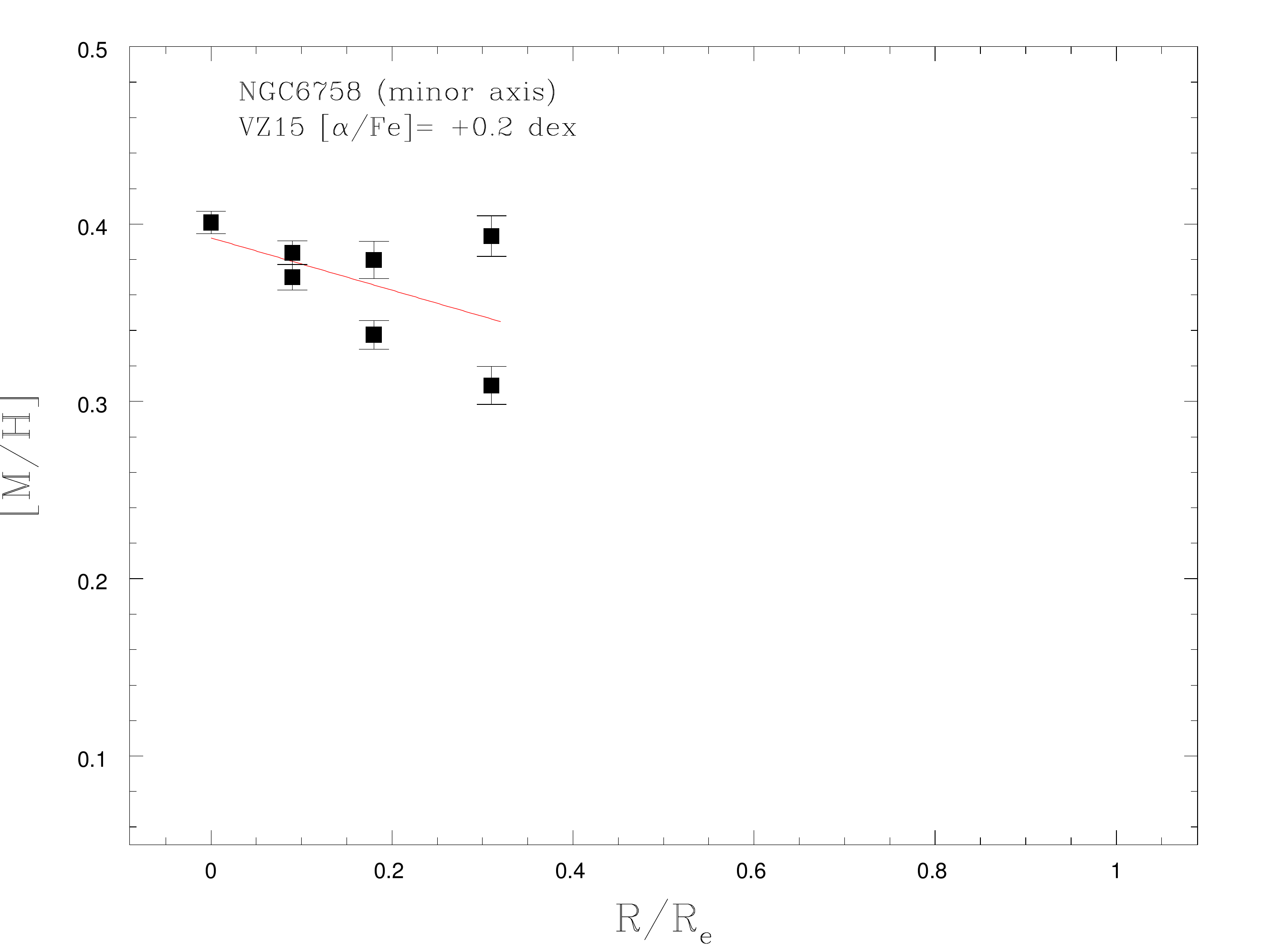}
\caption{Same as Fig.~\ref{age_all_a}, but for NGC\,5812 and NGC\,6758 as indicated.}
\label{age_all_b}
\end{figure*}

\begin{figure*}
\centering
\includegraphics*[angle=-90,width=0.8\columnwidth]{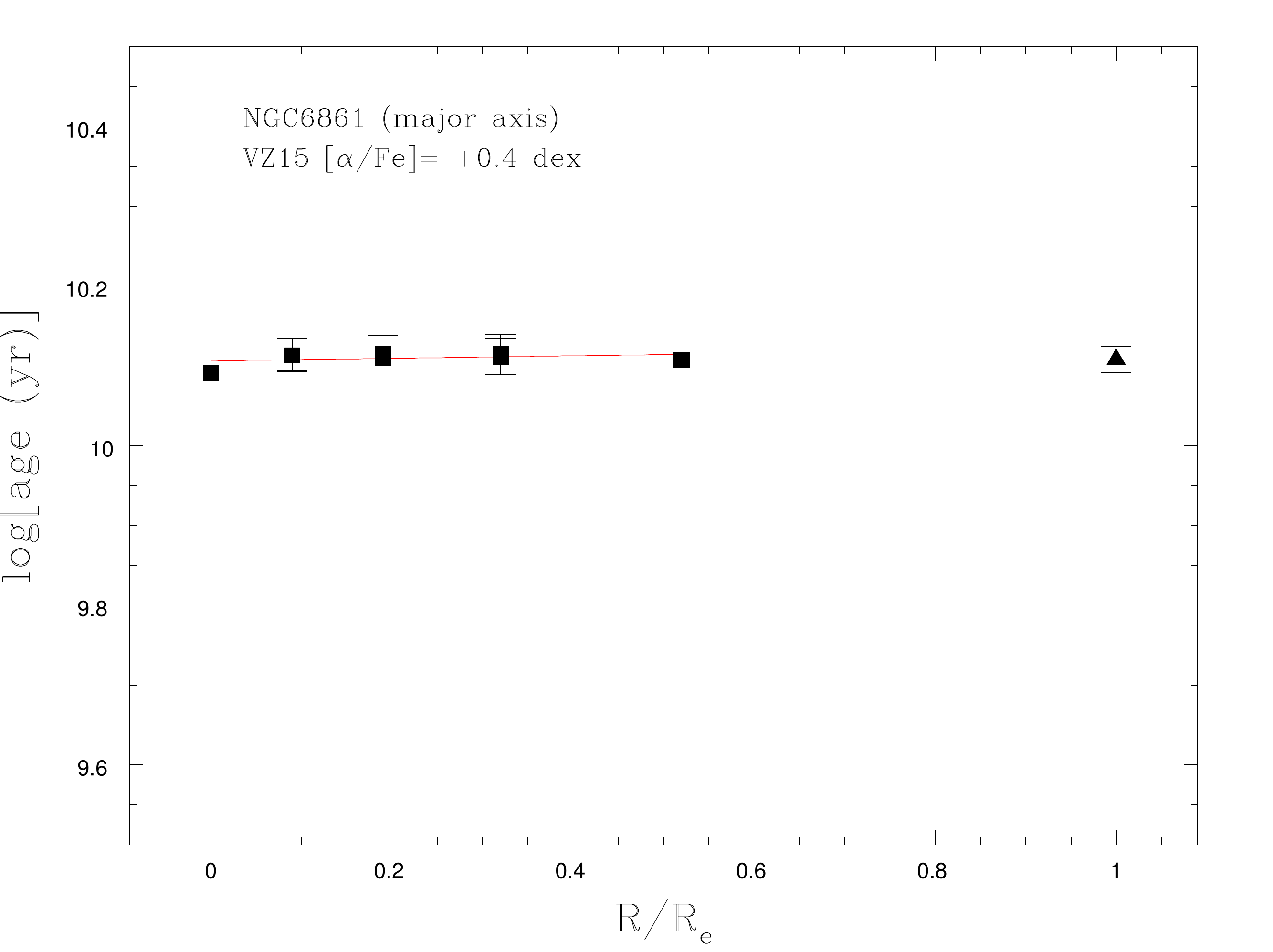}
\includegraphics*[angle=-90,width=0.8\columnwidth]{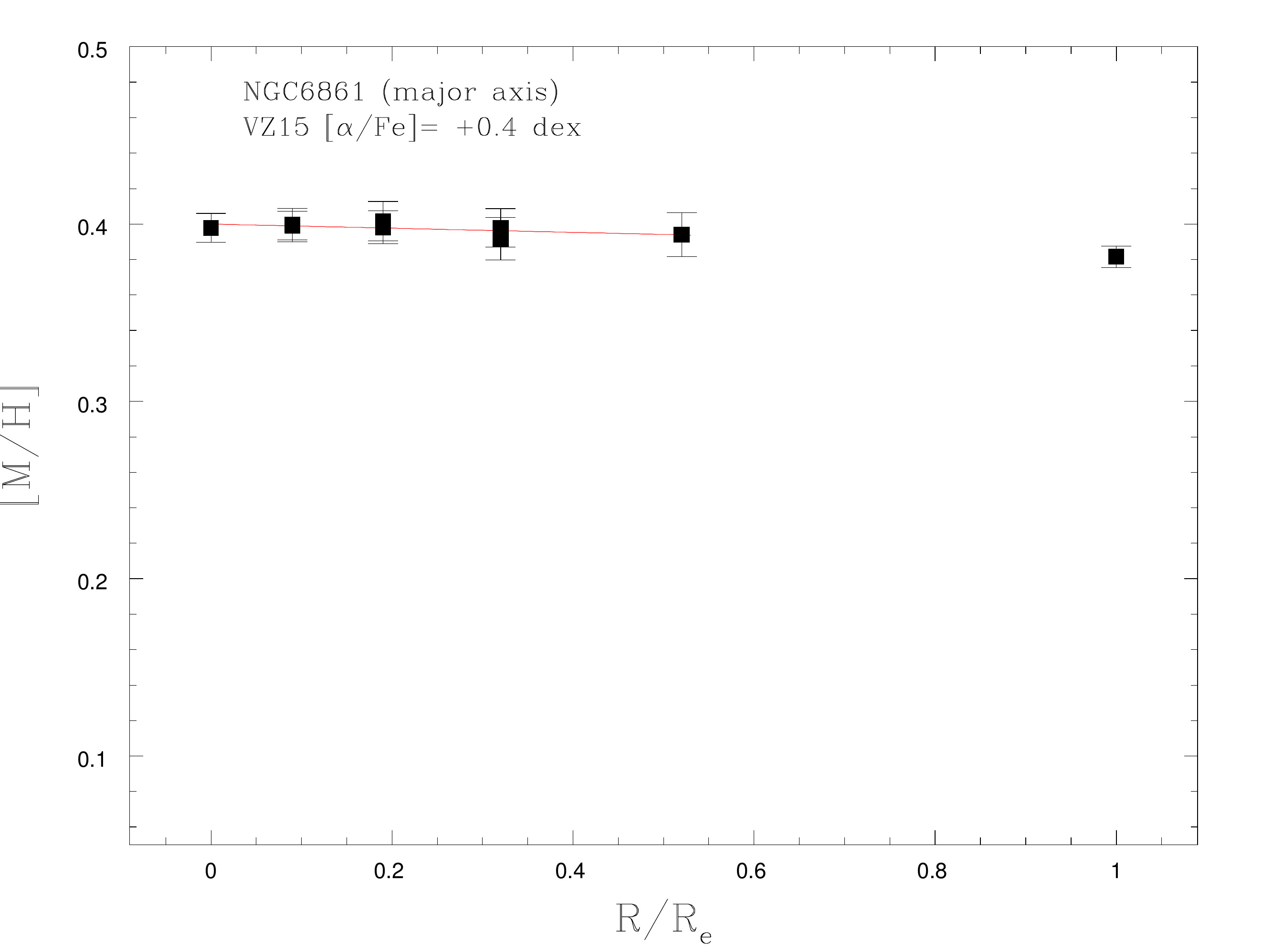}
\includegraphics*[angle=-90,width=0.8\columnwidth]{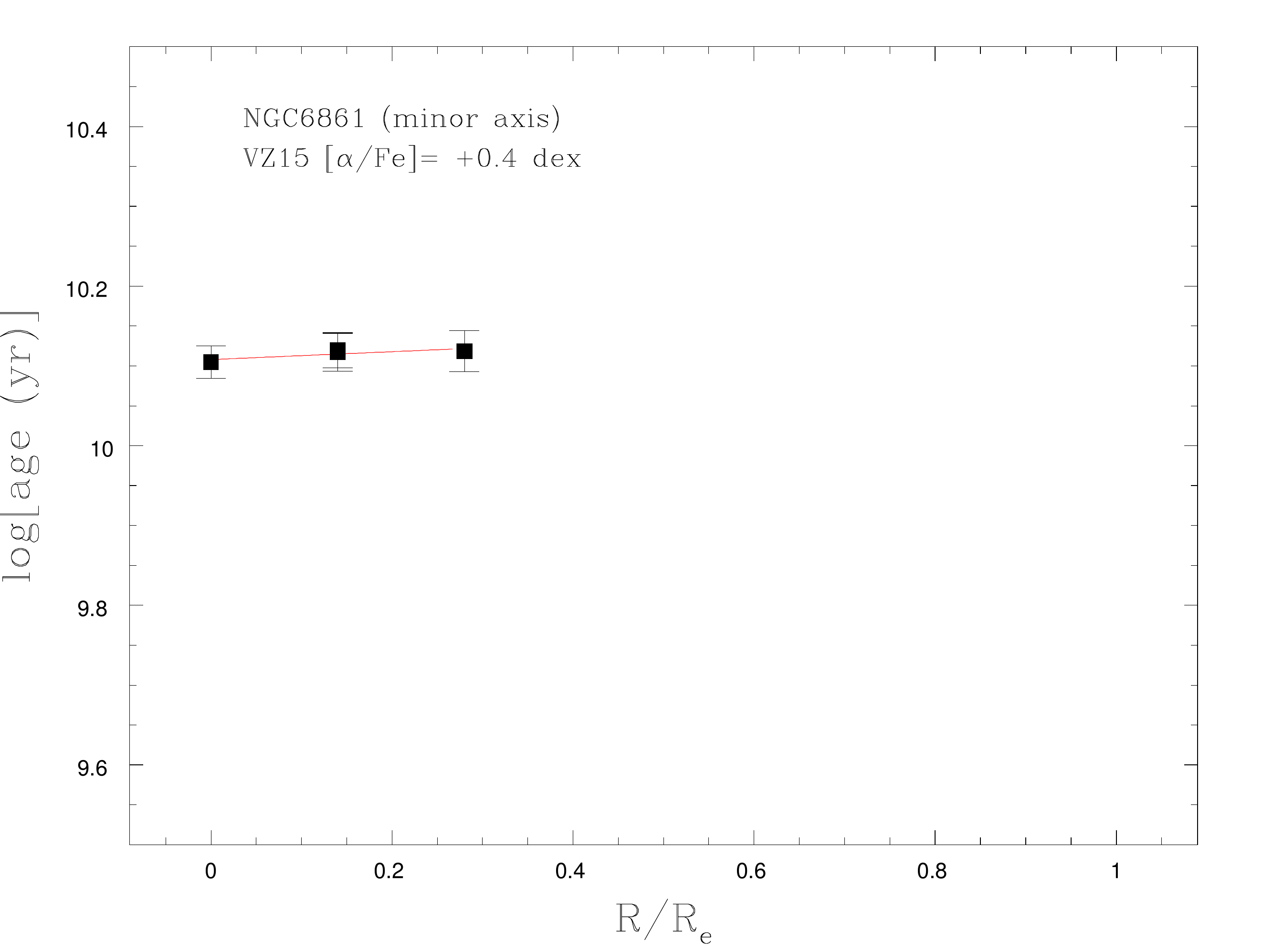}
\includegraphics*[angle=-90,width=0.8\columnwidth]{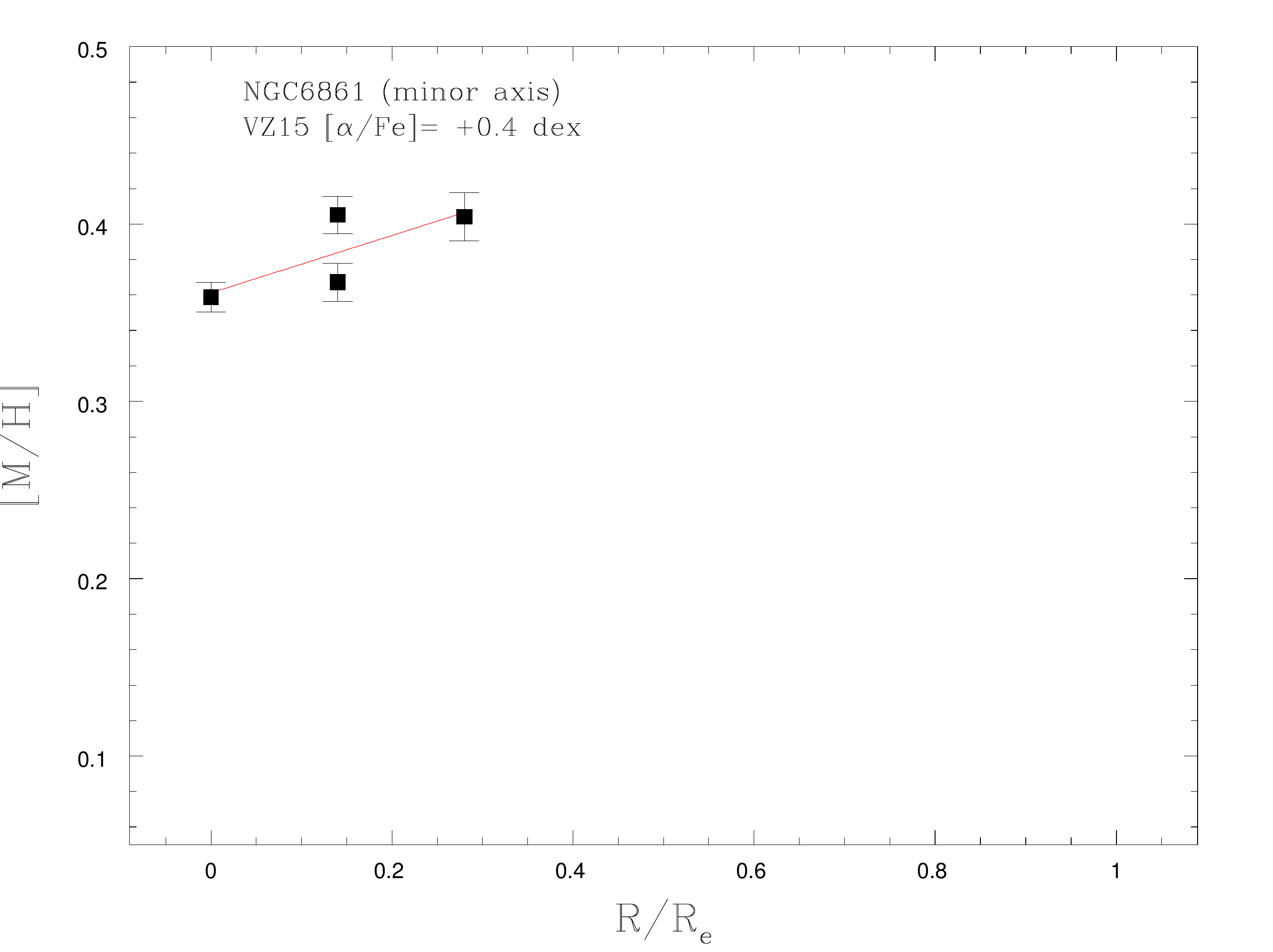}
\caption{Same as Fig.~\ref{age_all_a}, but for  NGC\,6861.}
\label{age_all_c}
\end{figure*}

The radial age and metallicity gradients,
measured up to 0.15\,--\,0.40\,$R/R_{\rm e}$ from the {\scriptsize\,STARLIGHT} stellar population synthesis,
indicate that the stellar populations are evolved, metal rich, and $\alpha$-enhanced.
The stellar age is radially uniform over the observed regions reaching a value around $12-13$\,Gyr
(except in NGC\,7507, for which we measured 10\,Gyr).
The stellar total metallicity exhibits
a negative radial gradient in IC\,5328, NGC\,1052, NGC\,5812, and NGC\,7507,
for which the variations in [M/H] are many times greater than its typical error;
a flat gradient in NGC\,1209, NGC\,6758, and NGC\,7796;
and a positive gradient over the minor axis only of NGC\,6861,
for which the variation in [M/H] is curiously three times greater than its typical error.
For [$\alpha$/Fe], our stellar population synthesis results denote an over-solar value for the
majority of sample galaxies such that some have a constant $\alpha$-enhancement
(IC\,5328, NGC\,5812, NGC\,6758, NGC\,7507, and NGC\,7796 with [$\alpha$/Fe]$=+0.2$ dex and NGC\,6861 with +0.4 dex).
Only two galaxies have variable $\alpha$-enhancement:
NGC\,1052 with a high negative gradient and NGC\,1209 with a moderate positive gradient.
In terms of the radial variation in [Fe/H],
IC\,5328, NGC\,1209, NGC\,5812, and NGC\,7507 have a negative gradient;
NGC\,6758 and NGC\,7796 have a constant value;
NGC\,1052 has a positive gradient;
and NGC\,6861 has a positive gradient over the minor axis only and a constant value over the major axis.

\begin{figure*}
\centering
\includegraphics*[angle=0,width=\columnwidth]{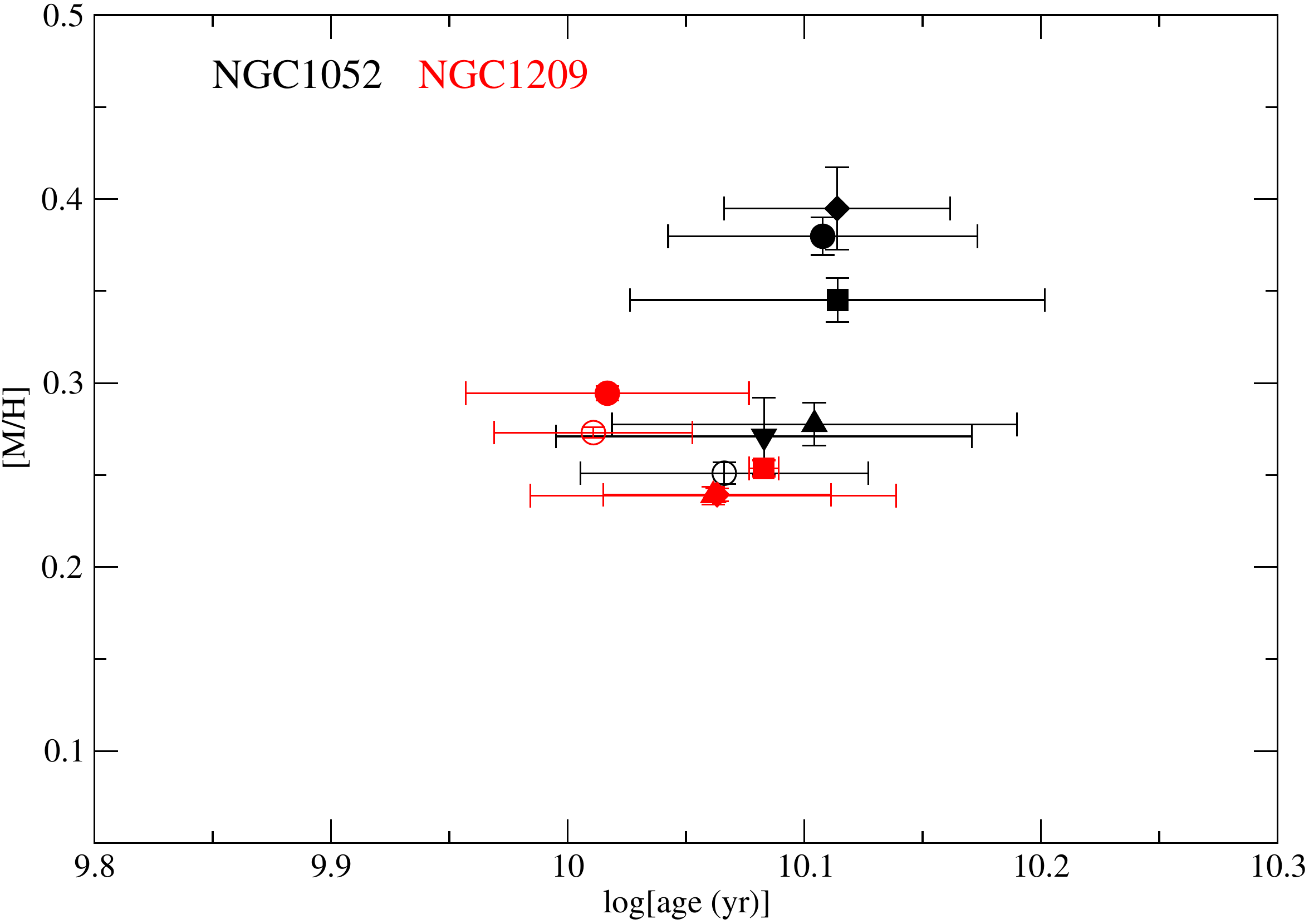}
\caption{Luminosity-weighted average global metallicity [M/H] as a function of luminosity-weighted average age
as derived from the {\scriptsize\,STARLIGHT} stellar population synthesis in different observed radial distances
for NGC\,1052 and NGC\,1209 across major axis.
Different radial distances are assigned distinct symbols:
(i) nuclear region is represented by filled circles;
(ii) the closest apertures to the nucleus r$\sim$0.1 $R/R_{\rm e}$ by filled diamonds;
(iii) the region r$\sim$0.2 $R/R_{\rm e}$ by filled squares;
(iv) the region r$\sim$0.3 $R/R_{\rm e}$ by pointed side up filled triangles;
(v) the region r$\sim$0.4 $R/R_{\rm e}$ by upside down filled triangle; and
(vi) the integrated region inside 1\,$R/R_{\rm e}$ by an open circle.}
\label{zsolar}
\end{figure*}

\subsection{Chemical enrichment in observed regions of the sample galaxies}
\label{star_2}

With the results from the stellar population synthesis, we were also able to investigate the relation
between age and metallicity measured over different radial distances
to obtain an overall picture of the chemical enrichment
in the region inside one  R$_{\rm e}$  of each galaxy
as traced by the spatial distribution of stellar populations properties.
Figure~\ref{zsolar} shows the luminosity-weighted average global metallicity [M/H] as a function of luminosity-weighted average age
for two sample galaxies as an example (NGC\,1052 and NGC\,1209),
by splitting the observed region of each galaxy up to five radius bins:
$r$\,$\leq$\,1 arcsec (assumed as galactic nucleus for all objects),
$r\,\sim$0.1\,$R/R_{\rm e}$, $r\,\sim$0.2\,$R/R_{\rm e}$, $r\,\sim$0.3\,$R/R_{\rm e}$, and $r\,\sim$0.4\,$R/R_{\rm e}$.
These two galaxies exhibit distinct chemical enrichments 
between each other. The uncertainties in age and [M/H] of each aperture are estimated as a function of the spectral S/N. 
Age and metallicity of radially symmetric apertures are averaged.
The integrated age and metallicity measured inside one R$_{\rm e}$ 
are also presented on the plot [M/H] vs. age.
These stellar population parameters are separately shown, galaxy by galaxy, on similar plots of Fig.~\ref{che2} of Appendix~\ref{ap_synth1}.
The possible relations between age and metallicity over the observed radial distances are individually analysed in the following paragraphs.

For NGC\,1052, there is a narrow radial trend between the global metallicity and age
suggesting a well-defined spatial history for the chemical enrichment that correlates four parameters among each other
(age, global metallicity, iron abundance, and alfa enrichment).
Although the star formation occurred in a relative short time-scale ($\sim$ 1 Gyr),
it seems to have been through an inside-out process that induced a decrease of global metallicity (always over-solar)
together with a decrease of [$\alpha$/Fe] and an increase of iron abundance (major axis direction); see also Fig.~\ref{gra_1052_1209}.
The integrated age and metallicity inside one  R$_{\rm e}$  are very close to the data of other observed apertures.

The case of NGC\,1209 is in some sense unique.
It has a very weak radial tendency between global metallicity and age that can be marginally suggested
despite the spatially homogeneous old age and over-solar global metallicity.
[M/H] is weakly anti-correlated with age such that the stars in the nucleus seem to have been formed
just after the stars in the external regions (a kind of outside-in process)
that could explain the inward decrease of [$\alpha$/Fe] associated with the inward increase of [Fe/H] (see also Fig.~\ref{gra_1052_1209}).
The integrated age and metallicity inside one R$_{\rm e}$  are very close to the data of other observed apertures.

The time-scale of star formation in IC\,5328 was very short as indicated by the small spread in age ($\sim$ 1 Gyr),
which can be associated to the spatially uniform and moderate [$\alpha$/Fe] (+0.2 dex).
We also measured a decrease of global metallicity of around 1 dex ([M/H] scales)
from the nucleus up to r$\sim$0.4 $R/R_{\rm e}$ during this time-scale (major axis direction).
The integrated age and metallicity inside one R$_{\rm e}$  are a kind of mean over all observed apertures.

The cases of NGC\,5812, NGC\,6758, NGC\,6861, NGC\,7507, and NGC\,7796
are quite similar.
The lack of correlation between over-solar global metallicity and old age indicates a rapid star formation with a
strong and fast chemical enrichment.
We measured a spatially homogeneous $\alpha$-enhancement for all of them:
moderate in NGC\,5812, NGC\,6758, and NGC\,7507 ([$\alpha$/Fe]$=+0.2$ dex), and high in NGC\,6861 ([$\alpha$/Fe]$=+0.4$ dex).
The integrated age and metallicity inside one R$_{\rm e}$  are very close to the data of other observed apertures in every galaxy.

In the following paragraphs, we compile our analysis relative to the stellar content
against other literature works in a galaxy-by-galaxy base.

\textbf{IC\,5328}. The integrated stellar populations inside one R$_{\rm e}$ of this galaxy
are 90\,per\,cent old (8\,Gyr\,$<t\leq13$\,Gyr)
and 10\,per\,cent young (100\,Myr\,$<$\,$t$\,$\leq$\,1\,Gyr)
with an average age of 11\,Gyr (light-weighted mean age)
and also metal rich
(light-weighted means
of [M/H]$=+0.19$\,dex, [Fe/H]$=+0.09$\,dex, and 
[$\alpha$/Fe]$=+0.14$\,dex).
The respective stellar population contributions in mass for age
are 99.5\,per\,cent (old) and 0.5\,per\,cent (young).
There are negative radial gradients of [M/H] and [Fe/H]
associated to a homogeneous moderate $\alpha$-enhancement ([$\alpha$/Fe]$=+0.14$\,dex).
Specifically, we found a light fraction contribution of 33\,per\,cent (20\,per\,cent in mass fraction)
from an intermediate-young stellar population (1\,Gyr\,$<$\,$t$\,$\leq$\,4\,Gyr) at $r=0.15$\,$R_{\rm e}$ in
the NE direction (major axis slit direction). Across the minor axis, the population synthesis revealed
light fractions around 14\,per\,cent by a young population at $R/R_{\rm e}=0.24$ and $0.41$
(SE direction, 0.5\,per\,cent in mass fraction) and intermediate-old at $R/R_{\rm e}=0.41$
(NW direction, 6\,per\,cent in mass fraction).
Except for [$\alpha$/Fe],
the estimates of age and [Fe/H] at nucleus by \citet{2002MNRAS.330..547T} ($R_{\rm e}$/8)
do not agree with our measured light-weighted means
(12\,Gyr, [M/H]$=+0.37$\,dex, [Fe/H]$=+0.27$\,dex, and [$\alpha$/Fe]$=+0.14$\,dex).
The current work was the first to measure the radial gradients of stellar population properties for this galaxy.
As perspective, we could investigate whether there is a relation
between the presence of young/intermediate-young populations
and the nuclear gas emission.

\textbf{NGC\,1052}. The integrated stellar populations within one  R$_{\rm e}$  for this object
are 12\,Gyr old and have [M/H]$=+0.25$, [Fe/H]$=+0.25$, and [$\alpha$/Fe]$=0.00$\,dex
(light-weighted means).
The radial gradients measured up to about 0.3\,$R_{\rm e}$
are negatively steep in [M/H] and [$\alpha$/Fe]
(outward variations of -0.24\,dex and -0.4\,dex respectively),
associated with a positive gradient in [Fe/H] (+0.2\,dex of outward increase).
Age shows a shallow negative gradient (variation of only -1\,Gyr).
Our measurements of stellar population properties in the nuclear region of {\bf NGC\,1052}
corroborate previous estimates, for example, those by \citet{2007A&A...463..455A}
for the $R_{\rm e}$/8 aperture.
We speculate that the AGN activity in this elliptical
may have been strong in the past
to quench the star formation in the nucleus
and to justify the great nuclear $\alpha$-enhancement ([$\alpha$/Fe]$=0.40$\,dex)
that rapidly decreases outwards, as also identified by \cite{2007A&A...469...89M} 
based on comparison of Lick indices against SSP models.
The radial decrease of [$\alpha$/Fe] and increase of [Fe/H] across the major axis
might be explained by an inside-out star formation process.

\textbf{NGC\,1209}. The integrated stellar populations within one  R$_{\rm e}$ of this galaxy 
are 10\,Gyr old and have [M/H]$=+0.27$\,dex,
[Fe/H]$=+0.25$\,dex, and [$\alpha$/Fe]$=0.00$\,dex (light-weighted means).
The radial gradient, measured up to about 0.3\,$R_{\rm e}$,
is negative in [Fe/H] and positive in [$\alpha$/Fe]
(outward variation of -0.2\,dex and +0.2\,dex respectively from $r=0$),
associated with a mull gradient in [M/H].
Age does not show any change.
Our measurements of total metallicity [M/H] in the nuclear region of {\bf NGC\,1209}
agree with \citet{2007A&A...463..455A} ($R_{\rm e}$/8 aperture),
but our age quantification differs from their estimate by 5\,Gyr. 
Other previous investigations obtained some contradictory results. On the one hand, 
\citet{2002MNRAS.330..547T}, by analysing high quality line index data, derived stellar populations
with mean age around 15\,Gyr and solar metallicity [Fe/H]$=-0.01$\,dex. 
\citet{2005AJ....130.2065H} found that old metal-rich populations dominate the nucleus
(age$=15.6$\,Gyr, [Z/H]$=+0.28$ dex, [$\alpha$/Fe]$=0.23$ dex).
On the other hand, \citet{2007MNRAS.381.1711I}, by applying parametric calibrations of Lick indexes,
concluded that the mean stellar population age in NGC\,1209 is about 9.3\,Gyr associated to an over-solar metallicity.
The radial distribution of stellar populations in this galaxy is indeed peculiar.
This behaviour could be due to an outside-in star formation process, which would
explain the gradual decrease of $\alpha$-enhancement inwards.

\textbf{NGC\,5812}. The  integrated stellar populations within one  R$_{\rm e}$ of this object
are 13\,Gyr old and have [M/H]$=+0.37$\,dex, [Fe/H]$=+0.20$\,dex,
and [$\alpha$/Fe]$=+0.23$\,dex
(light-weighted means).
The radial gradient, measured up to about 0.3\,$R_{\rm e}$,
is slightly negative in age (outward variation of -1\,Gyr from $r=0$),
negative in [M/H] (variation of -0.07\,dex outwards from $r=0$),
and positive in [Fe/H] (variation of +0.07\,dex outwards from $r=0$).
There is no variation in [$\alpha$/Fe].
Only in the nuclear region ($r$\,$\leq$\,1\,arcsec),
the stellar population synthesis revealed a contribution of
24\,per\,cent in flux fraction (20\,per\,cent in mass fraction)
by an intermediate-old population (4\,Gyr\,$<$\,$t$\,$\leq$\,8\,Gyr),
such that the light-weighted age reaches a smaller value of 10 Gyr.
Our results in stellar population properties agree quite well with those measured by \citet{2007A&A...463..455A}
for the $R_{\rm e}$/8, but marginally for age (8.5\,Gyr was their estimate). 
The stellar content of NGC\,5812 has also been analysed by some works 
\citep{2000AJ....119.1645T,2000AJ....120..165T,2005ApJ...621..673T,2006A&A...457..809S,2007A&A...463..455A,2013MNRAS.431..440F}.
However, the nuclear aperture is excluded to compute the age gradient
by a two $\sigma$ clipping, which is why the age gradient is estimated as negative.
If this intermediate-old age contribution is real,
some more extended star formation could have occurred in this ETG, which would
justify its homogeneous moderate $\alpha$-enhancement ([$\alpha$/Fe]$=+0.2$\,dex).

\textbf{NGC\,6758}. The integrated stellar populations within one  R$_{\rm e}$ of this galaxy
are 11\,Gyr old and have [M/H]$=+0.32$\,dex, [Fe/H]$=+0.20$\,dex,
and [$\alpha$/Fe]$=+0.14$\,dex (light-weighted means).
There is no radial gradient in age, as measured up to about 0.3\,$R_{\rm e}$.
The gradient is negative in [M/H] (variation of -0.15\,dex outwards from $r=0$)
and positive in [Fe/H] (variation of +0.15\,dex outwards from $r=0$).
A moderate $\alpha$-enhancement ([$\alpha$/Fe]$=+0.14$\,dex)
is radially constant.
Just in the nuclear aperture of minor axis exposure, the stellar population synthesis indicated
a contribution of 19\,per\,cent in flux fraction (or 9\,per\,cent in mass fraction)
by an intermediate-young population,
4\,per\,cent by an intermediate-old population
and 77\,per\,cent in flux fraction by an old population.
The major axis of the nuclear aperture provides a contribution of 6\,per\,cent in flux fraction
(or 3\,per\,cent in mass fraction) by this intermediate-young population
and 94\,per\,cent in flux fraction by an old population.
Our results in stellar population properties compared with those measured by \citet{2007A&A...463..455A}
for the $R_{\rm e}$/8 only agree marginal because they estimated
16\,Gyr, solar metallicity, and a greater $\alpha$-enhancement ([$\alpha$/Fe]$=+0.32$\,dex).
\citet{2007A&A...463..455A}, for instance, measured for the $R_{\rm e}$/8 aperture, 
as luminosity-weighted average stellar population properties,
an extremely old age (16.0$\pm$2.5\,Gyr), a solar metallicity ($Z=0.016\pm0.002$),
and an over-solar $\alpha$/Fe ratio ([$\alpha$/Fe]$=+0.32\pm0.05$ dex).
If the contribution of an intermediate-young population at the centre is actually valid,
a long star formation process could have occurred in this ETG nuclei, which would 
justify its homogeneous moderate $\alpha$-enhancement ([$\alpha$/Fe]$=+0.14$\,dex).

\textbf{NGC\,6861}. For this galaxy the integrated stellar populations within one  R$_{\rm e}$ 
are 13\,Gyr old and have [M/H]$=+0.38$\,dex, [Fe/H]$=+0.08$\,dex,
and [$\alpha$/Fe]$=+0.40$\,dex (light-weighted means).
All gradients in stellar population properties (age, [M/H], [Fe/H], [$\alpha$/Fe]) are consistent with zero.
Our results from the stellar population synthesis agree
with those estimated by \citet{2001MNRAS.324.1087R,2005MNRAS.364.1239R},
who stated that representative stellar population properties should be measured at $r=r_{\rm e}$.
The stars in this ETGs must have been formed very fast
through efficient star formation episode(s) before the system mass assembly,
despite the existence of stellar rotating disc
that could have involved some energy dissipation
and/or could have been formed by a secular process.

\textbf{NGC\,7507}. Its integrated stellar populations within one  R$_{\rm e}$ 
are 10\,Gyr old and have [M/H]$=+0.29$\,dex, [Fe/H]$=+0.19$\,dex,
and [$\alpha$/Fe]$=+0.13$\,dex (light-weighted means).
The globally a contribution by young population
is 9\,per\,cent in light fraction.
\citet{2007A&A...464..853L} found a shorter age (4\,Gyr),
but the over-solar metallicity measured by them is in agreement with our estimate.
The current work was the first that measured the radial gradients of stellar population properties for this galaxy.
There is no radial gradient in age, as measured up to about 0.15\,$R_{\rm e}$.
The gradient is negative in [M/H] (variation of -0.07\,dex outwards from $r=0$)
and positive in [Fe/H] (variation of +0.07\,dex outwards from $r=0$).
A moderate $\alpha$-enhancement ([$\alpha$/Fe]$=+0.13$\,dex)
is radially constant.
The bulk of stars in this ETG nuclei must have been formed
under an efficient process,
but a young population has a small contribution (negligible in mass fraction),
which could be related to some star formation induced by the fusion of two spirals
involving such a small gas mass.

\textbf{NGC\,7796}. Its integrated stellar populations within one  R$_{\rm e}$ 
are 13\,Gyr old and have [M/H]$=+0.39$\,dex, [Fe/H]$=+0.27$\,dex,
and [$\alpha$/Fe]$=+0.17$\,dex (light-weighted means).
The current work was the first to measure the radial gradients of stellar population properties for this galaxy.
All gradients in stellar population properties (age, [M/H], [Fe/H], [$\alpha$/Fe]) are consistent with zero
(radial distance up to about 0.35\,$R_{\rm e}$).
\citet{2005ApJ...621..673T} and \citet{2007A&A...469...89M} had estimated old age and metal rich populations for this ETG.
However, \citet{2007A&A...469...89M} wrongly suggested negative gradients
in age, total metallicity, and [$\alpha$/Fe].
The stars in this ETGs must have been formed very fast
through efficient star formation episode(s) before the system mass assembly.

\subsection {Warm ionized gas detection in five sample galaxies}
\label{cine3}

We detected  emission lines for five galaxies of our sample after the subtracting of the underlying stellar population spectrum, which was derived 
from the stellar population synthesis.  For IC\,5328, NGC\,1052, NGC\,1209, and  NGC\,6758, we
identified the extended emission lines H\,$\beta$, [O{\scriptsize\,III}]$\lambda$4959{\AA}
and/or [O{\scriptsize\,III}]$\lambda$5007{\AA}. For NGC\,6861,
we only measured emission at H\,$\beta$ for its nuclear aperture spectrum ($r$\,$\leq$\,1 arcsec).
For IC\,5328 only, no previous work had identified these lines \citep[e.g.][]{1997ApJS..111..181D},
probably because the lines are very weak
and the stellar contribution had not been adequately considered.
From these galaxies, only IC\,5328 and NGC\,6861 did not have a classification in the literature for their gas emission.

Emission line  diagnostic diagrams have been extensively used to distinguish among H{\scriptsize\,II} regions 
(photoionization by stars), AGN, and LINER natures. The most widely adopted method is based on the BPT diagnostic 
diagrams proposed by \citet{1981PASP...93....5B},
which utilizes line ratios such as [O{\scriptsize\,III}]$\lambda$5007{\AA}/H\,$\beta$
versus [N{\scriptsize\,II}]$\lambda$6584{\AA} lines/H\,$\alpha$ and
[O{\scriptsize\,III}]$\lambda$5007{\AA}/H\,$\beta$
versus [S{\scriptsize\,II}]$\lambda$6717\,6731{\AA} lines/H\,$\alpha$.
Unfortunately, we were not able to  use this type of diagnostic diagrams   
because the spectral coverage of our data does not extend up to the region
of H\,$\alpha$ and [N{\scriptsize\,II}]$\lambda$6584{\AA} lines.

An alternative diagram using stellar mass Mass-Excitation (MEx) diagnostic diagrams proposed by 
\citet{2011ApJ...736..104J,2014ApJ...788...88J}  
combines [O{\scriptsize\,III}]$\lambda$5007{\AA}/H\,$\beta$ and the stellar mass and 
successfully distinguishes between star formation and AGN emission.
This MEx diagnostic diagram can be used to classify the ionization source
of IC\,5328, NGC\,1052, NGC\,1209, and  NGC\,6758,
since we measured the [O{\scriptsize\,III}]$\lambda$5007{\AA}/H\,$\beta$ ratio.

\begin{figure*}
\centering
\includegraphics*[angle=270,width=\columnwidth]{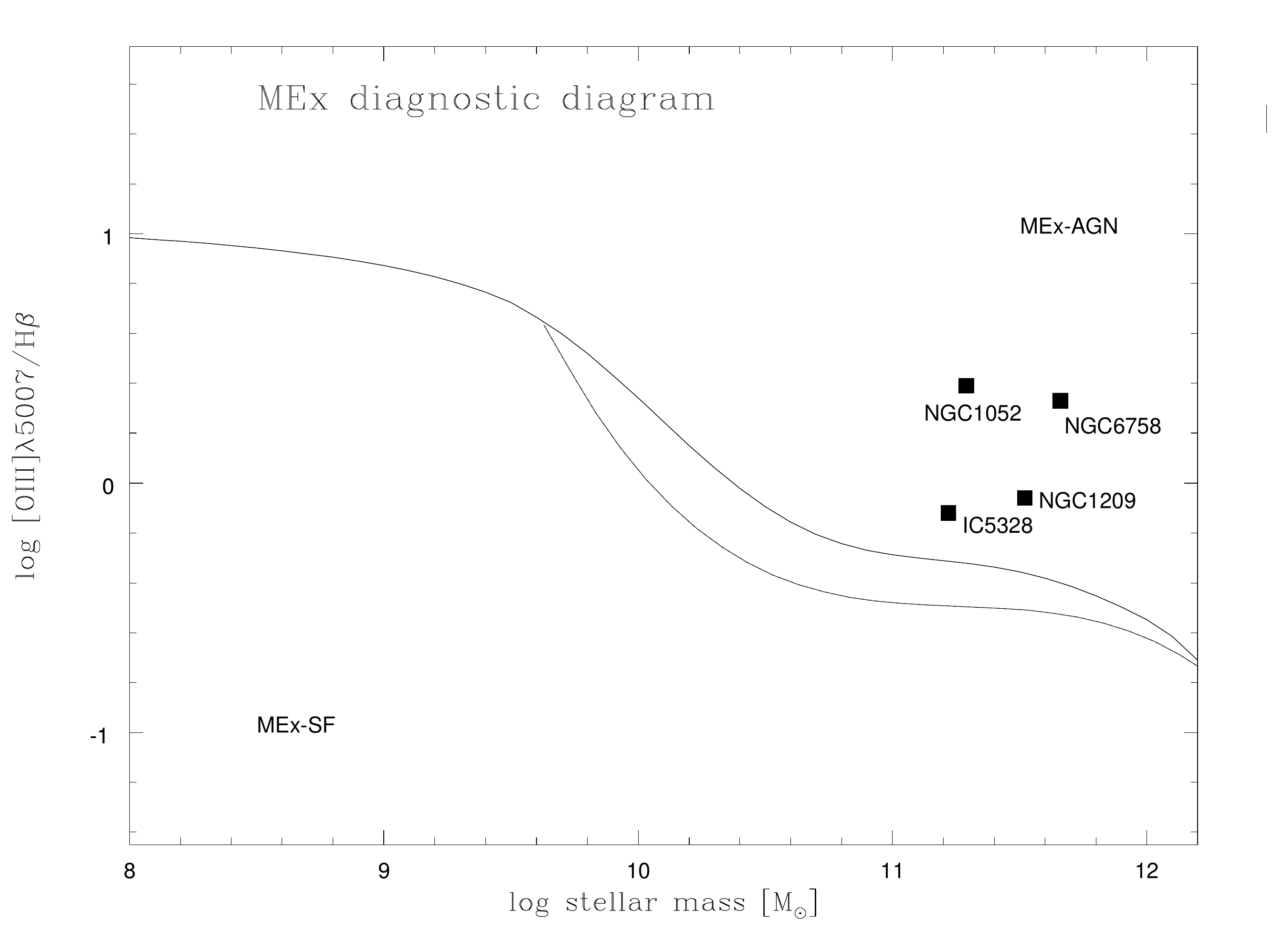}
\caption{Stellar mass as a function of the line ratio [O{\scriptsize\,III}]$\lambda$5007{\AA}/H\,$\beta$. 
The  empirical curves were  taken from \cite{2014ApJ...788...88J} and  the upper mark is the main division between 
star forming galaxies from  AGN.
  The region between the two empirical curves contains a mix of star forming galaxies and AGN. }
\label{bpt_sig} 
\end{figure*}

The MEx diagram has the advantage of requiring
[O{\scriptsize\,III}]$\lambda$5007{\AA}/H\,$\beta$ emission line ratio and, therefore, is  suitable for our cases. 
Fig.~\ref{bpt_sig} shows this diagram for our sample of objects. The stellar mass was estimated by
the equation given by  \cite{2011ApJ...736..104J}, which was established by relating the  stellar masses
derived from SED fitting and rest-frame K-band absolute magnitudes.

As can be seen in this figure, all galaxies occupy the AGN region. To better assess the classes 
of the AGNs, we used the code given by \cite{2011ApJ...736..104J}, 
and calculated the probability that the galaxies belong to the following classes: star-forming, 
composite (mixed star-formation, and AGN), LINER, and Seyfert 2.
According to \cite{2011ApJ...736..104J}, this probabilistic AGN classification scheme has a built-in uncertainty, 
but this is useful to utilize in combination with other alternative diagrams, for which the classification is not certain. 
We found that IC\,5328, NGC\,1052, NGC\,1209, and  NGC\,6758 are probably LINER. 

\section{Discussion}
\label{discute}

One of the open questions
about the formation and evolution of massive early-type galaxies,
even under the paradigm of hierarchical formation scenario,
is which main factor plays a determinant role:
the progenitor mass or the environment  
\citep[the known debate ``nature versus nurture'', e.g.][]{2007MNRAS.376.1445C, 2009A&A...503..379T, 2012MNRAS.427.1530M}.
Specifically, the SFH, chemical evolution, and merger history of a given ETG
are imprinted in the spatial distribution of their stellar populations.

The kinematics of stars and gas can also provide some clues
about the more recent phase of the merger history of a massive ETG.
It is also pertinent to investigate possible relations
among the galaxy, the dynamical state, and the stellar population properties
as well as between the stellar population properties and 
the stellar/gas kinematics (this analysis will be carry out in a future work).

The current sample of ETGs includes slow and fast rotators
distributed over groups and pairs (3 out of 8 galaxies under possible interaction). 
Optical emission lines from ionized gas are in 5 out of 8 objects, 
with 3 of them already confirmed as LINERs in the literature.
The sample galaxies also span non-negligible variations in total luminosity (1 mag in $M_{B}$)
and nuclear velocity dispersion (about 0.3\,dex).
We found that the mean stellar age of the sample objects is $11.4\pm1.3$\,Gyr.
[M/H] covers an interval of around 0.2\,dex,
[Fe/H] changes in about 0.2\,dex,
and [$\alpha$/Fe] ranges from the solar value up to +0.4\,dex.
Therefore, we can explore what the resulting stellar population properties means for the sample ETGs
as a function of the nuclear stellar velocity dispersion in some sense,
distinguishing the sample ETGs as group and pair galaxies,
as interacting and non-interacting galaxies,
and finally galaxies with or without warm gas emission.

We compiled our results about the spatial distribution of stellar population properties galaxy-by-galaxy
in Sect. \ref{stellar2} and Sect.~\ref{star_2}.
Other individual characteristics collected from the literature,
such as morphology, environment, presence/absence of emission lines, and interaction signature
can be found in Appendix~\ref{ap_sample}.

We noticed some trends and correlations
between light-weighted mean stellar population properties
(quantified in the nucleus and inside one  R$_{\rm e}$)
and the nuclear stellar velocity dispersion.
All parameters are dealt with in the logarithm scale.
Nuclear velocity dispersion 
(measurements corrected by the aperture effect and homogenized collected from 
HYPERLEDA\footnote{http://leda.univ-lyon1.fr/} database \citep{2014A&A...570A..13M}: 
is directly related
to the mass inside a circular radius of 813\,pc \citep[Coma normalization,][]{1995MNRAS.276.1341J} 
and indirectly quantifies the total mass of the galaxy.

\begin{figure}
\centering
\includegraphics*[angle=270,width=\columnwidth]{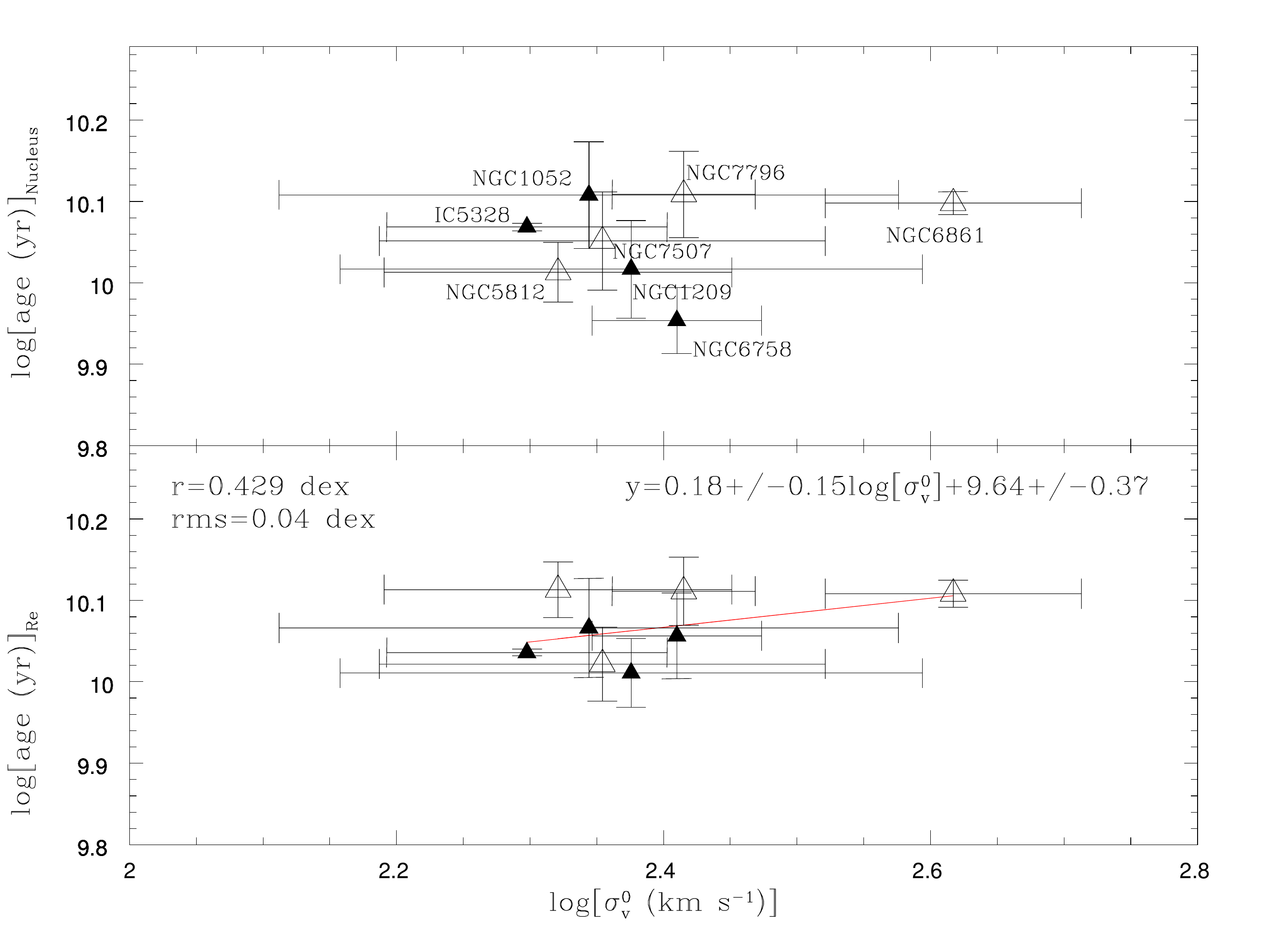}{}
\caption{log(age)$_{\rm Nucleus}$  and log(age)$_{R_{\rm e}}$ versus log($\sigma_{\rm v}^{\rm 0}$). 
ETGs are denoted by triangles. The ones that harbour LINER nuclei are plotted with filled triangles. 
A simple $lsq$ fit is represented by a solid red line, only when the 
linear correlation coefficient $r$ exceeds 0.400. A two-sigma clipping is also adopted to exclude outliers. 
The derived linear correlation is shown in the right top corner, and the fit $r$ and $rms$ are in the 
left top corner, in the case when the fit $r$ is greater than 0.400.}
\label{y}
\end{figure}

\begin{figure}
\centering
\includegraphics*[angle=270,width=\columnwidth]{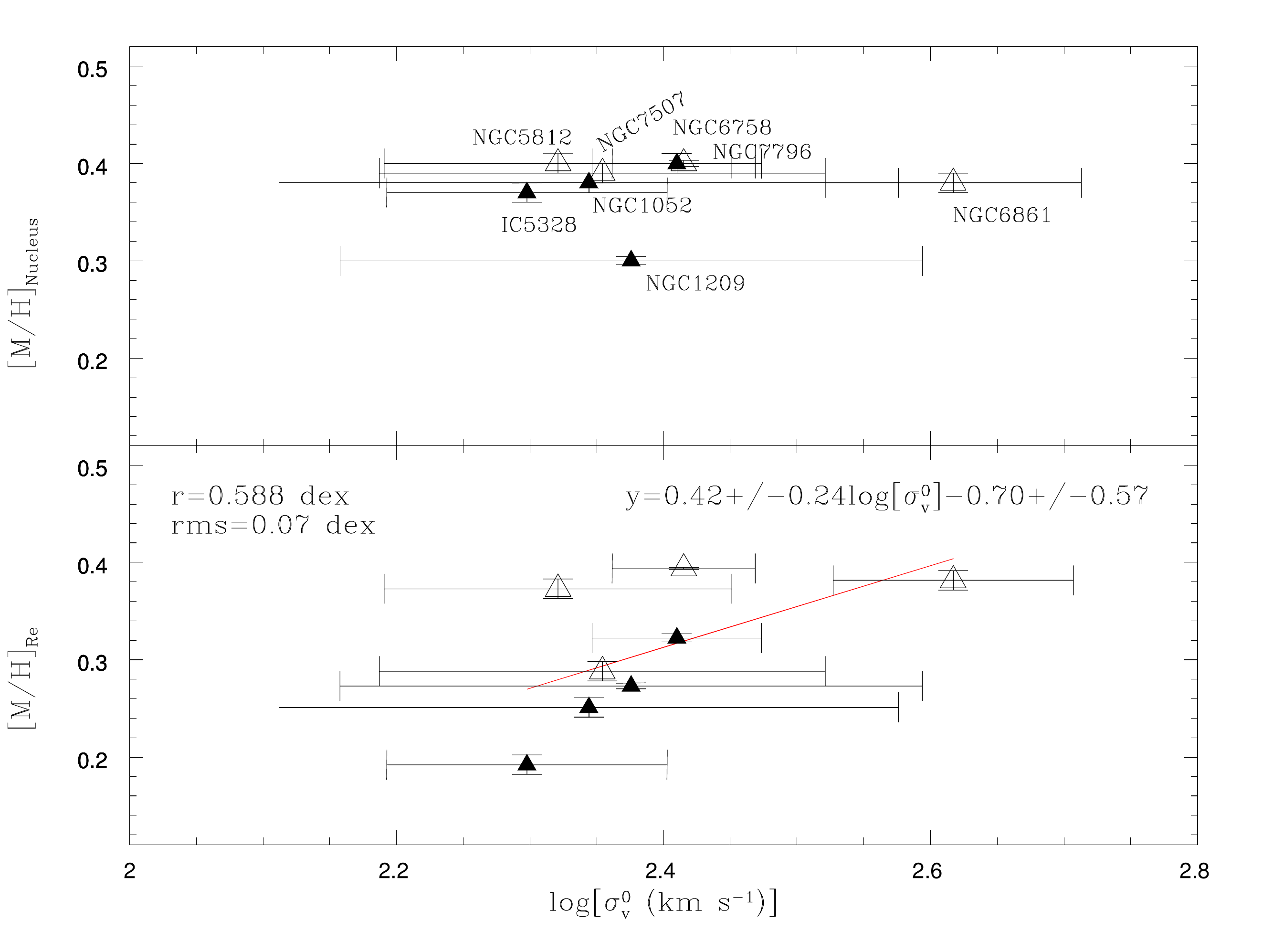}{}
\caption{[M/H]$_{\rm Nucleus}$ and [M/H]$_{R_{\rm e}}$ versus log($\sigma_{\rm v}^{\rm 0}$). Same notation as Figure~\ref{y}.}
\label{w}
\end{figure}

\begin{figure}
\centering
\includegraphics*[angle=270,width=\columnwidth]{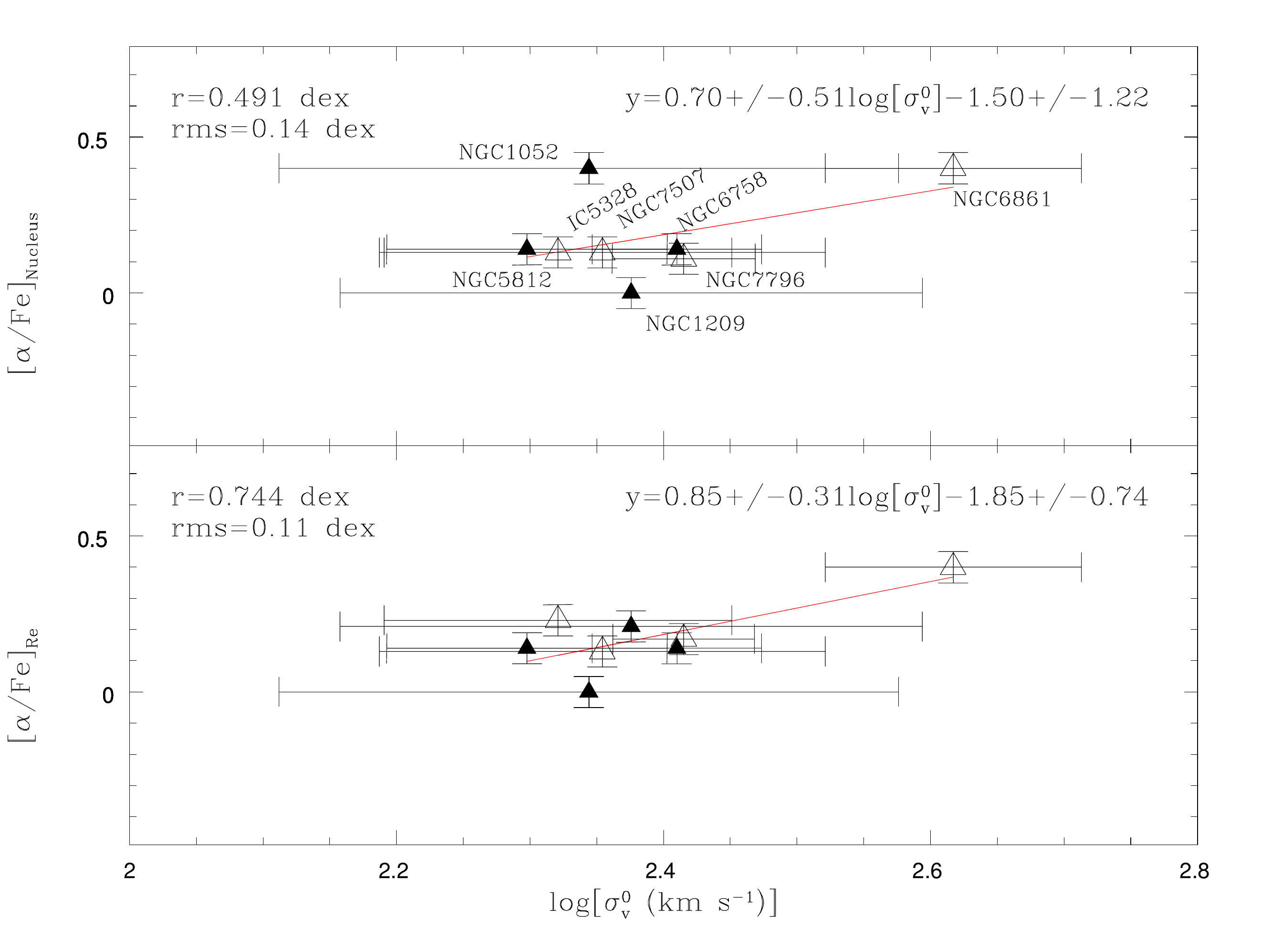}{}
\caption{[$\alpha$/Fe]$_{\rm Nucleus}$ and [$\alpha$/Fe]$_{R_{\rm e}}$ versus log($\sigma_{\rm v}^{\rm 0}$). Same notation as Figure~\ref{y}.}
\label{Z}
\end{figure}

\begin{figure}
\centering
\includegraphics*[angle=270,width=\columnwidth]{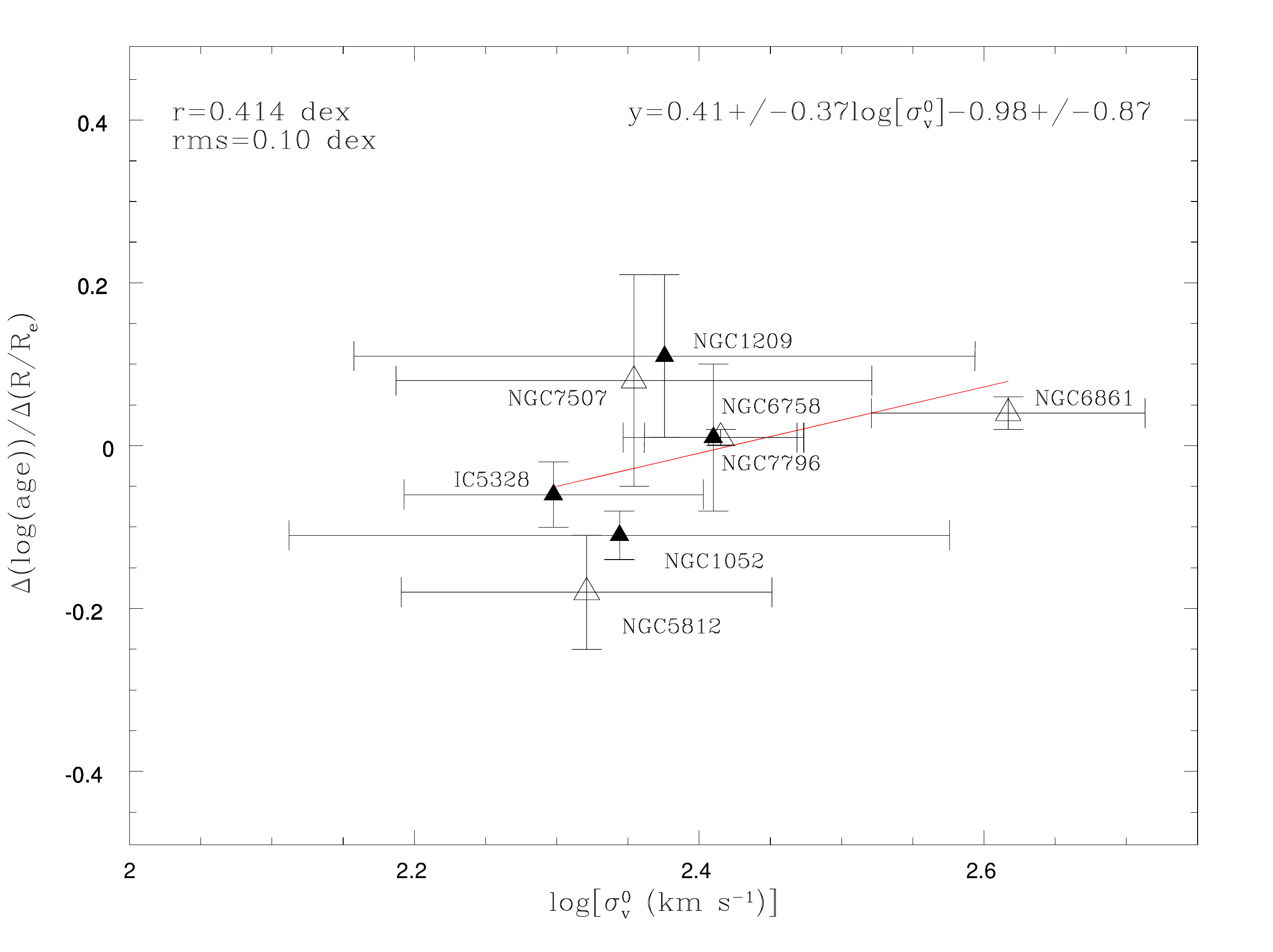}{(a)}
\includegraphics*[angle=270,width=\columnwidth]{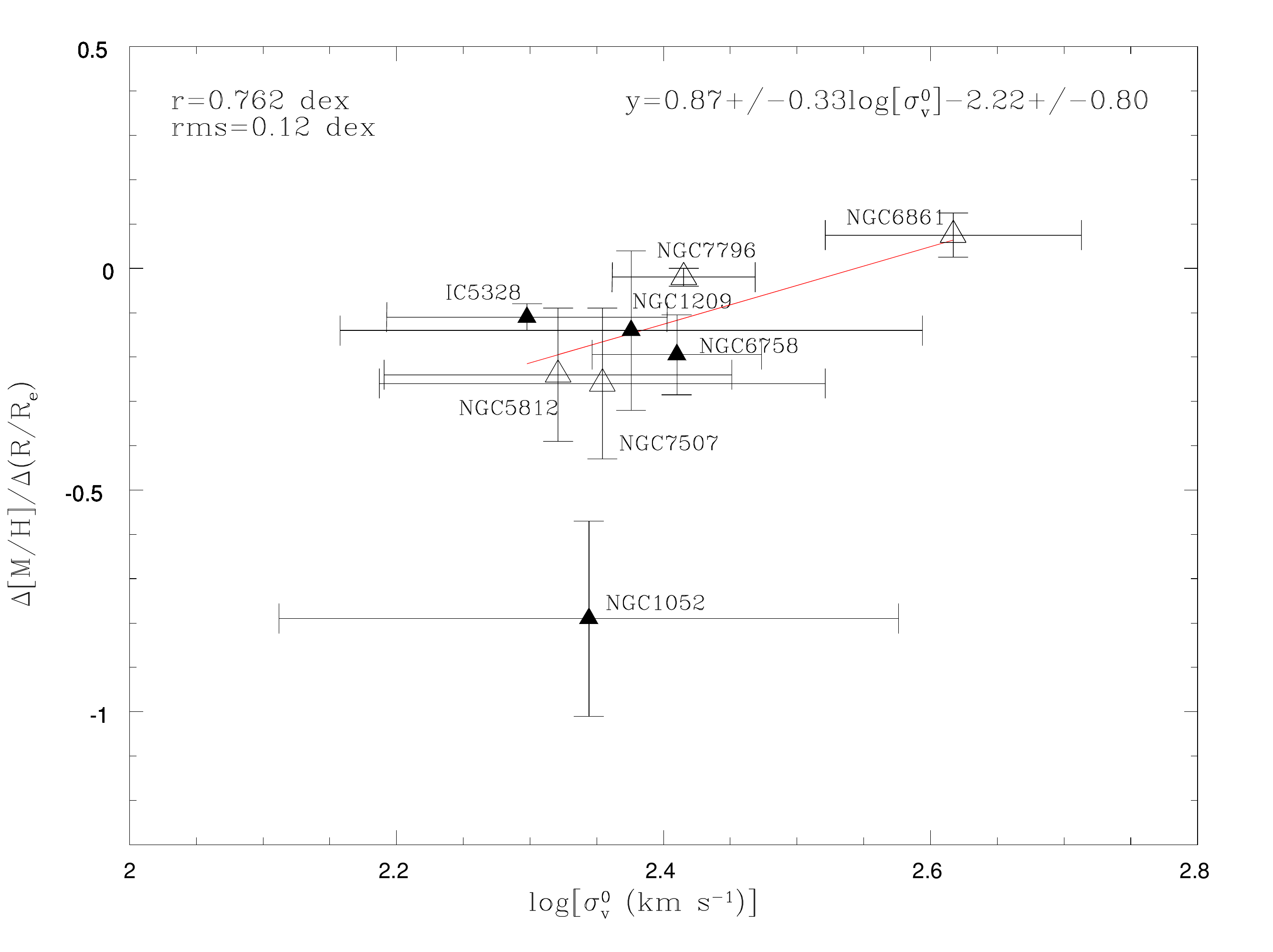}{(b)}
\caption{Age ($\Delta$(log(age))/$\Delta$($R/R_{\rm e}$) and metallicity ($\Delta$[M/H]/$\Delta$($R/R_{\rm e}$)) radial gradients  
versus log($\sigma_{\rm v}^{\rm 0}$). 
Same notation as Figure~\ref{y}.}
\label{A}
\end{figure}

A positive trend exists between log(age) and log($\sigma_{\rm v}^{\rm 0}$)
for the region inside 1 $R_e$ only,
i.e. more massive galaxies have older stars as also found by
\citet{2015yCat..74483484M} and \citet{2000AJ....119.1645T, 2000AJ....120..165T}
(see Fig.~\ref{y}).
There is also a positive correlation
between the integrated [M/H] inside one $R_{\rm e}$ and velocity dispersion,
but this correlation is also erased for the nuclear [M/H] ($r$\,$\leq$\,1\,arcsec);
see Fig.~\ref{w}.
Our results also prove the well-known correlation
between $\alpha$-enhancement and velocity dispersion.
We obtained positive correlations
between them for both estimates of [$\alpha$/Fe]
(nuclear and integrated in 1\,$R_{\rm e}$);
reported in Fig.~\ref{Z}.
Finally, the ETGs of our sample follow a strong positive correlation
between the radial gradient of [M/H] (scale dex/$R/R_{\rm e}$)
and log($\sigma_{\rm v}^{\rm 0}$),
such that more massive galaxies have shallower gradients
or even null gradients in metallicity
(see Fig.~\ref{A}).
Although this relation has a high linear correlation coefficient ($r=0.762$),
we suppose that some bias effect is involved,
especially because we would expect the opposite as quantified by
\citet{2015A&A...581A.103G} and \citet{2010MNRAS.408...97K},
who noticed that the stellar metallicity gradient is roughly dependent on the stellar mass M$_{\odot}$,
up to around 10$^{11}$\,M$_{\odot}$.
Perhaps, one reason is that we are dealing with only luminous ETGs in sparse groups and interacting pairs.

{\bf

Our results corroborate the stellar population analysis of the CALIFA survey from \citep{2015A&A...581A.103G} that included
40 ellipticals (E0-E7) and 32 lenticular (S0) nearby galaxies, which was also carried out with {\scriptsize\,STARLIGHT}. 
The CALIFA observations show that massive ellipticals  exhibit negative mass-weighted metallicity radial gradients,
similar to what we detected in luminosity-weighted  metallicity (CALIFA gradients were measured up to almost 
three effective radius in the case of Es). This CALIFA analysis confirms that  ellipticals are composed of older and more metal 
rich stellar populations than spirals at fixed total stellar mass $M_{\star}$, in agreement with our results. 
It indicates that the star formation quenching does not only depend on the galaxy mass, but also on the morphology. 
We obtained flat radial profiles of age in our sample of ETGs,  in agreement with CALIFA results  \citep{2018MNRAS.475.3700M}, 
which found rather flat age and [Mg/Fe] radial gradients, weakly dependent on the effective velocity dispersion of the
galaxy within half-light radius.}

Furthermore, we found for this sample of ETGs
no distinction between the radial gradients of stellar population properties in ETGs in sparse groups and ETGs in interacting pairs.
{\bf This result agrees with that found by \citet{2017MNRAS.465..688G}, who studied the internal radial gradients of stellar population 
properties within
1.5 $R_{\rm e}$ and analysed the impact of galaxy environment, using  a representative sample of 721 galaxies. These authors suggested 
that galaxy mass is the main driver of stellar population gradients in both
early and late-type galaxies, and any environmental dependence, if present at all, must be very
subtle.}

Additionally, we detected in three of the sample ETGs 
(IC\,5328, NGC\,6758, and NGC\,5812) the presence of intermediate-young and intermediate-old populations. 
Moreover, using the MEx diagnostic diagram, we verified the LINER emission in   
IC\,5328, NGC\,1052, NGC\,1209, and NGC\,6758,
and no correlation was found between the stellar population properties and the LINER presence.

\section{Summary and Conclusions}
\label{conclude}

This work presents detailed results about stellar populations properties
in the region up to one R$_{\rm e}$ 
in a sample of eight massive nearby ETGs; 
all but one are members of sparse groups (three of them in pairs under possible interaction and other three in pairs with weak interaction).
Five galaxies are classified as ellipticals (IC\,5328, NGC\,1052, NGC\,1209, NGC\,5812, and NGC\,7507), two are simultaneously 
classified as ellipticals and giant ellipticals (NGC\,6758 and NGC\,7796) and one as lenticular (NGC\,6861).
Five of them (IC5328, NGC1052, NGC1209, NGC6861, and NGC6758) have LINER nuclei, such that all sample galaxies have
$M_{B}$ $<$\,$M_{B}^{\star}$ at $z=0$.
Long-slit spectra at $\lambda\lambda$4320\,--\,6360\,{\AA}
were analysed across both apparent photometric axes, when possible.

The regions up to one effective radius of all sample galaxies are basically dominated
by alpha-enhanced metal-rich old stellar populations,
likely due to rapid star formation episodes that induced efficient chemical enrichments.
On average, the age and [$\alpha$/Fe] gradients are null, and the [M/H] gradients are negative,
although discordant cases were found
splitting the galaxies with common characteristics in three groups as listed ahead.
We found no correlation between the stellar population properties
and the LINER presence as well as between the stellar properties and environment or gravitational interaction,
suggesting that the influence of progenitor mass can not be discarded in the formation and evolution of
massive ETGs.

Our results on the stellar population synthesis approach
and warm ionized gas
are summarily presented as follows.

\begin{enumerate}

\item The results of stellar population synthesis,
split the galaxies into:
(i) IC\,5328 and NGC\,1052 that have some stellar properties in common as described,
(ii) NGC\,1209 being the only one with a positive [$\alpha$/Fe] gradient, and
(iii) NGC\,5812, NGC\,6758, NGC\,6861, NGC\,7507, and NGC\,7796,
which have uniform old metal-rich stellar populations in the observed regions
in terms of derived age and chemical composition
indicating a rapid star formation with a strong and fast chemical enrichment.
A common stellar property is that all sample galaxies
are dominated by old stellar populations (8$\times10^{9}<\,t\,\leq13\times10^{9}$ years).
IC\,5328 and NGC\,1052 both have a negative gradient of metallicity that changes in the over-solar regime.
While IC\,5328 has a [Fe/H] negative gradient, NGC\,1052 has a positive one.
Specifically, NGC\,1052 and NGC\,1209 have variable [$\alpha$/Fe]
(decrescent outwards for NGC\,1052 and crescent outwards for NGC\,1209).
IC\,5328, NGC\,5812, NGC\,6758, NGC\,7507, and NGC\,7796
have homogeneous moderate $\alpha$-enhancement ([$\alpha$/Fe]$=+0.2$\,dex),
and NGC\,6861 has homogeneous high $\alpha$-enhancement
([$\alpha$/Fe]$=+0.4$\,dex). 
The stellar population synthesis detected contributions greater 
than 20\,per\,cent (flux scale) of intermediate-young populations in IC\,5328 and NGC\,6758 
(age between 1 and 4 Gyr) as well as contributions of intermediate-old populations in NGC\,5812 ($4-8$\,Gyr).
IC\,5328 and NGC\,5812 are galaxies in distinct pairs under possible interaction (both pairs inside groups) 
and NGC\,6758 is galaxy in a group.

\item Concerning the detection of ionized gas,
we can state that extended gas is present in IC\,5328, NGC\,1052, NGC\,1209, and NGC\,6758.
The greatest radial extension of this emitting gas
is across the major axis of NGC\,6758, reaching a radial distance of 0.4\,R/R${e}$.
NGC\,6861 likely owns a very concentrated nuclear ionized gas emission.
The presence of a LINER is identified in IC\,5328, NGC\,1052, NGC\,1209, and  NGC\,6758.

Our results on stellar populations properties (up to 0.5 $R_{\rm e}$)
and ionized gas (up to 0.4 $R_{\rm e}$),
empirically characterize luminous ETGs,
which were selected from nearby pairs and sparse groups including interacting galaxies and LINERs.
These results add detailed information to recover the SFH and chemical enrichment imprinted in
the stellar populations of each individual galaxy as well as to investigate the influence of galaxy mass,
environment, interaction with other galaxy and AGN feedback in the galaxy formation and evolution.

\end{enumerate}

\section*{Acknowledgements}

D. A. Rosa is grateful for the scholarship from CAPES and CNPq 
foundations for the PCI/MCTIC/INPE posdoc fellowship, (process $300082/2016-9$). 
A. Milone and I. Rodrigues thank the Brazilian foundation 
CNPq/MCTIC (grants number $309562/2015-5$, 
$311920/2015-2$, respectively). 
A. Krabbe thanks FAPESP (grant number $2016/21532-9$).
We all acknowledge the usage of the HyperLeda database (http://leda.univ-lyon1.fr).
This research made use of the NASA/IPAC Extragalactic Database (NED), which is operated by the Jet Propulsion Laboratory, 
California Institute of Technology, under contract with the National Aeronautics and Space Administration.
{\scriptsize\,IRAF} is written and supported by the National Optical Astronomy Observatories (NOAO) in Tucson, Arizona.
NOAO is operated by the Association of Universities for Research in Astronomy (AURA), Inc.
under cooperative agreement with the National Science Foundation.

\bibliographystyle{aastex}
\bibliography{publicacao}

\appendix

\section{The early-type galaxies sample } 
\label{ap_sample}

{\bf IC\,5328} is an E4 galaxy, member of a sparse group, with a total of four galaxies based on our criteria (see Section 2),
in which IC\,5328 is the unique ETG. It is the main galaxy of the closest pair of our sample, along with IC\,5328\,A, at a projected 
distance of 8.2\,kpc. IC\,5328\,A is a SB galaxy, much smaller than IC\,5328 with a pure exponential brightness profile.
It shows clear signs of tidal interaction \citep{2004A&A...428..837F} and weak emissions in H$\alpha$ and [N\,{\scriptsize\,II}] lines
possibly due to two identified H\,{\scriptsize\,II} regions
\citep{1997ApJS..111..181D,2004A&A...428..837F,2008A&A...481..645F}.
IC\,5328 is a globally dynamic hot system, but the stars show slow rotation around a minor axis.
Its stellar velocity dispersion profile exhibits irregularities
\citep[as observed by][]{1998A&AS..130..267L},
possibly indicating the presence of a dynamical double core.
We have not found any work relating to radial gradients of stellar population properties for this galaxy.

{\bf NGC\,1052} is a LINER E4 galaxy and the third brightest member
of the sparse group NOGG\,165 with eleven galaxies in total as identified by \citet{2000ApJ...543..178G} (NOGG: Nearby Optical Galaxy Group),
whose brightest galaxy is NGC988 (SB),
or with twelve galaxies according to our criteria.
There are only two ETGs in this group.
This LINER in particular shows variability in the ultraviolet \citep{2005ApJ...625..699M},
near-infrared \citep{2015ApJS..217...13M},
radio \citep{2003A&A...401..113V},
and X-ray \citep{2013A&A...556A..47H}.
Many other works have confirmed the presence of emission lines in the optical and mid-infrared.
Recently, \citet{2014xru..confE..84G}, using an X-ray catalogue with 40 LINERs,
figured out that this galaxy seems to be more like a Seyfert than a LINER.
Its stellar kinematics along both photometric axes has been extensively measured by several works;
\citet{2015ApJS..218...10V} is an example.
Tidal tails have been identified in NGC\,1052 as interaction sign
\citep{1986AJ.....91..791V, 2000AJ....120.1946S}.
Its closest companion is 2MASX\,J02413514-0810243, which makes with NGC\,1052 a weak pair, according to our criteria (see Sec. \ref{amostra}).

{\bf NGC\,1209} is an E6? galaxy in a weak pair with NGC\,1231 (Sc), based on 
our criterion.
However, \citet{2000ApJ...543..178G} identified it in a E+E pair with NGC\,1199 
(E3) (which seems to be the central galaxy of a compact group), NGC\,1209 being 
the brightest one (NOGG\,183). 
\citet{2010A&A...519A..40A} and  \citet{2011A&A...528A..10P} supposed that NGC\,1209 is a LINER based on 
optical and mid-infrared emission lines.
\citet{2012MNRAS.419..687R} revealed evidence of disturbed elliptical morphology 
in NGC\,1209 at relatively high surface brightness levels
by investigating the effects of galaxy interactions in the triggering of 
powerful radio galaxies (PRGs),
which are likely the result of past or on-going gravitational interaction.
\citet{2009AJ....138.1417T} identified X-like isophotal structure and a NW 
linear feature as tidal signatures.
We found stellar kinematics measurements across major photometric axis for 
{\bf NGC\,1209} indicating that elliptical is a fast rotator 
\citep{1993MNRAS.265..553C, 1997A&AS..126..519S}.

{\bf NGC\,5812} is an E0 galaxy in the group NOGG\,849, which contains three 
galaxies \citep{2000ApJ...543..178G}, and is the brightest member and the only ETG.
However, according to our criteria it belongs to a sparse group of six galaxies. 
The closest companion is  IC\,1084, in a projected distance of 38.2\,kpc.
An interacting signature was found in NGC\,5812,
i.e. a tidal tail due to a gravitational interacting with a dwarf companion
\citep{2009AJ....138.1417T}. 
No emission lines in the optical and mid-infrared have been detected in the spectrum of NGC\,5812
\citep{2010A&A...519A..40A, 2011A&A...528A..10P}.
Its stellar kinematics was first quantified by \citet{1994A&A...292..381B} across a single direction.
Following our criteria, NGC\,5812 is member of a sparse group and makes a pair with IC\,1084.

{\bf NGC\,6758} is an E+ (late-type elliptical) or cD galaxy. It is, the sixth brightest member 
of a group with 22 or 30 galaxies \citep{2000ApJ...543..178G},
the NOGG\,940 group, whose brightest galaxy is IC\,4837\,A.
This galaxy is a LINER host \citep{2010A&A...519A..40A, 2010ApJ...711.1316M}.
However, according to our criteria, it belongs to a sparse group of seven galaxies in total, with no physical companion.
The galaxy shows some evidence of past interaction \citep{2015MNRAS.449..612E, 2010ApJ...711.1316M}.
Its stellar kinematics in NGC\,6758 has been analysed by several works.
We consider NGC\,6758 as a sparse group member.

{\bf NGC\,6861} is a SA0$^{-s}$ galaxy in the NOGG\,962 group with 10 members and is, 
the second brightest galaxy (the brightest is NGC\,6868)
\citep{2000ApJ...543..178G}.
According to our criteria, it belongs to a sparse group of 17 galaxies, making a weak pair with IC\,4943;
although Chandra X-ray observations reveal that NGC\,6861 probably undergoes an on-going interaction with NGC\,6868
that could induce a merger between two sub-structures of the group in a near future \citep{2010ApJ...711.1316M}.
Optical emission lines from warm gas have already been measured in the spectrum of this galaxy
\citep{1986AJ.....91.1062P, 1996A&AS..120..463M}.
However, there is no classification for this emission yet.
Its stellar kinematics was, for instance, investigated by \citet{2000A&AS..145...71K}, and classified as a fast rotator.
We consider NGC\,6861 as a sparse group member in a weak pair.

{\bf NGC\,7507} is an E0 galaxy in a sparse group with seven members in total based on our criterion,
under possible interaction with GALEXASC\,J231146.81-283145.2.
There are no clear signs of tidal distortions in NGC\,7507 \citep{2013A&A...555A..56C},
although photometric shells have been identified \citep{2009AJ....138.1417T, 2012MNRAS.419..687R}
and a moderately high index of peculiarity $\Sigma_{2}$ has been measured \citep{2004A&A...423..833M}.
It also has two stellar haloes with opposite small rotations
confirmed as a S+S merger remnant \citep{2015A&A...574A..93L, 2012A&A...538A..87S}.
Its stellar kinematics suggests it is a slow rotator.
\citet{2003A&A...405....5B} did not compile ionized gas emission for the galaxy.
We consider NGC\,7507 as an interacting pair galaxy in a sparse group.

{\bf NGC\,7796} is a cD galaxy that is isolated, based on our criteria,
although it has NGC\,7796\,1 as a dwarf satellite, which exhibits a tidal tail and multiple nuclei \citep{2015A&A...574A..21R}.
On the other hand, there is no photometric interaction sign in NGC\,7796 \citep{2002A&A...391..531R}.
\citet{1996A&AS..120..463M} did not measure optical emission lines for NGC\,7796.
Several works quantified the stellar kinematics of {\bf NGC\,7796} across different directions.
We consider NGC\,7796 as a field galaxy.

\section{Radial gradients of stellar population properties analysed galaxy-by-galaxy}
\label{ap_radial_grad}

For \textbf{NGC\,5328}, the age and global metallicity radial distributions across both photometric axes of IC\,5328 
are shown in Fig.~\ref{age_all_a}.
While we found a flat age gradient across major axis and a nearly flat age gradient across minor axis (providing a uniform age of
about 12\,$\pm$\,1\,Gyr), the metallicity gradient is negative across both axes (-0.36\,$\pm$\,0.10\,dex/($R/R_{\rm e}$) for major axis
and -0.52\,$\pm$\,0.19\,dex/($R/R_{\rm e}$) for minor axis) indicating a nuclear over-solar global metallicity ([M/H]$=+0.32$\,dex)
and a decrease of around $\sim$+0.2\,dex up to about 0.4\,$R/R_{\rm e}$. The integrated age and metallicity inside one  R$_{\rm e}$ 
(quantified across major axis only) are about 11 Gyr and [M/H]$=+0.20$ dex, respectively, which are in agreement with their radial
variations within the observational errors.

For \textbf{NGC\,1052}, the age and global metallicity radial distributions across major axis are shown in Fig.~\ref{age_all_b}.
The radial gradients of both parameters are negatives (see Table~\ref{tab33}),
with the metallicity gradient the steepest of all sample galaxies.
However the change in age from the nucleus up to about 0.3\,$R/R_{\rm e}$ is still compatible with a constant value
(13\,$\pm$\,1 Gyr downing up to 12\,$\pm$\,1 Gyr).
The metallicity changes from [M/H]$=+0.42$ dex at the nucleus downing up to [M/H]$=+0.17$ dex
(a decrease of 0.25 dex that is seven times greater than the average error of 0.04 dex).
The integrated age and metallicity inside one  R$_{\rm e}$  are about 12 Gyr and [M/H]$=+0.25$ dex, respectively,
which are in agreement with their radial variations within the observational errors.

For \textbf{NGC\,1209}, the age and global metallicity radial distributions across major axis are shown in Fig.~\ref{age_all_c}.
The radial gradients of both parameters are compatible with zero, taking into account their errors
that are equal or greater than their gradient values (see Table~\ref{tab33}).
Age and [M/H] from the nucleus up to about 0.23\,$R/R_{\rm e}$ change around constant values
(11$\pm$1 Gyr and +0.27$\pm$0.03 dex).
The integrated age and metallicity inside one R$_{\rm e}$ 
are about 10$\pm$1 Gyr and $Z=+0.27\pm0.01$ dex, respectively, 
which are in agreement with their radial variations within the observational errors.

For \textbf{NGC\,5812}, the age and global metallicity radial distributions across EW direction are shown in Fig.~\ref{age_all_b}.
The radial gradients of both parameters are negative (see Table~\ref{tab33}).
However, the change in age from the nucleus up to about 0.27\,$R/R_{\rm e}$ is still compatible with a constant value
(13\,$\pm$\,1 Gyr downing up to 12\,$\pm$\,1 Gyr).
The metallicity changes from [M/H]$=+0.41$ dex at the nucleus downing up to [[M/H]$=+0.34$ dex
(a decrease of 0.07 dex that is 2 times greater the average error of 0.03 dex).
The integrated age and metallicity inside one  R$_{\rm e}$ are about 13 Gyr and [M/H]$=+0.37$ dex, respectively,
which are in agreement with their radial variations within the observational errors.

For \textbf{NGC\,6758}, the age and global metallicity radial distributions across major and minor axes are shown in
Fig.~\ref{age_all_b}.
The radial gradients of both parameters are compatible with zero, taking into account their errors
that are equal or greater than their gradient values (see Table~\ref{tab33}).
In fact, age and  [M/H] from the nucleus up to about 0.32\,$R/R_{\rm e}$ change around constant values
(12$\pm$1 Gyr and +0.32$\pm$0.01 dex).
The integrated age and metallicity inside one  R$_{\rm e}$ 
are about 11$\pm$1 Gyr and [M/H]$=+0.322\pm0.004$ dex, respectively,
which are in agreement with their radial variations within the observational errors
(note that population synthesis over the one  R$_{\rm e}$  region was only applied by using the major axis data).

For \textbf{NGC\,6861}, the age and global metallicity radial distributions across major and minor axes are shown in
Fig.~\ref{age_all_c}.
The radial gradients of both parameters are null over the major axis and positive over the minor axis (see Table~\ref{tab33}).
However, the change in age across minor axis from the nucleus up to about 0.3\,$R/R_{\rm e}$ is still compatible with a constant value
(variation smaller than 0.3 Gyr !).
The age and [M/H] change across the major axis are around 13 Gyr and [M/H]$=+0.40$ dex 
(from the nucleus up to 0.52\,$R/R_{\rm e}$), respectively.
Over the minor axis, the metallicity changes from [M/H]$=+0.36$ dex at the nucleus downing up to [M/H]$=+0.40$ dex
(an increase of 0.06 dex that is 2 times greater than the average error of 0.02 dex).
The integrated age and metallicity inside one  R$_{\rm e}$  are about 13 Gyr and [M/H]$=+0.38$ dex, respectively,
which are in agreement with their radial variations within the observational errors
(note that population synthesis over the one  R$_{\rm e}$  region was only applied by using the major axis data).

For \textbf{NGC\,7507}, the age and global metallicity radial distributions across EW direction are shown in Fig.~\ref{age_all_h}.
While we have found a flat age gradient providing a uniform age of about 12\,$\pm$\,1 Gyr,
the metallicity gradient is negative (-0.46\,$\pm$\,0.39\,dex/($R/R_{\rm e}$)) despite its great error,
pointing to a nuclear over-solar global metallicity ([M/H]$=+0.41$\,dex)
that decreases down to about [M/H]$=+0.28$\,dex at 0.15\,$R/R_{\rm e}$
(a variation of 0.13 dex that is still greater than the metallicity error).
The integrated age and metallicity inside one R$_{\rm e}$  are about 10.5 Gyr and [M/H]$=+0.29$ dex, respectively,
which are in agreement with their radial variations within the observational errors.

For \textbf{NGC\,7796}, the age and global metallicity radial distributions across major and minor axes are shown in
Fig.~\ref{age_all_h}.
The radial gradients of both parameters are compatible with zero,
taking into account the radial variations of age and metallicity themselves and the gradient errors
that are comparable with the own gradient values (see Table~\ref{tab33}),
even though the metallicity gradient is slightly negative (but supplying a very small variation of 0.05 dex in [M/H]).
Age and [M/H] from the nucleus up to about 0.3\,$R/R_{\rm e}$ (major axis) and 0.2\,$R/R_{\rm e}$ change around constant values
(13$\pm$1 Gyr and +0.40$\pm$0.01 dex).
The integrated age and metallicity inside one  R$_{\rm e}$ 
are about 10$\pm$1 Gyr and [M/H]$=+0.40\pm0.01$ dex, respectively,
which are in agreement with their radial variations
(note that population synthesis over the one  R$_{\rm e}$ region was only applied by using the major axis data).

\section{Stellar population synthesis} 
\label{ap_synth1}

\begin{figure*}
\centering
\includegraphics*[angle=-90,width=0.4\textwidth]{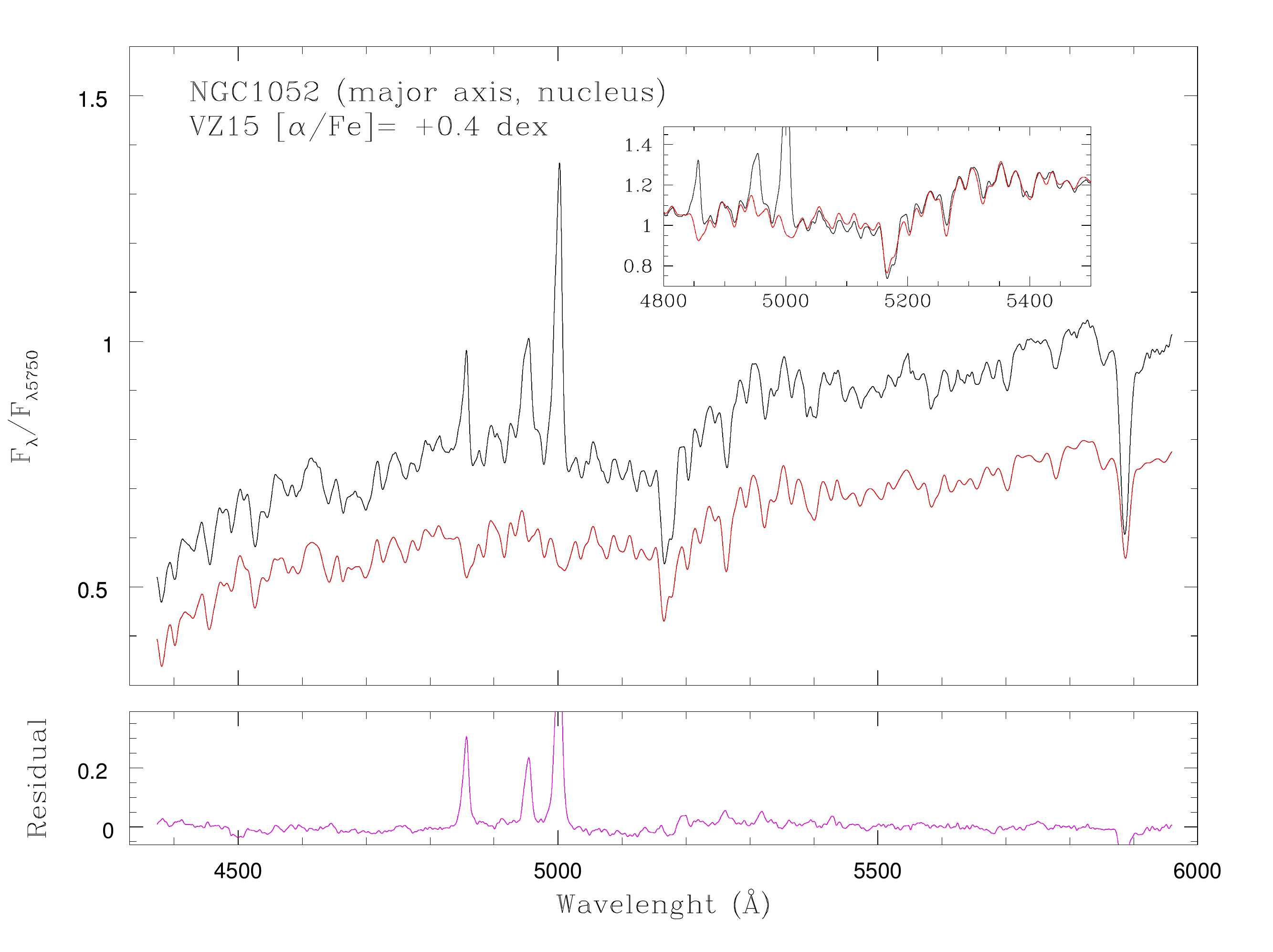}
\includegraphics*[angle=-90,width=0.4\textwidth]{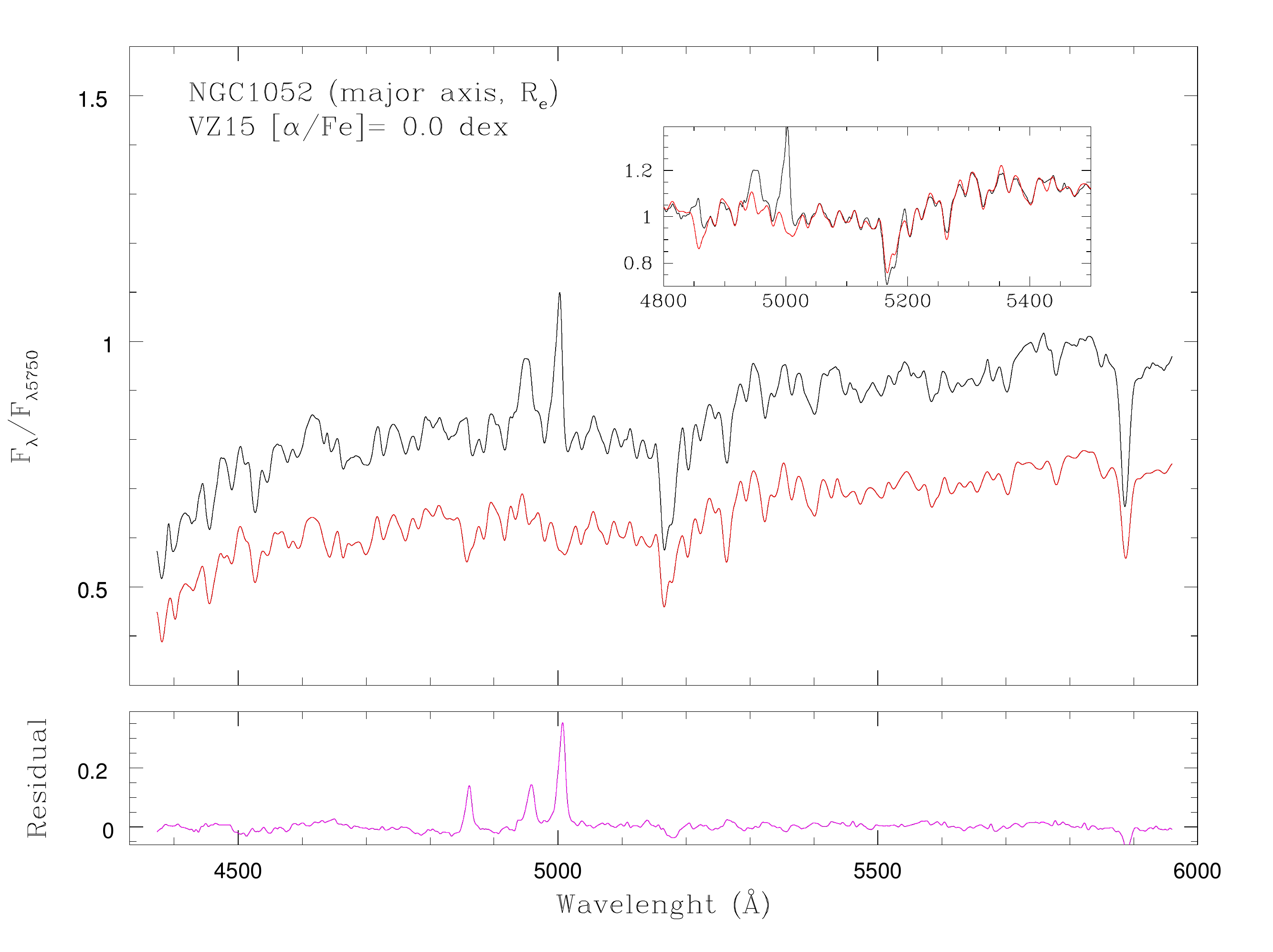}
\includegraphics*[angle=-90,width=0.4\textwidth]{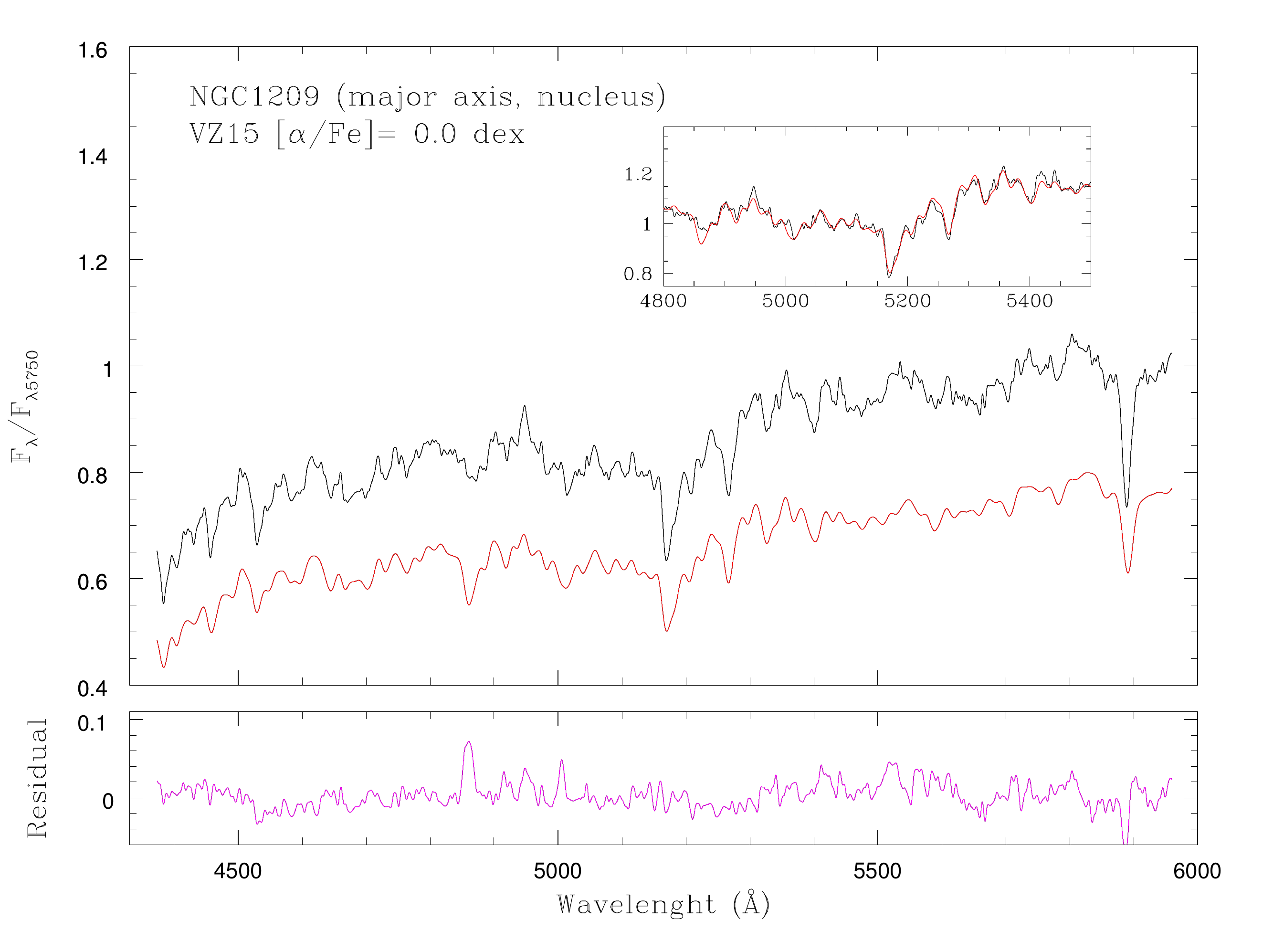}
\includegraphics*[angle=-90,width=0.4\textwidth]{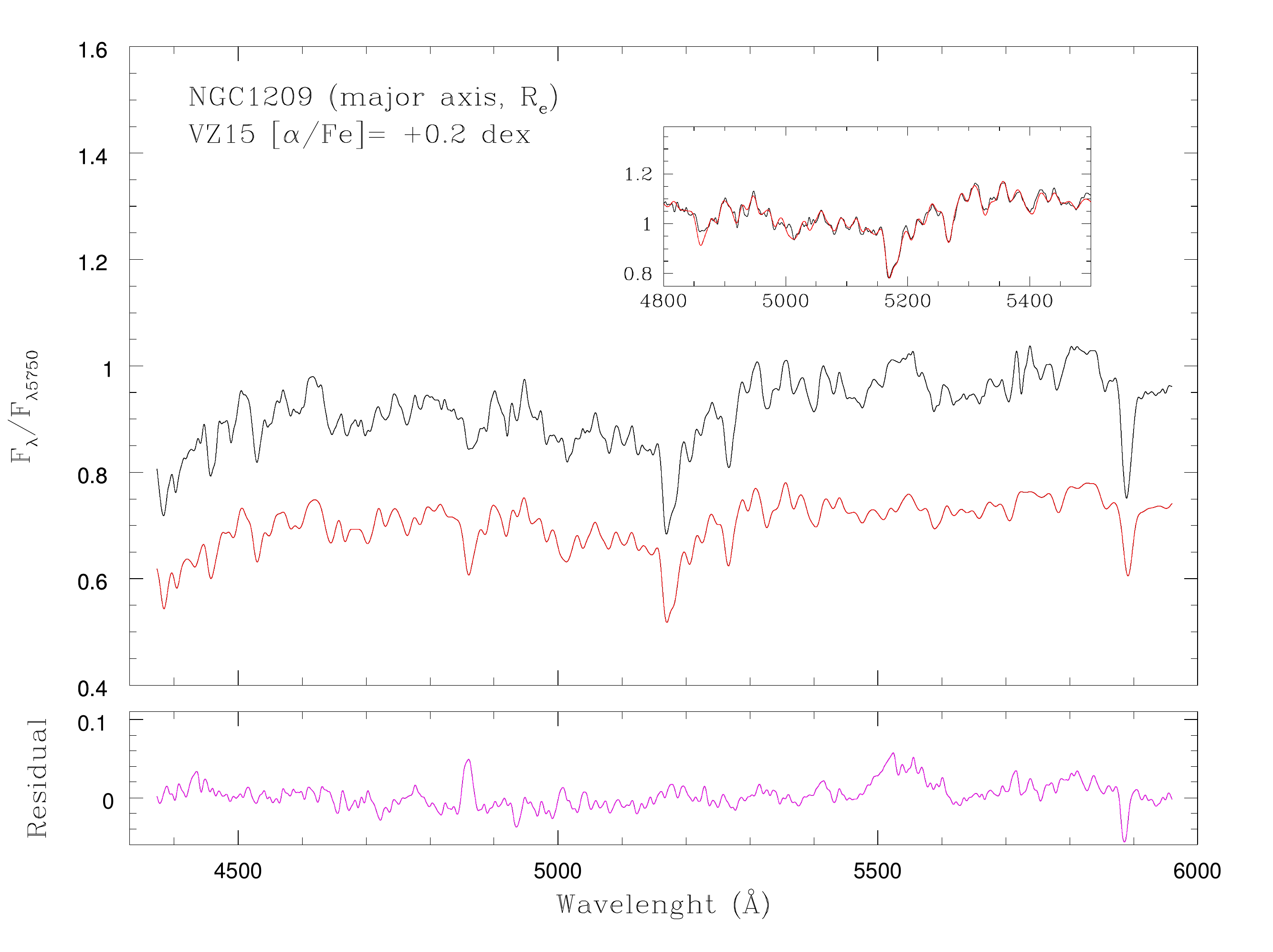}
\includegraphics*[angle=-90,width=0.4\textwidth]{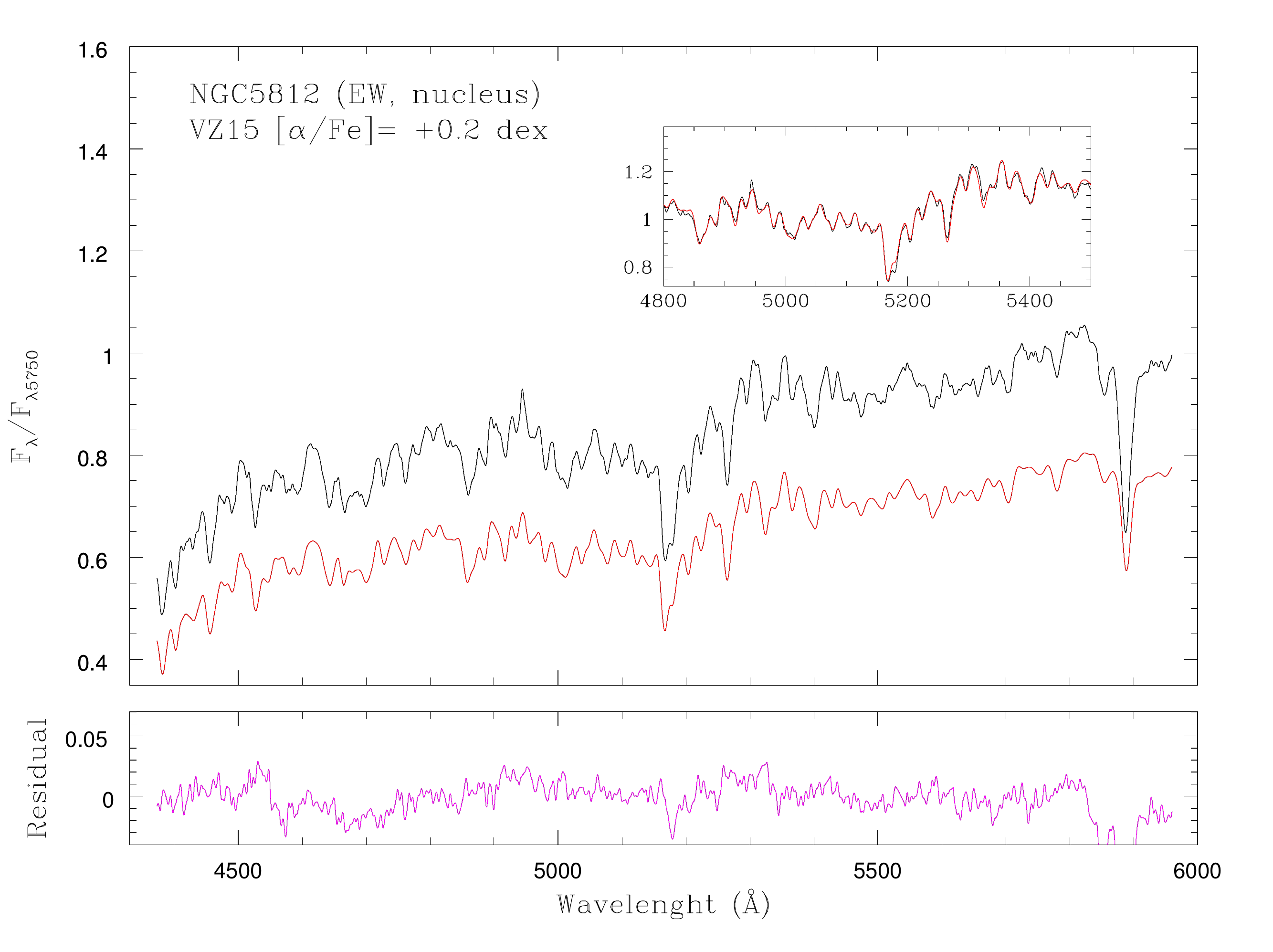}
\includegraphics*[angle=-90,width=0.4\textwidth]{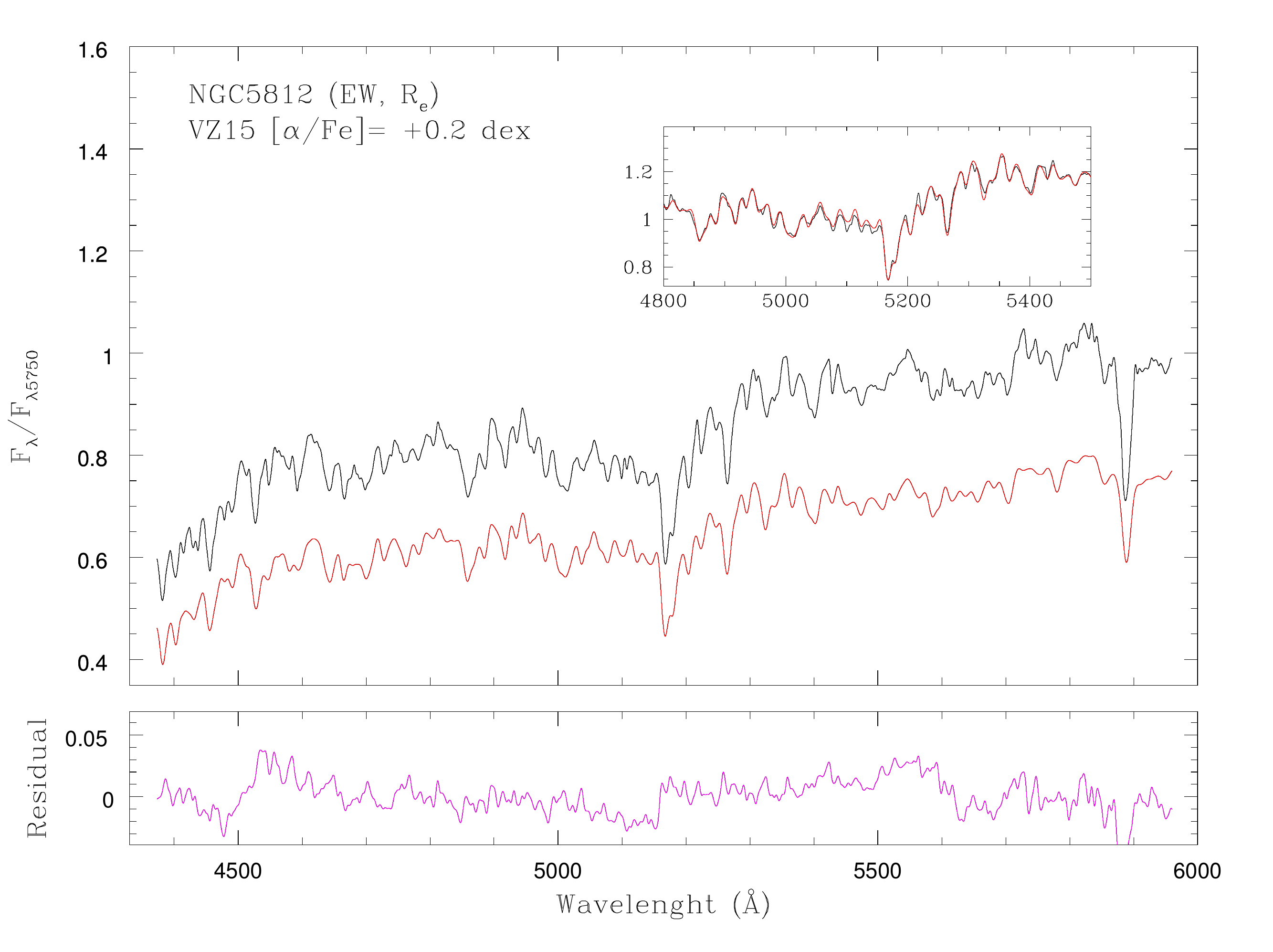}
\includegraphics*[angle=-90,width=0.4\textwidth]{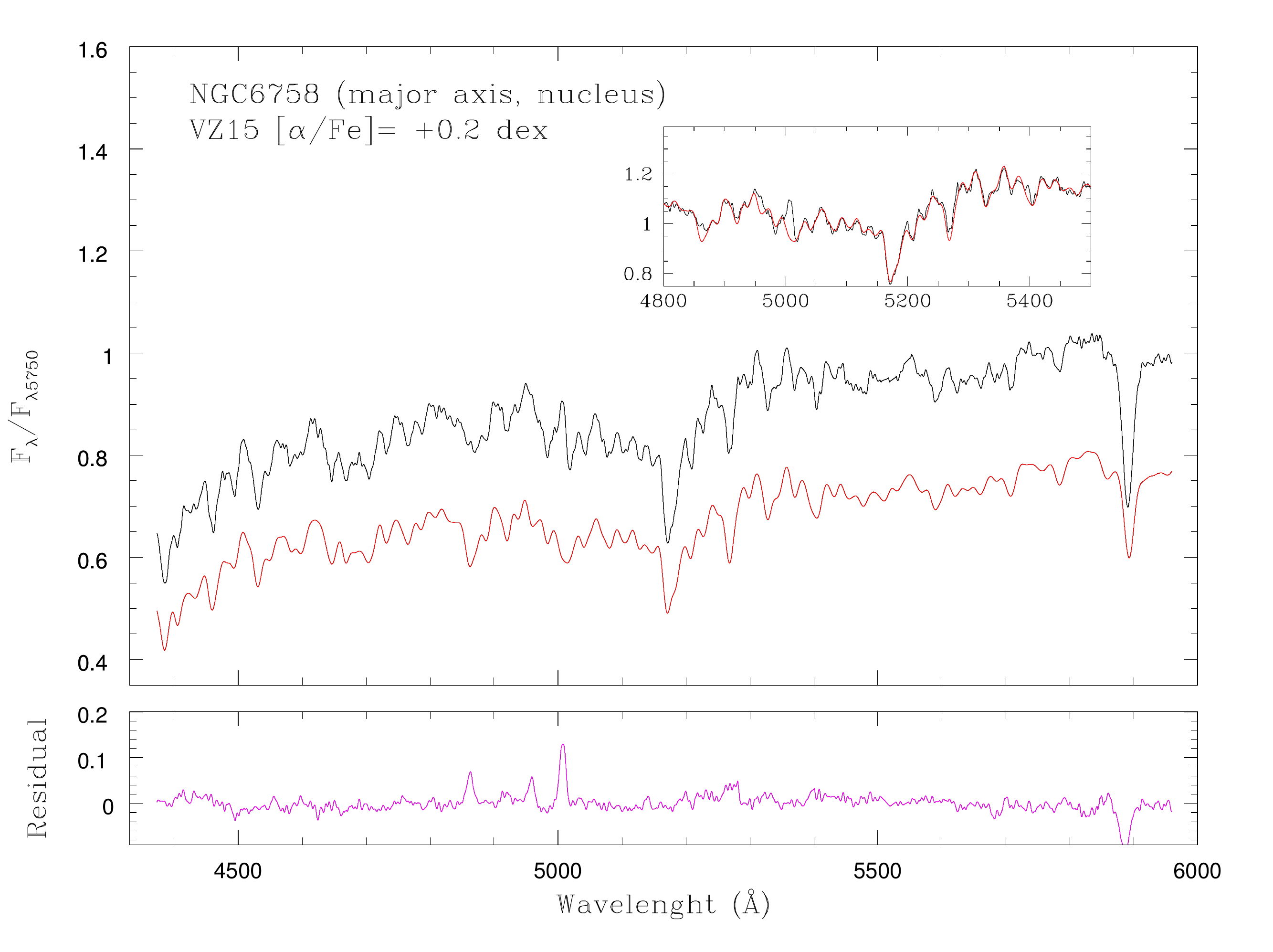}
\includegraphics*[angle=-90,width=0.4\textwidth]{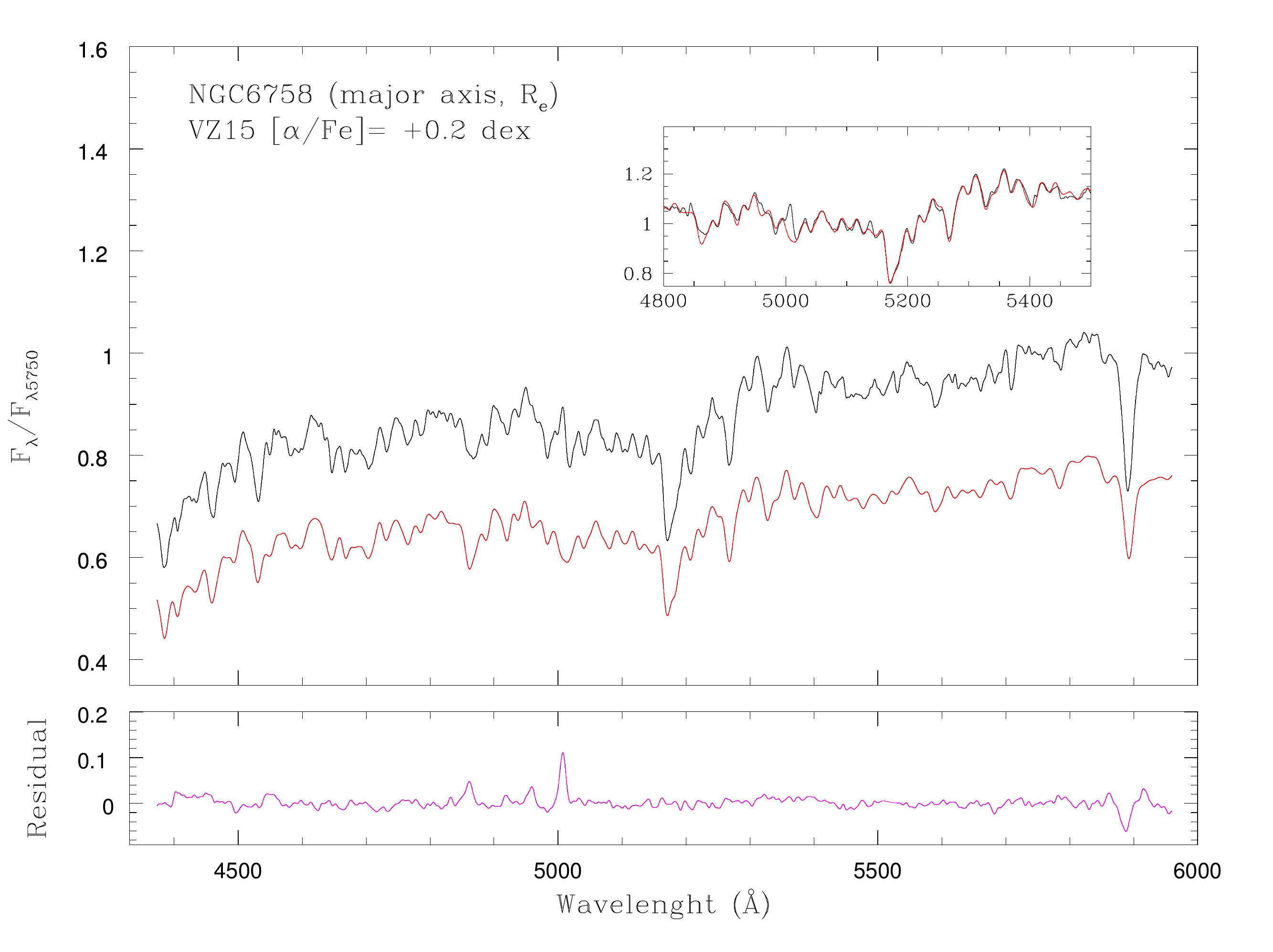}
\caption{Stellar population synthesis for the nuclear (left panel)  and one $R_{\rm e}$ (right panel) regions across the major axis  of 
IC\,5328, NGC\,1052, NGC\,1209, NGC\,5812, and NGC\,6758
(Notation as in Fig.~\ref{sintese_002}).}
\label{sintese_003}
\end{figure*}

\begin{figure*}
\centering
\includegraphics*[angle=-90,width=0.4\textwidth]{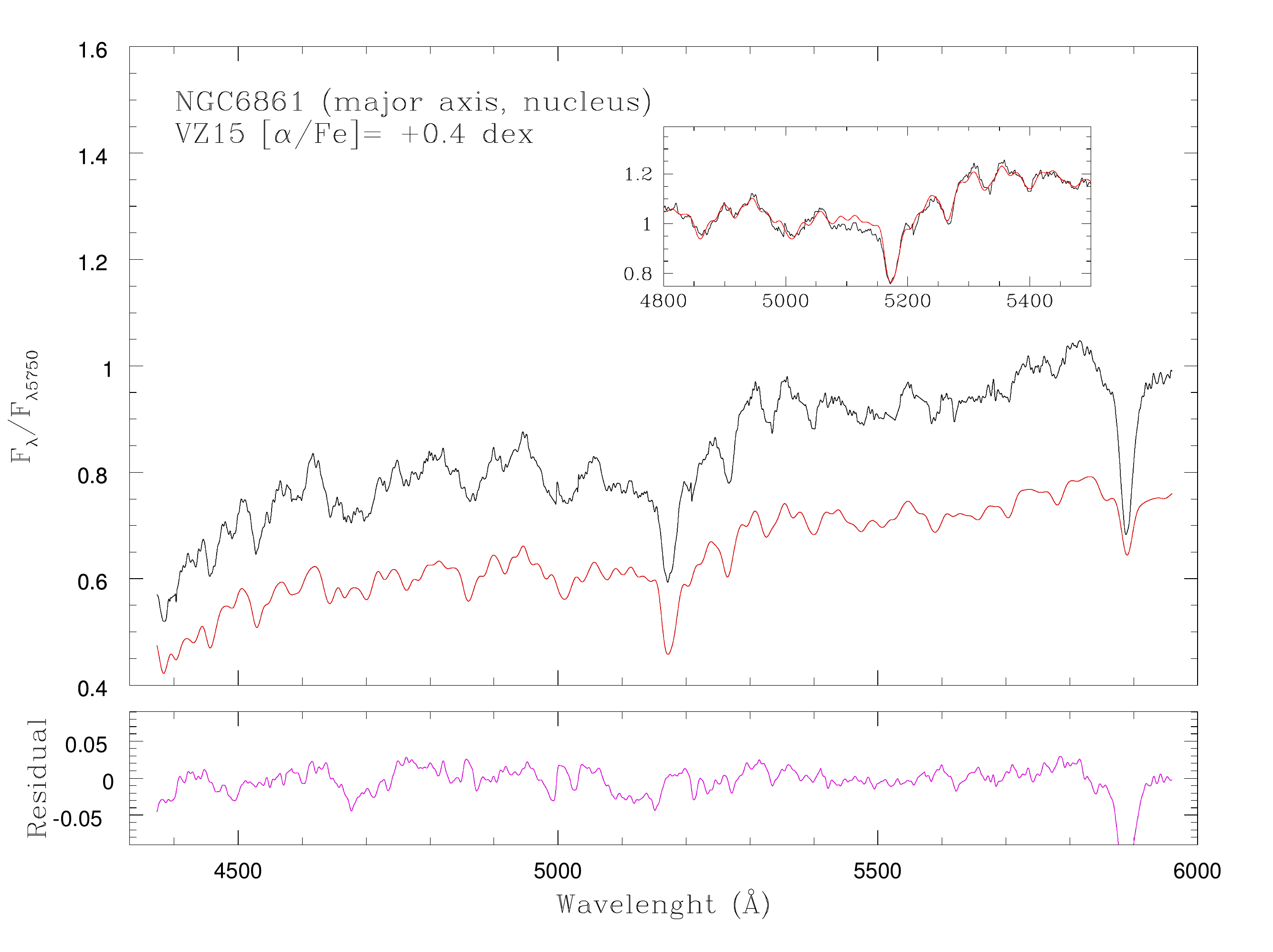}
\includegraphics*[angle=-90,width=0.4\textwidth]{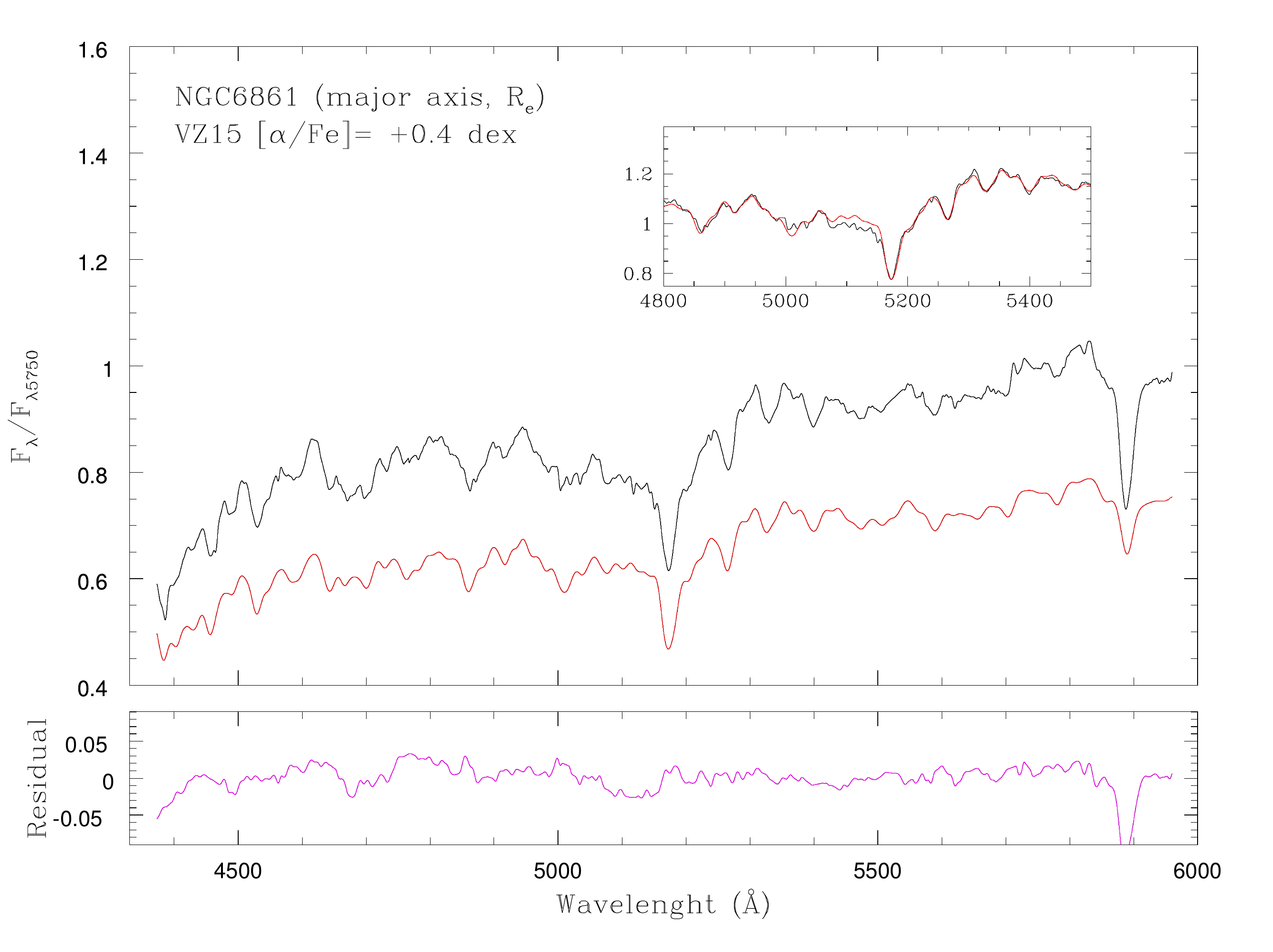}
\includegraphics*[angle=-90,width=0.4\textwidth]{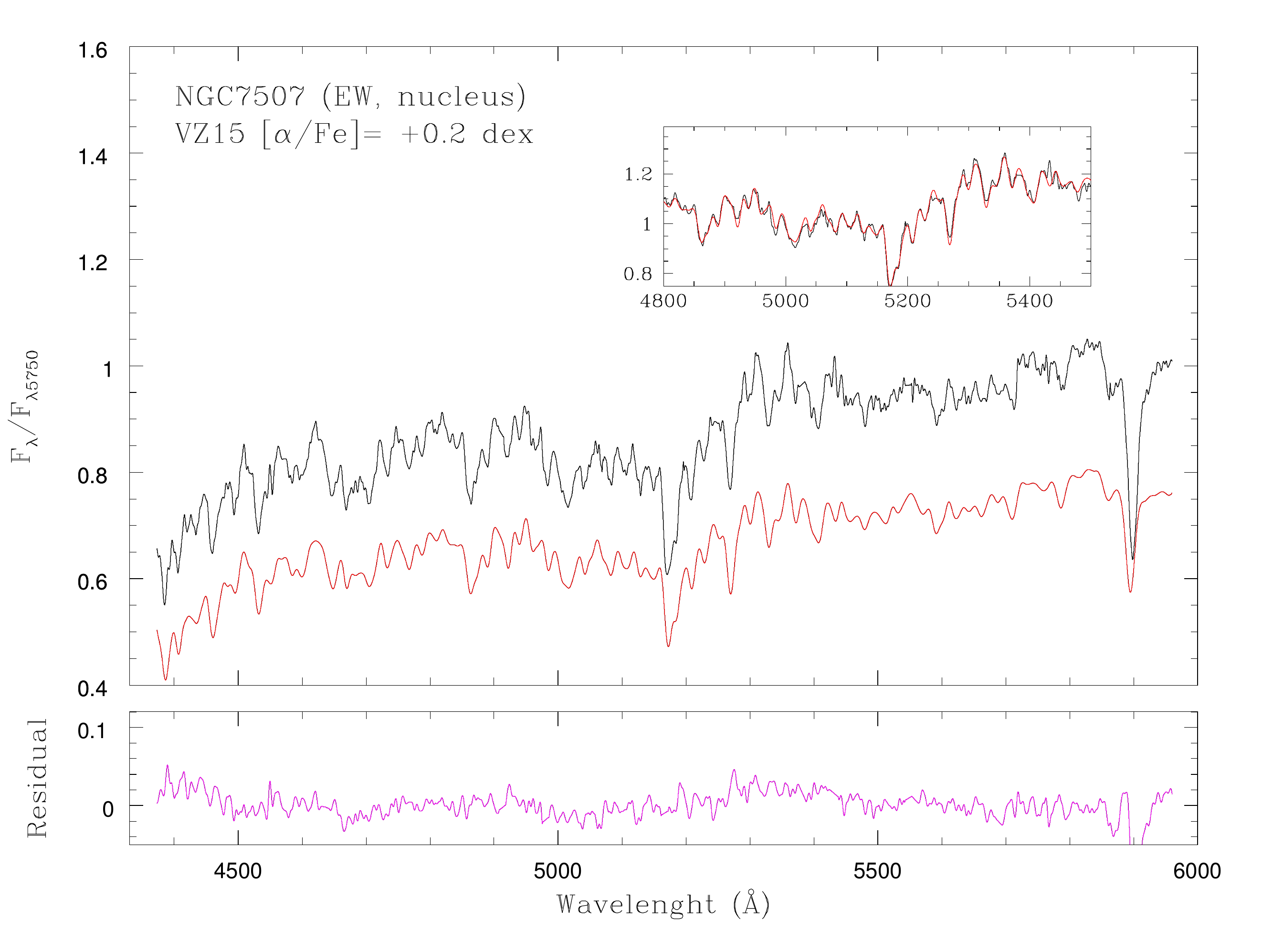}
\includegraphics*[angle=-90,width=0.4\textwidth]{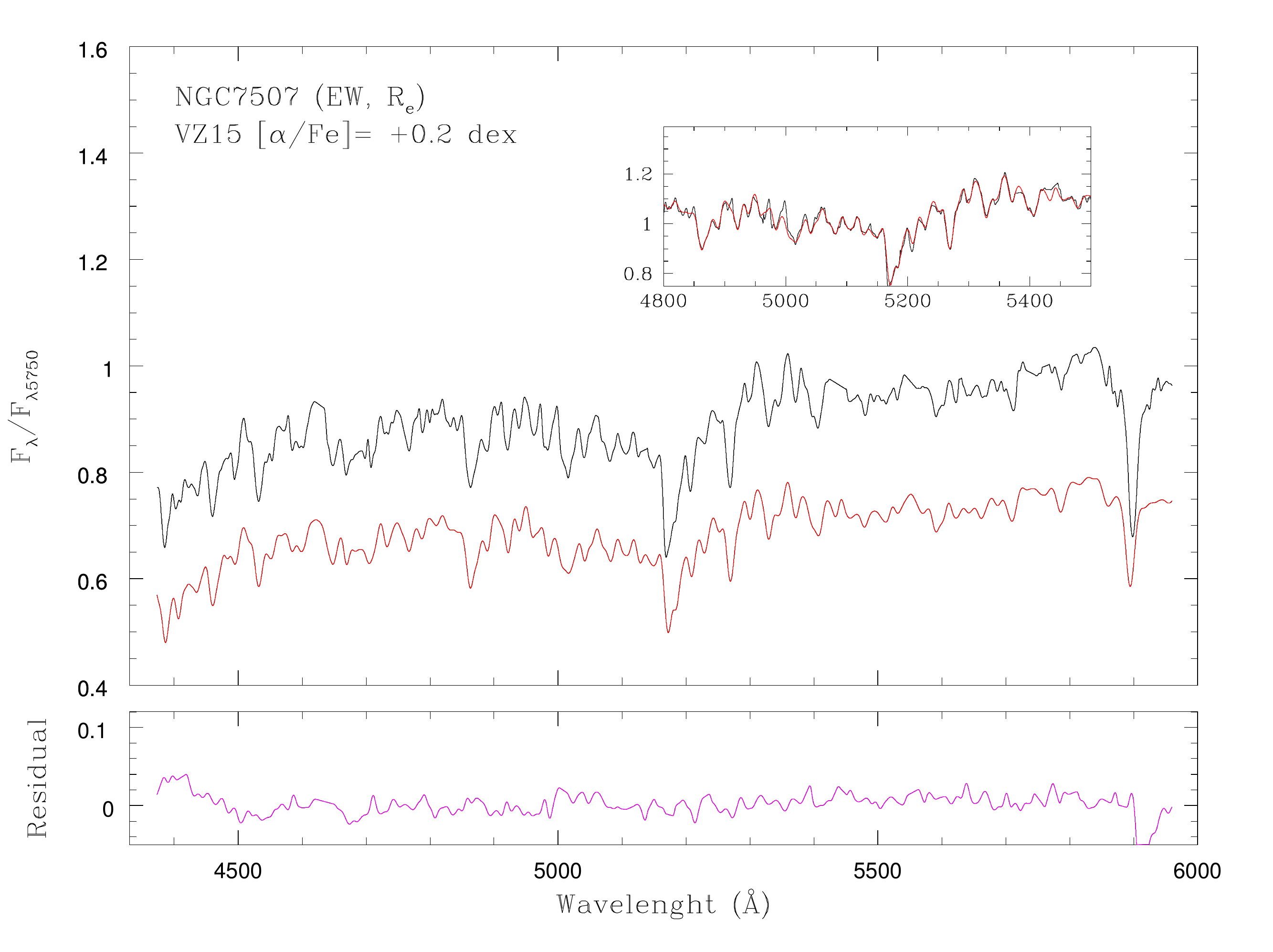}
\includegraphics*[angle=-90,width=0.4\textwidth]{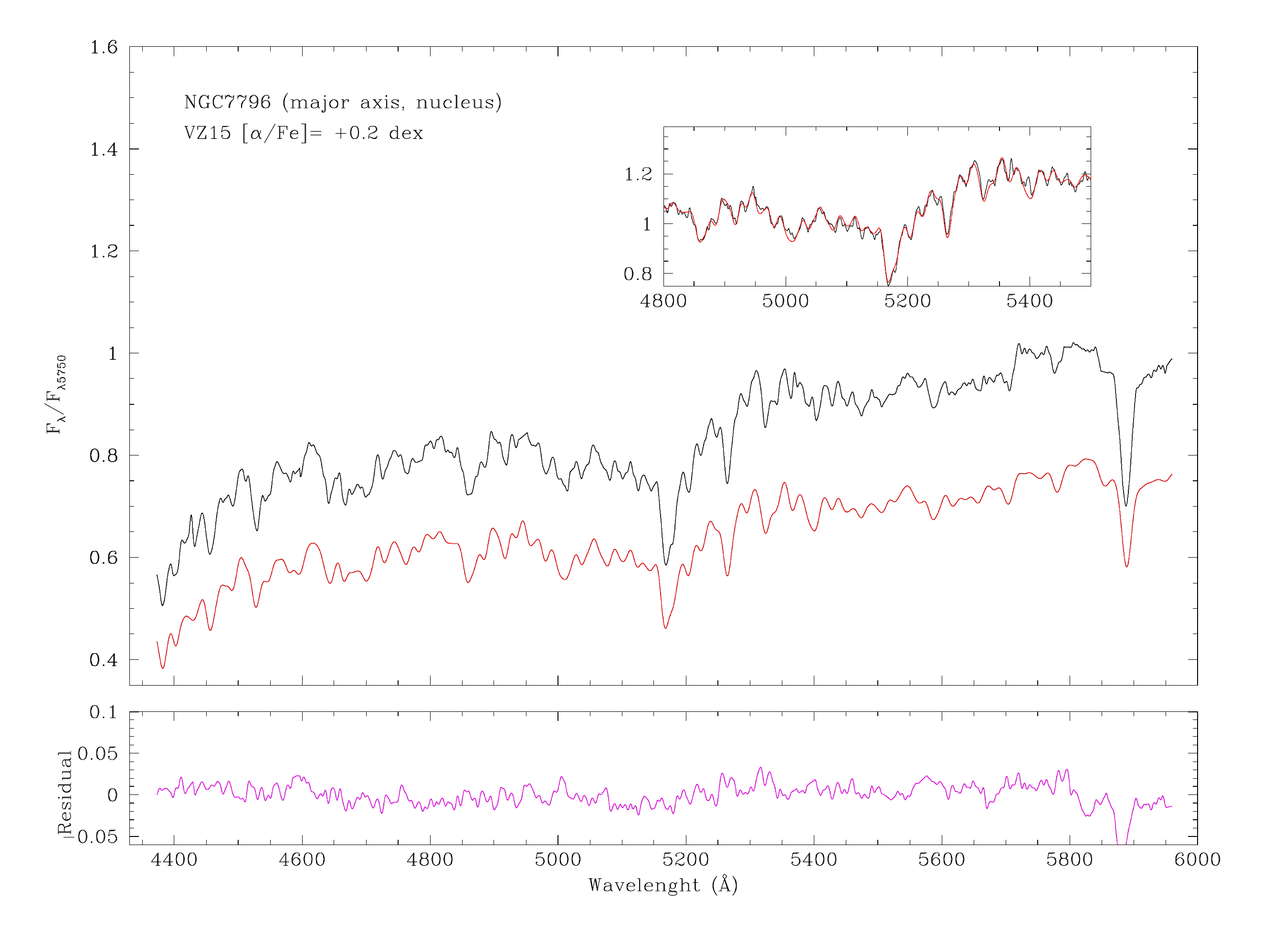}
\includegraphics*[angle=-90,width=0.4\textwidth]{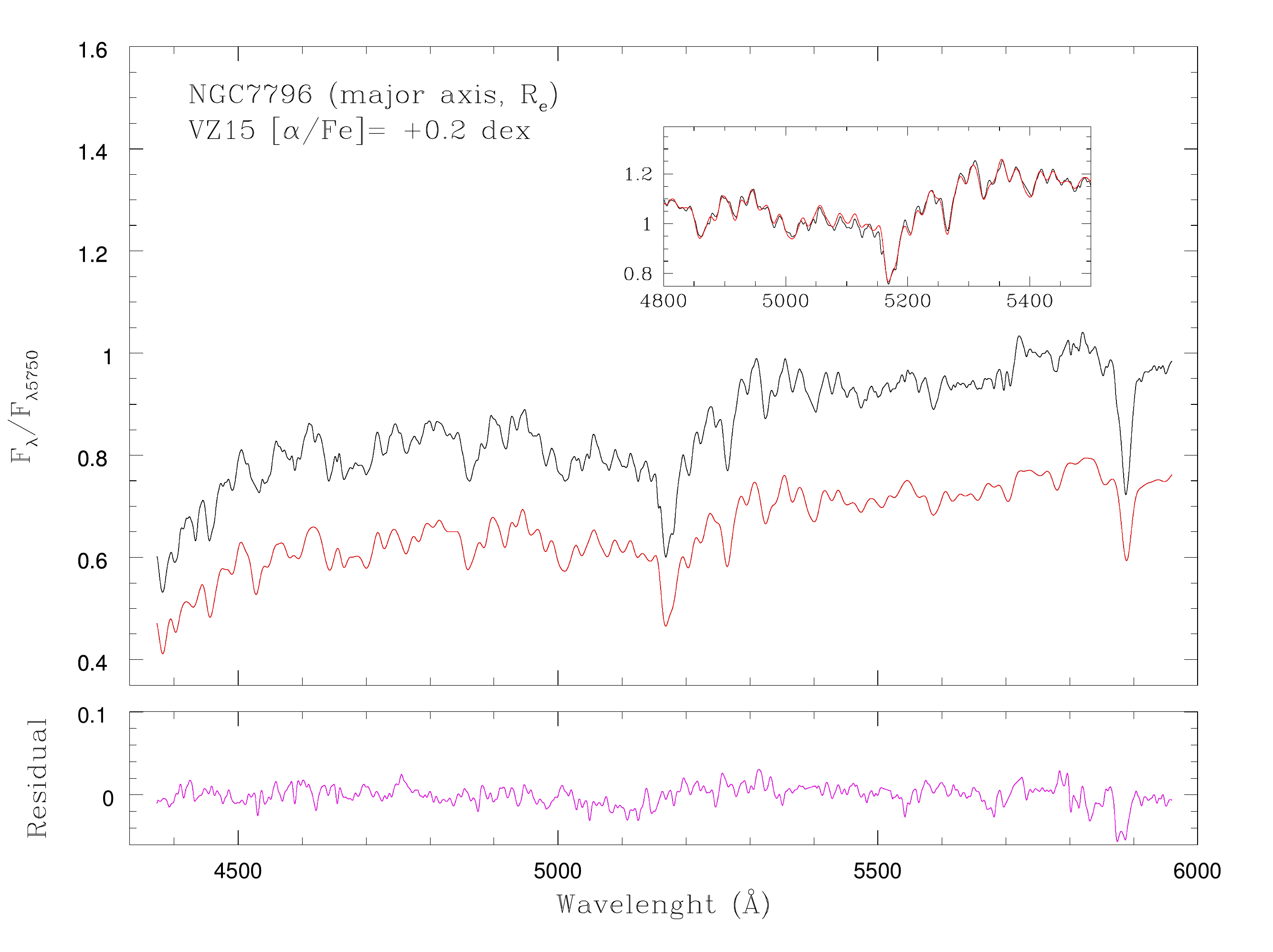}
\caption{The same as Fig.~\ref{sintese_003}, but for NGC\,6861, NGC\,7507, and NGC\,7796. }
\label{sintese_003b}
\end{figure*}

\begin{figure*}
\includegraphics*[angle=0,width=0.35\textwidth]{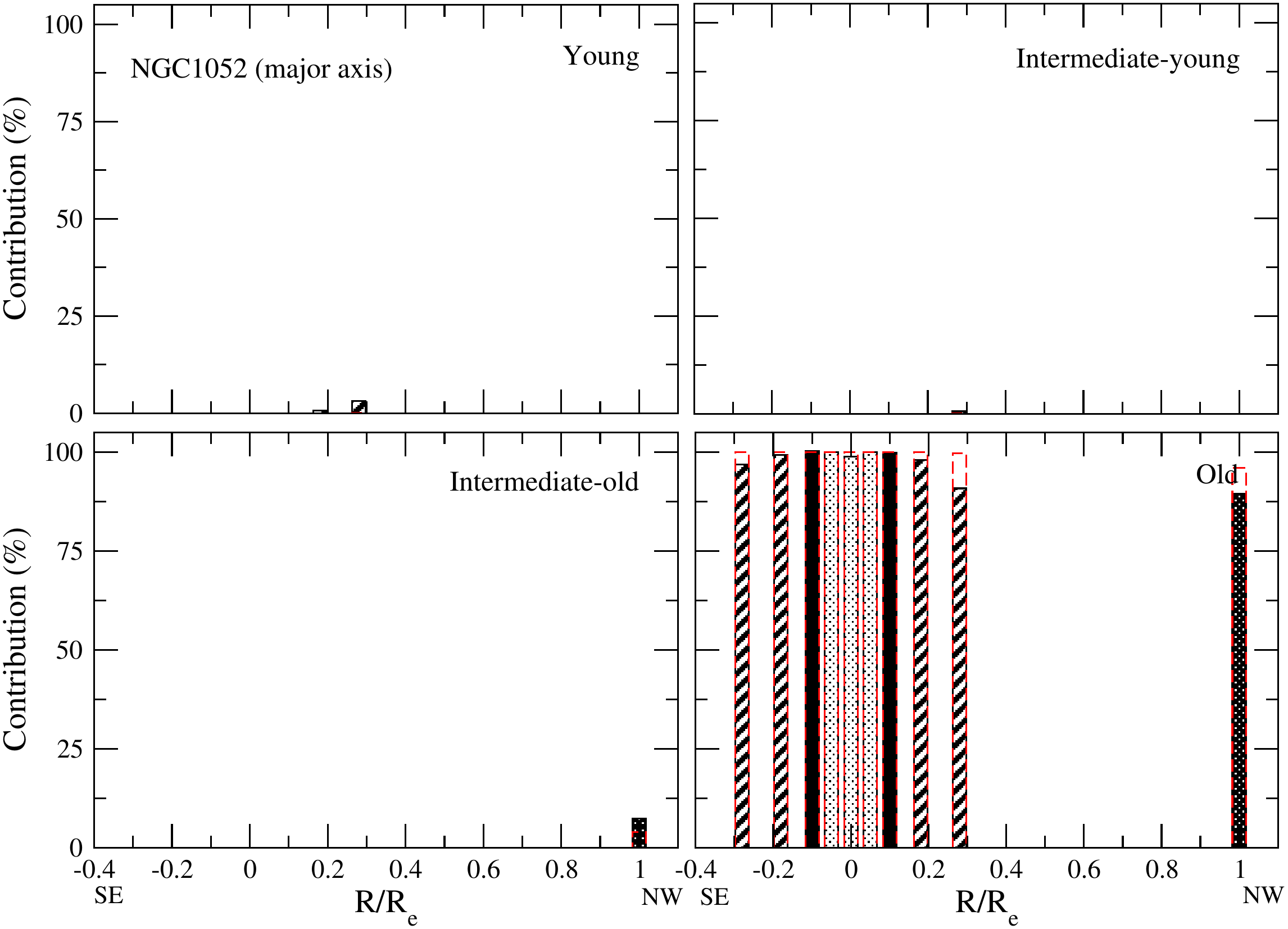} 
\includegraphics*[angle=0,width=0.35\textwidth]{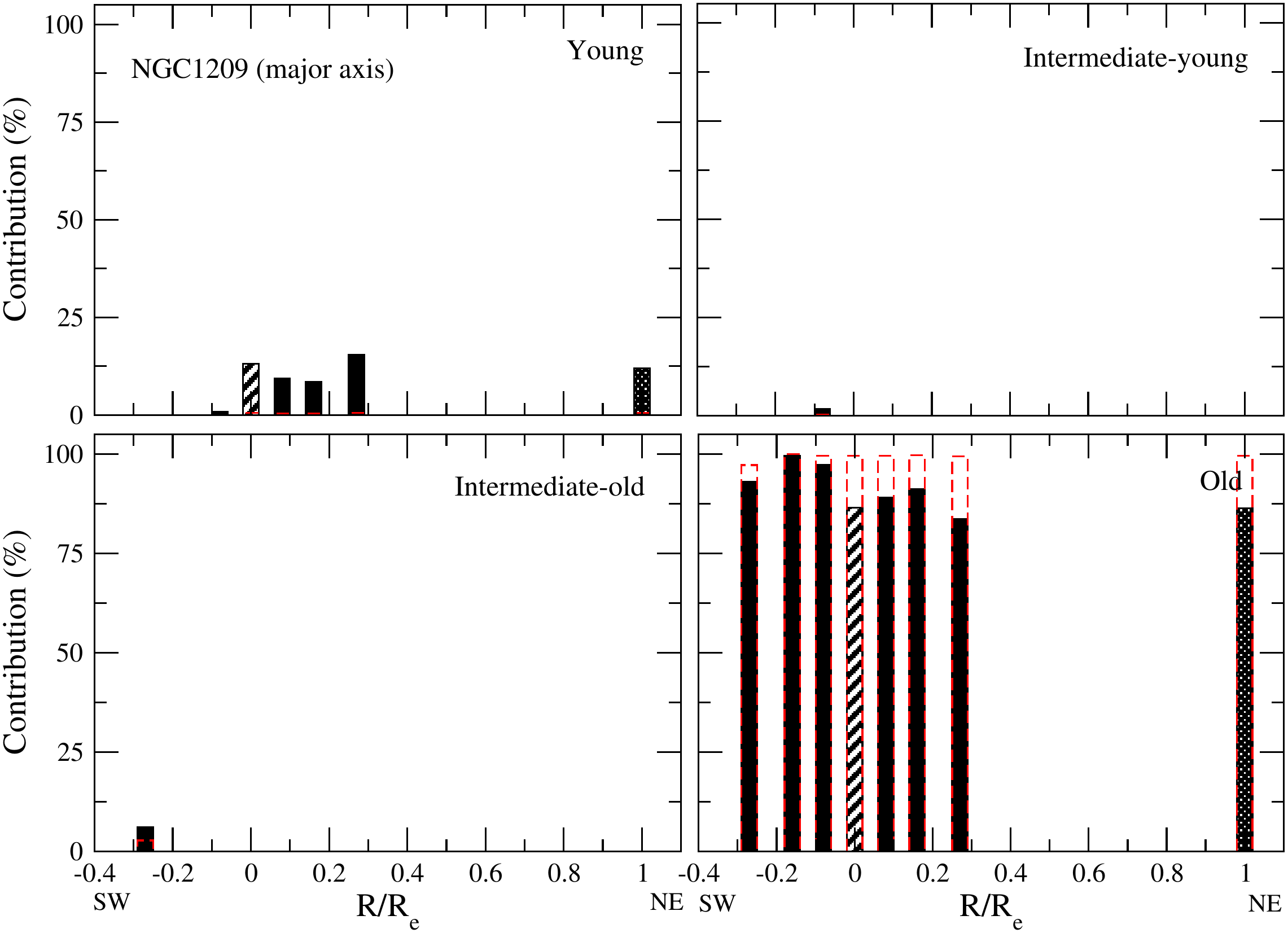}
\includegraphics*[angle=0,width=0.35\textwidth]{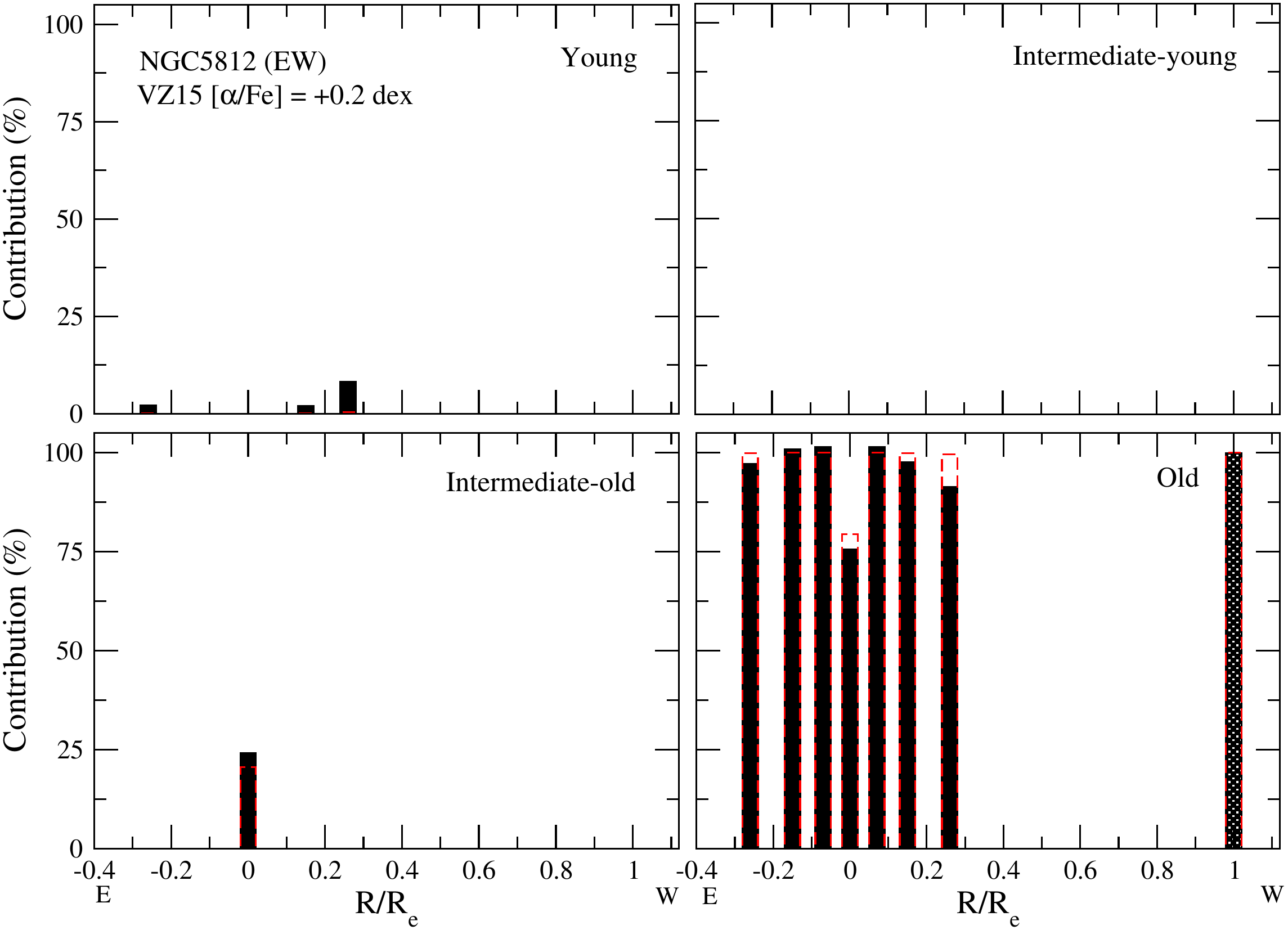}
\includegraphics*[angle=0,width=0.35\textwidth]{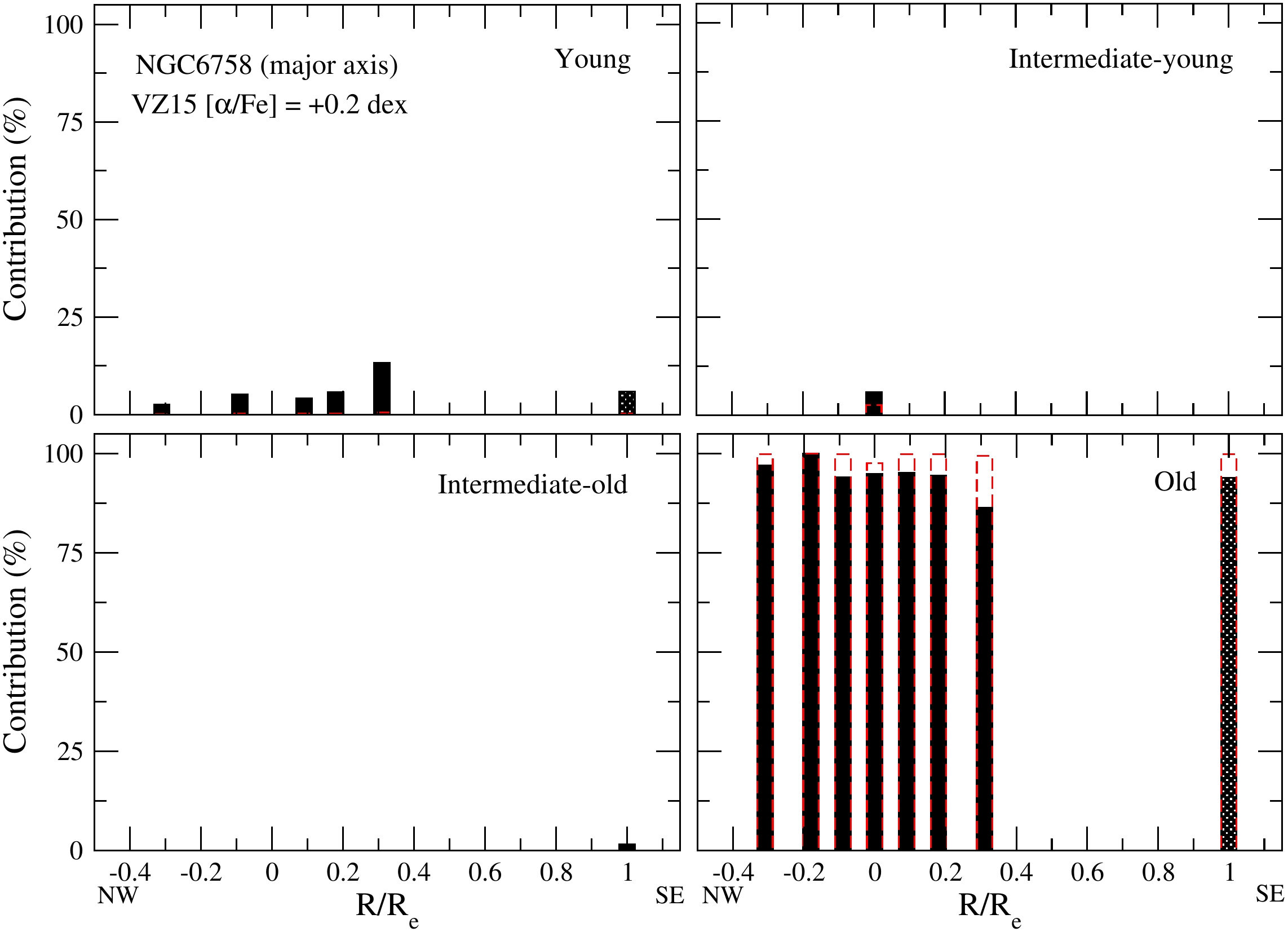}
\includegraphics*[angle=0,width=0.35\textwidth]{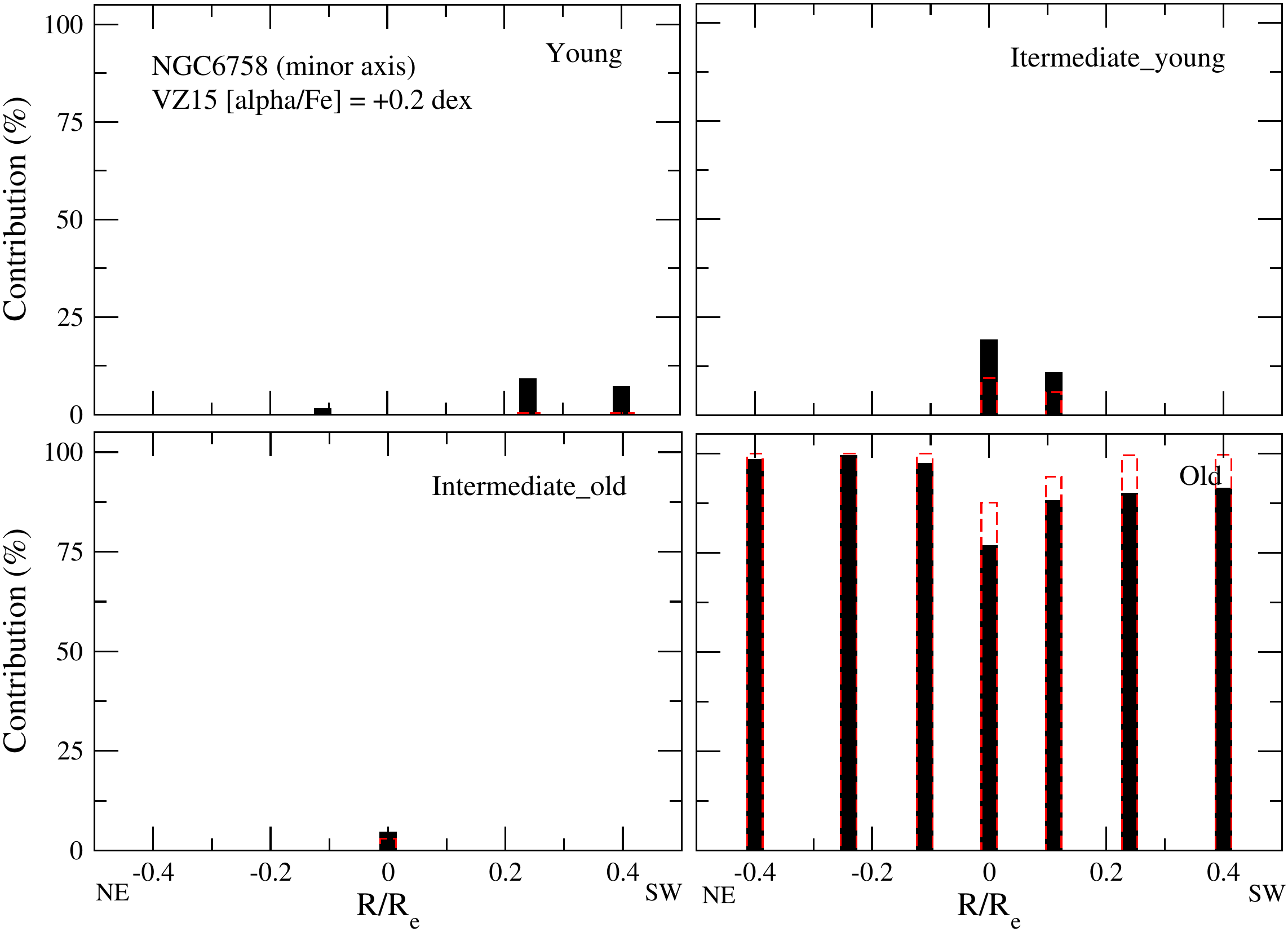}
\includegraphics*[angle=0,width=0.35\textwidth]{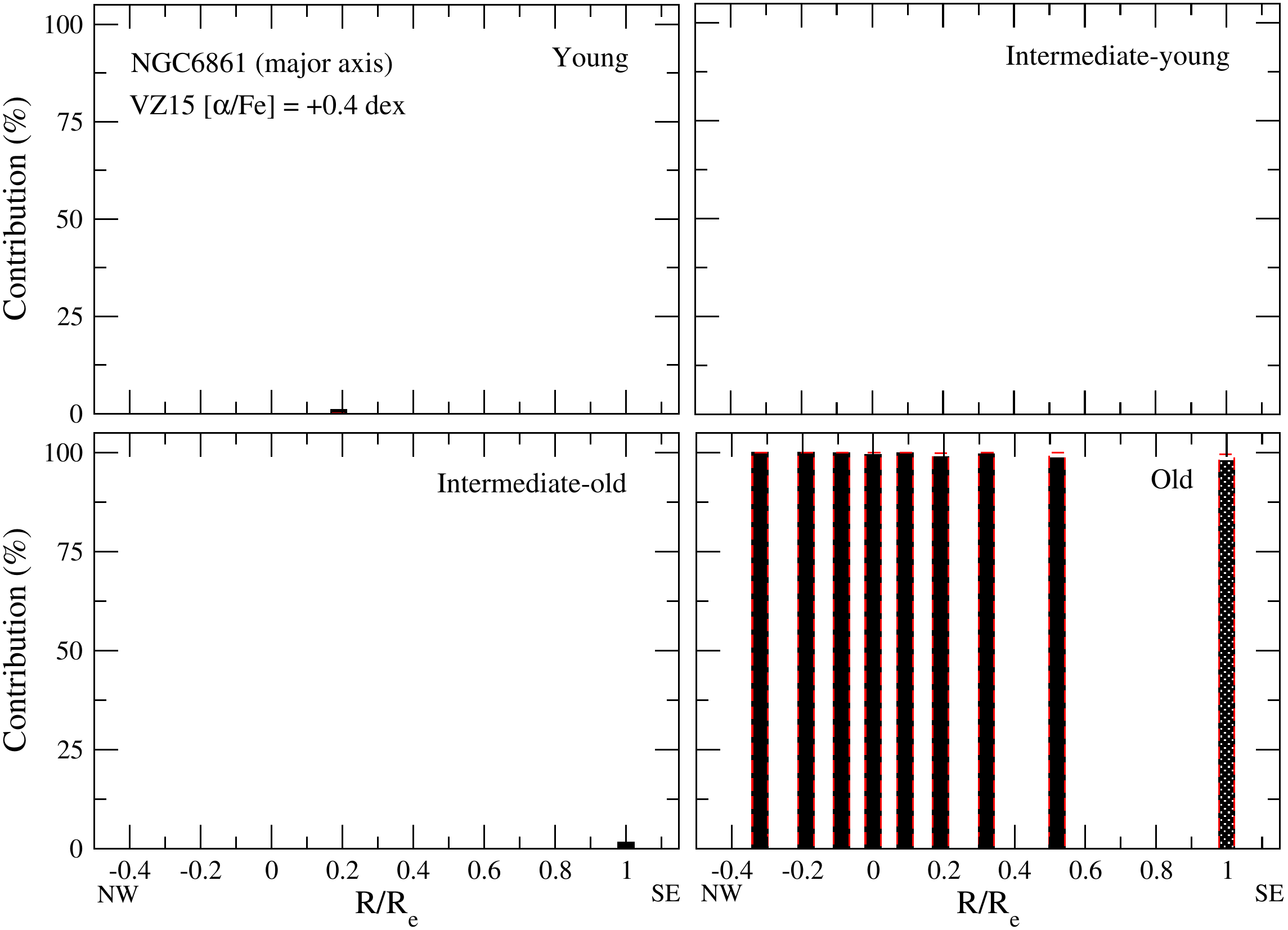}
\includegraphics*[angle=0,width=0.35\textwidth]{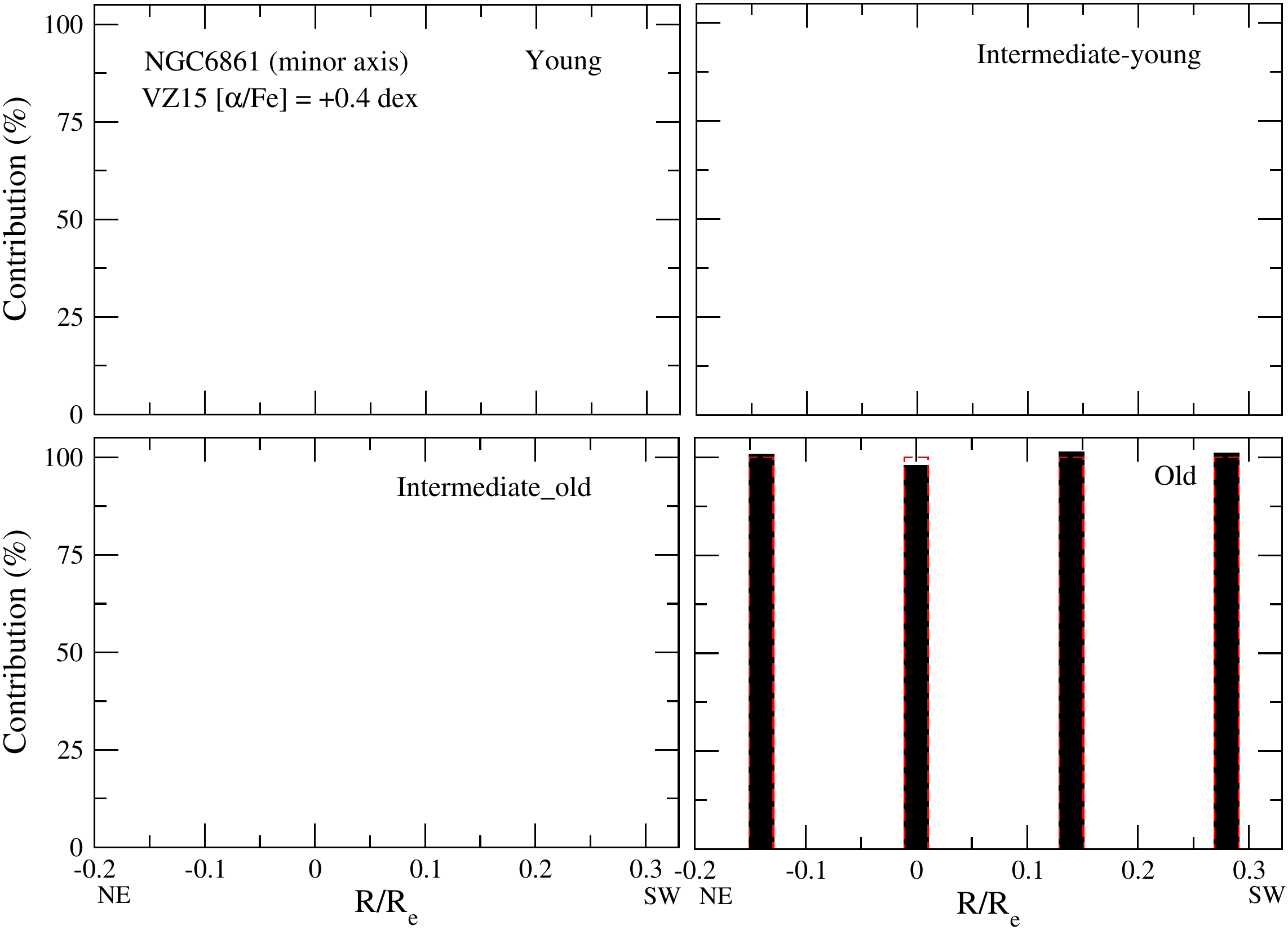}
\includegraphics*[angle=0,width=0.35\textwidth]{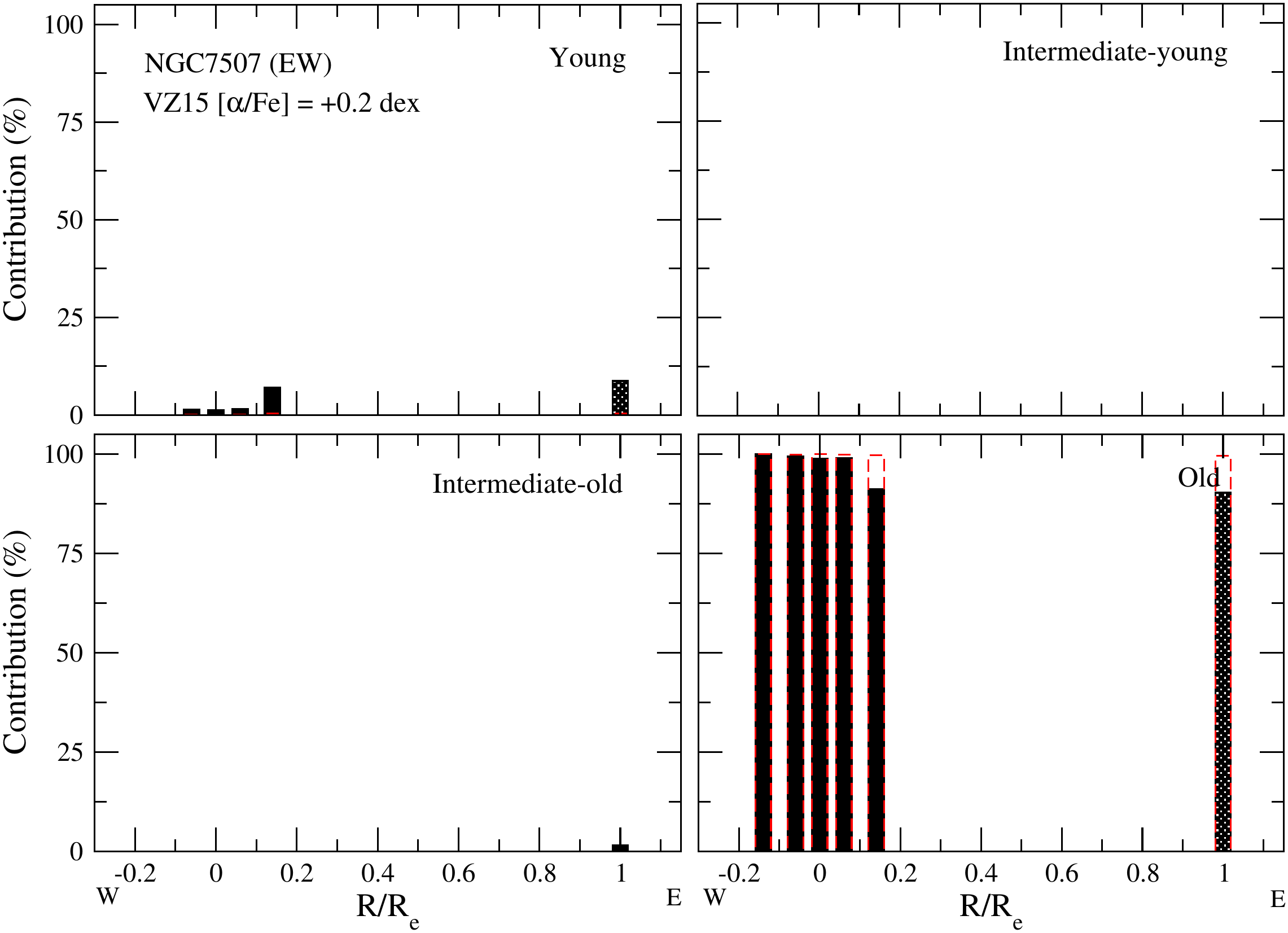}
\includegraphics*[angle=0,width=0.35\textwidth]{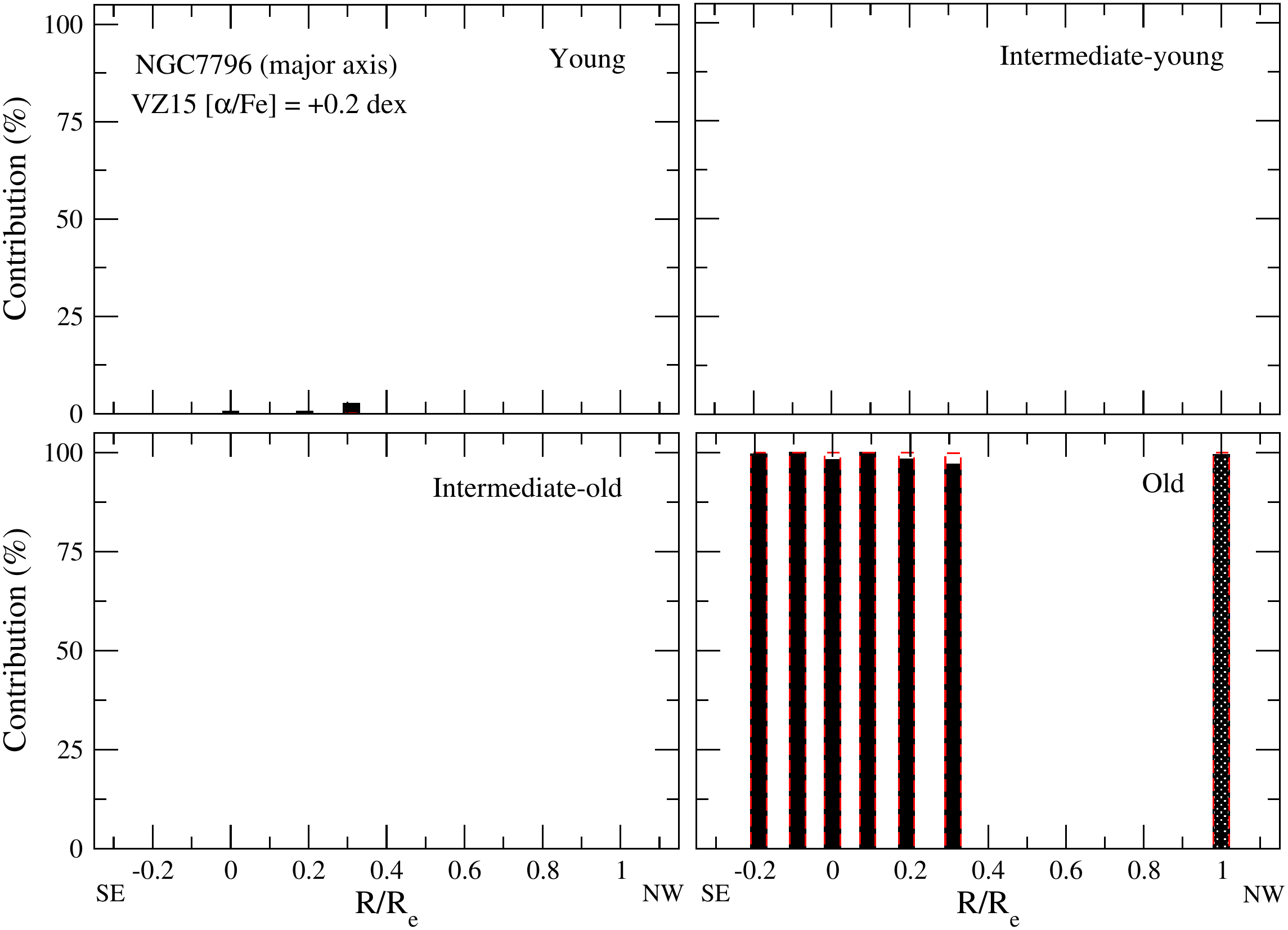}
\includegraphics*[angle=0,width=0.35\textwidth]{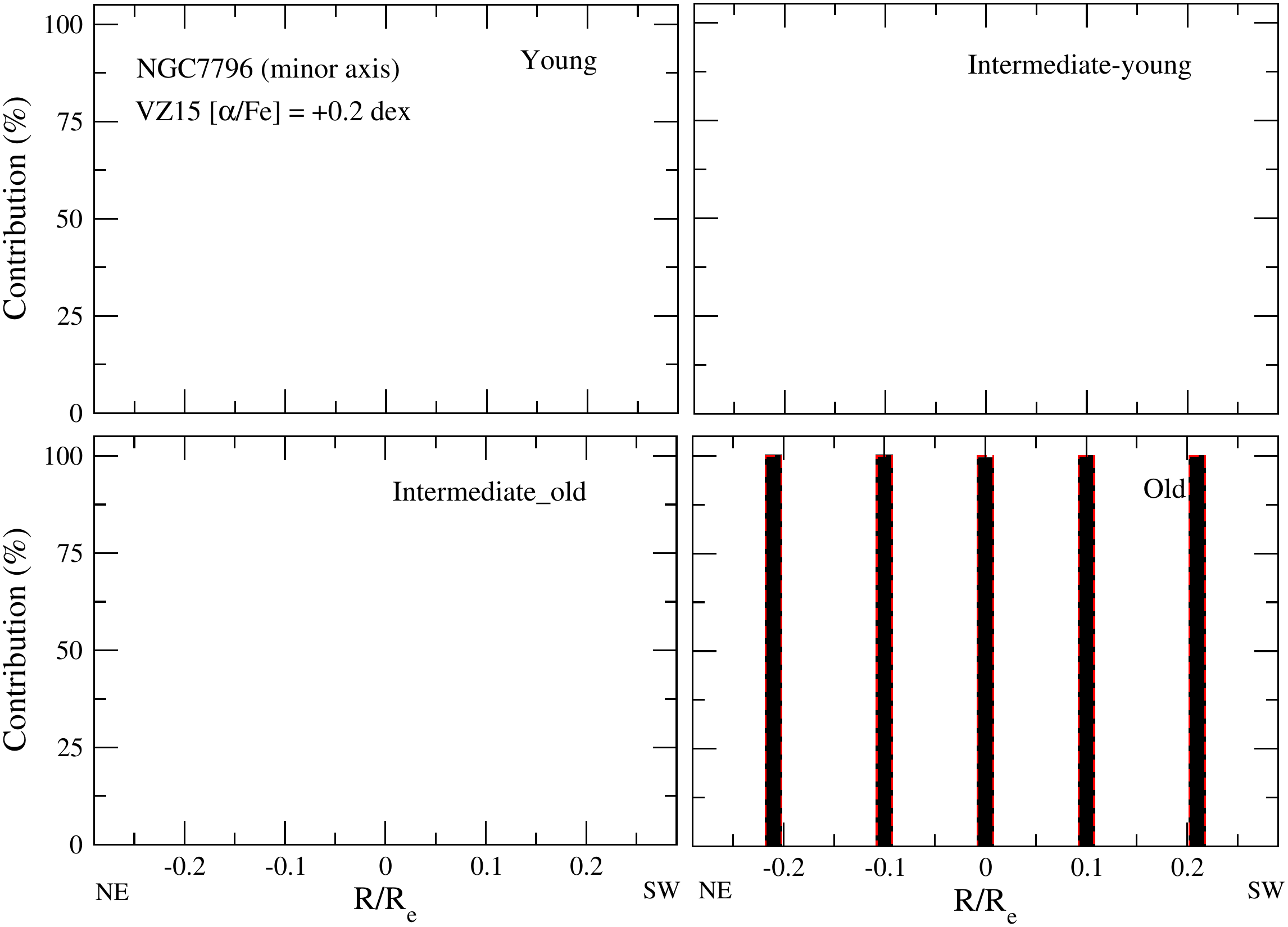}
\caption{Same as Fig.~\ref{ic5328_A}, but for NGC\,1052, NGC\,1209, NGC\,6758, NGC\,6861, and NGC\,7796 
on the photometric axes and EW for NGC\,5812, and NGC\,7507.}
\label{n_bins_nova}
\end{figure*}

\begin{figure*}
\centering
\includegraphics*[angle=0,width=0.43\textwidth]{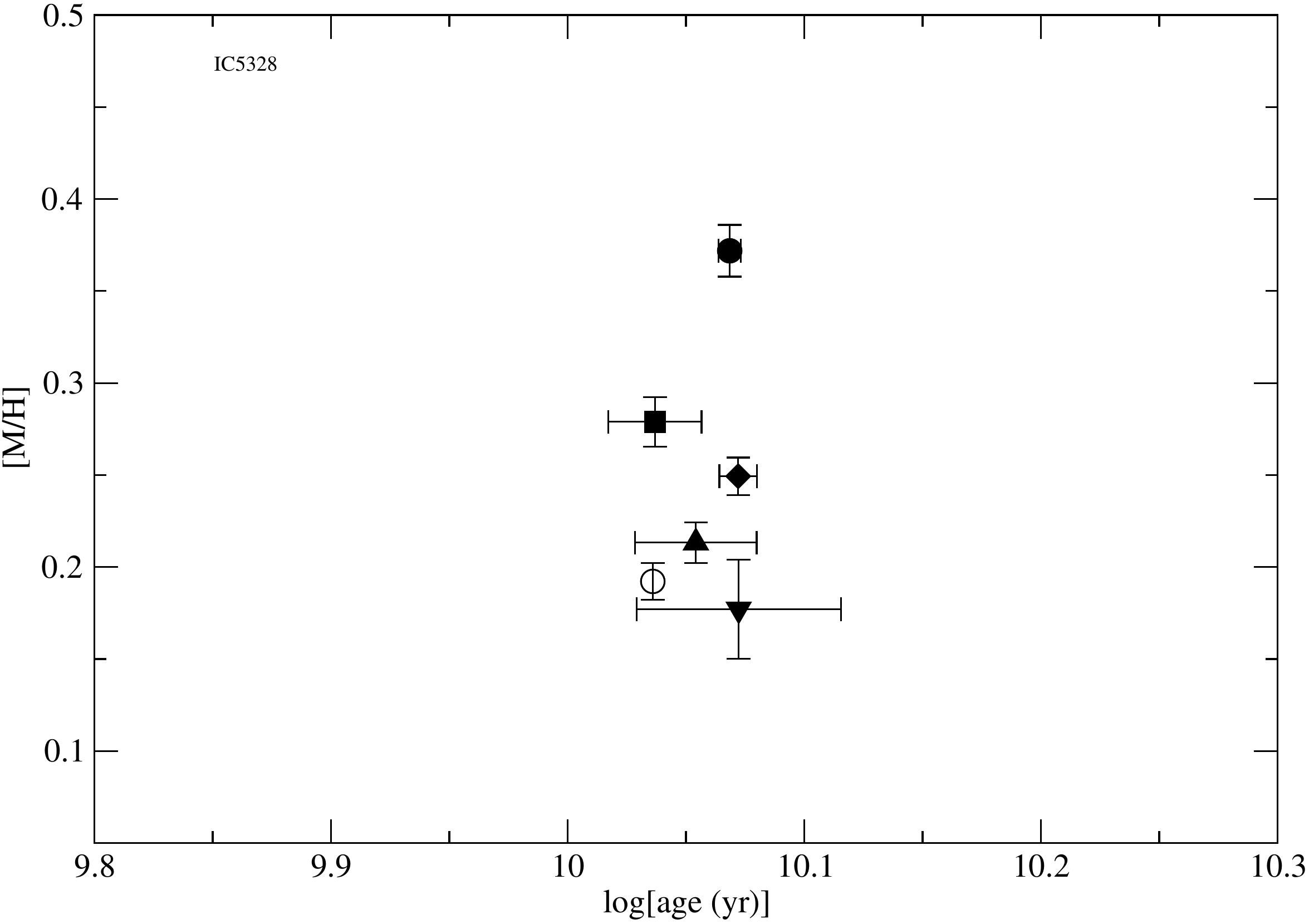}
\includegraphics*[angle=0,width=0.43\textwidth]{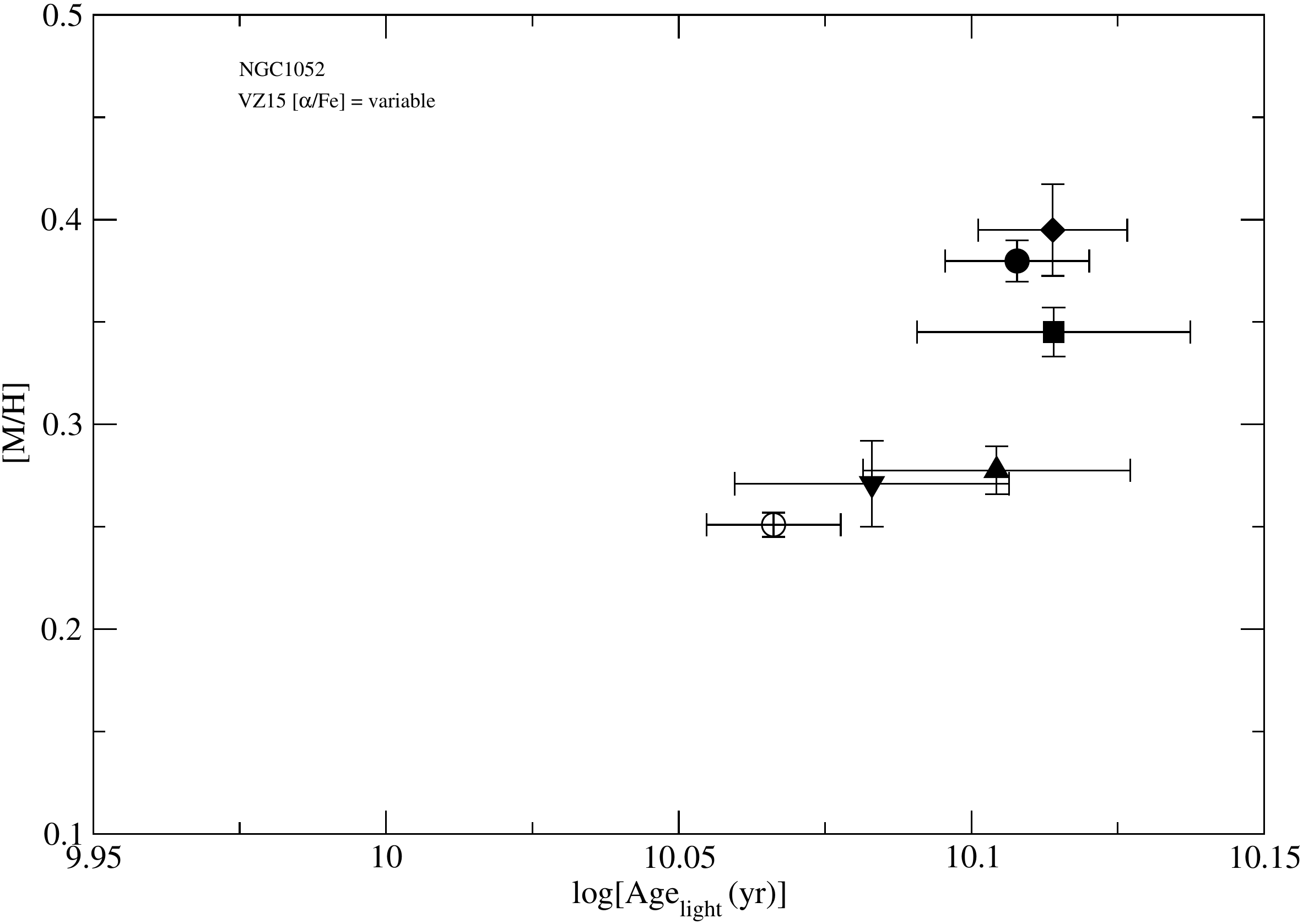}
\includegraphics*[angle=0,width=0.43\textwidth]{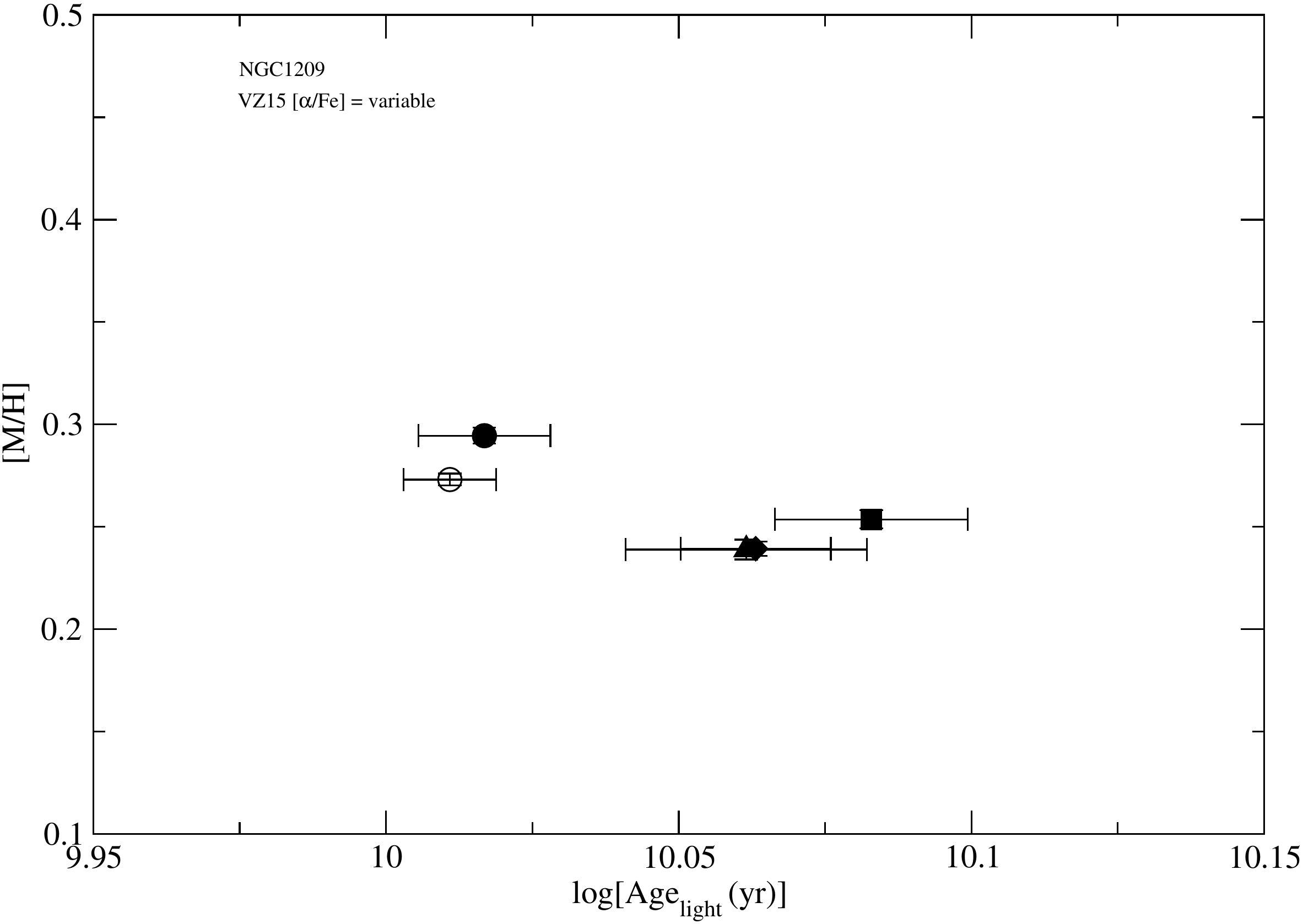}
\includegraphics*[angle=0,width=0.43\textwidth]{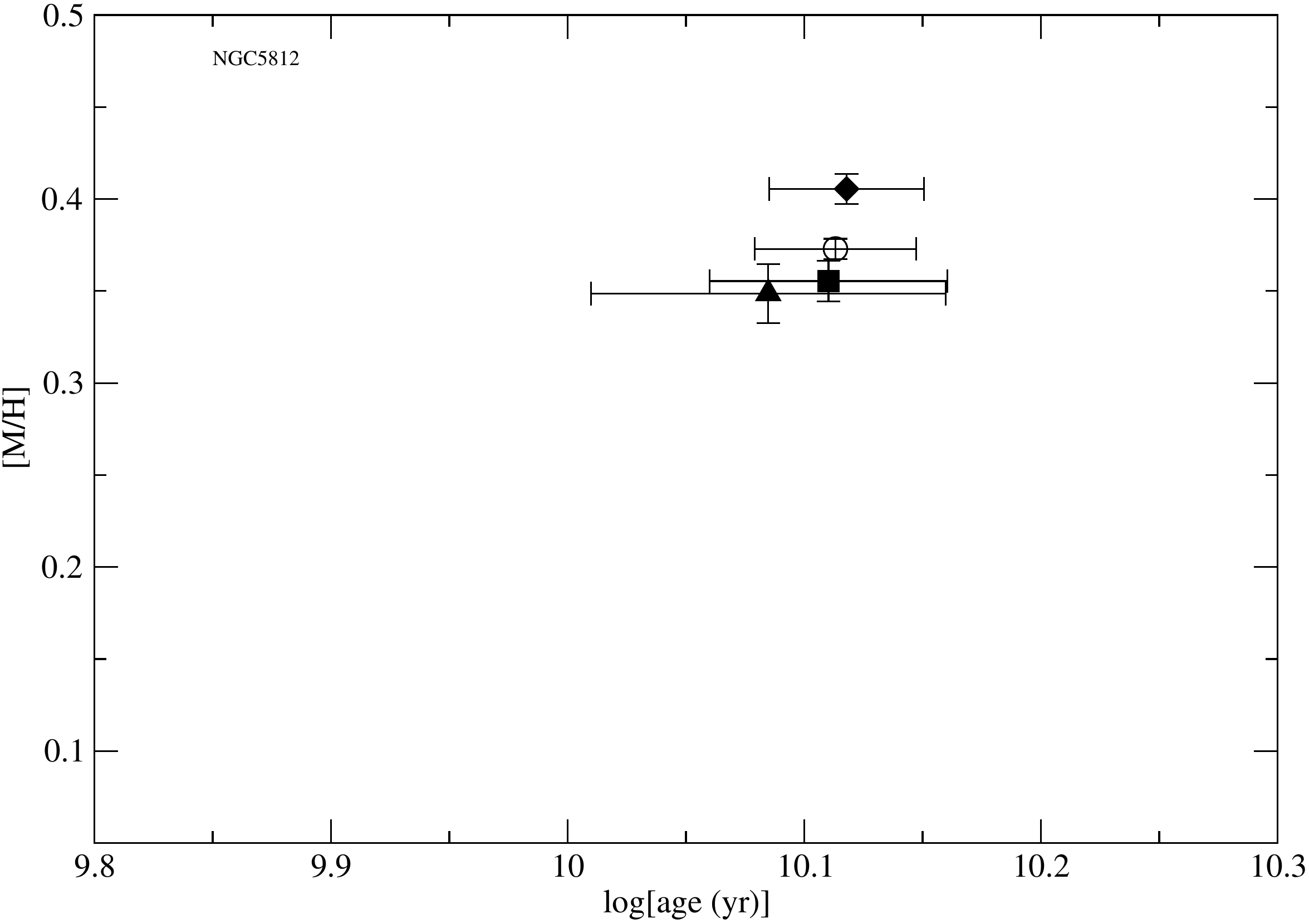}
\includegraphics*[angle=0,width=0.43\textwidth]{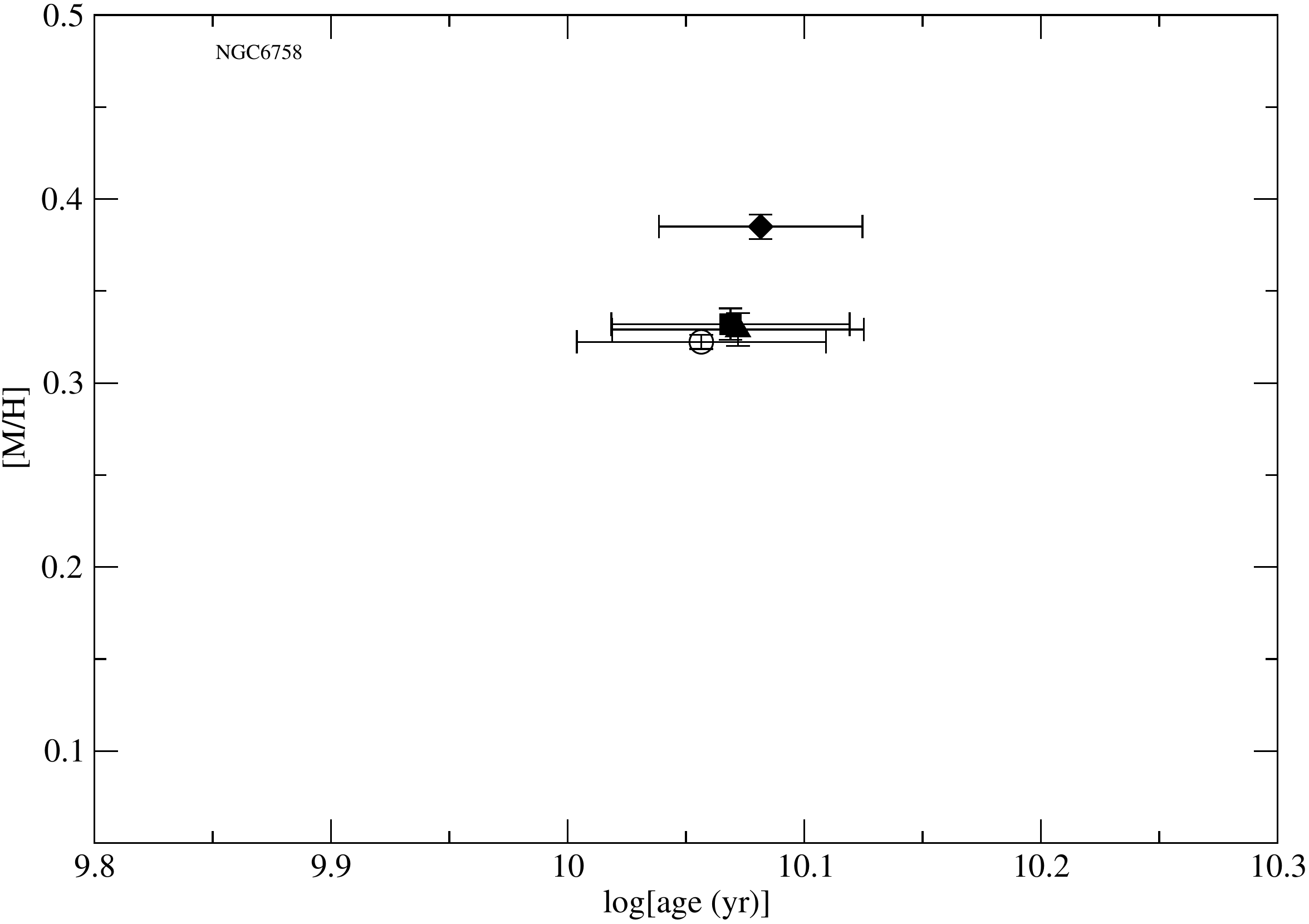}
\includegraphics*[angle=0,width=0.43\textwidth]{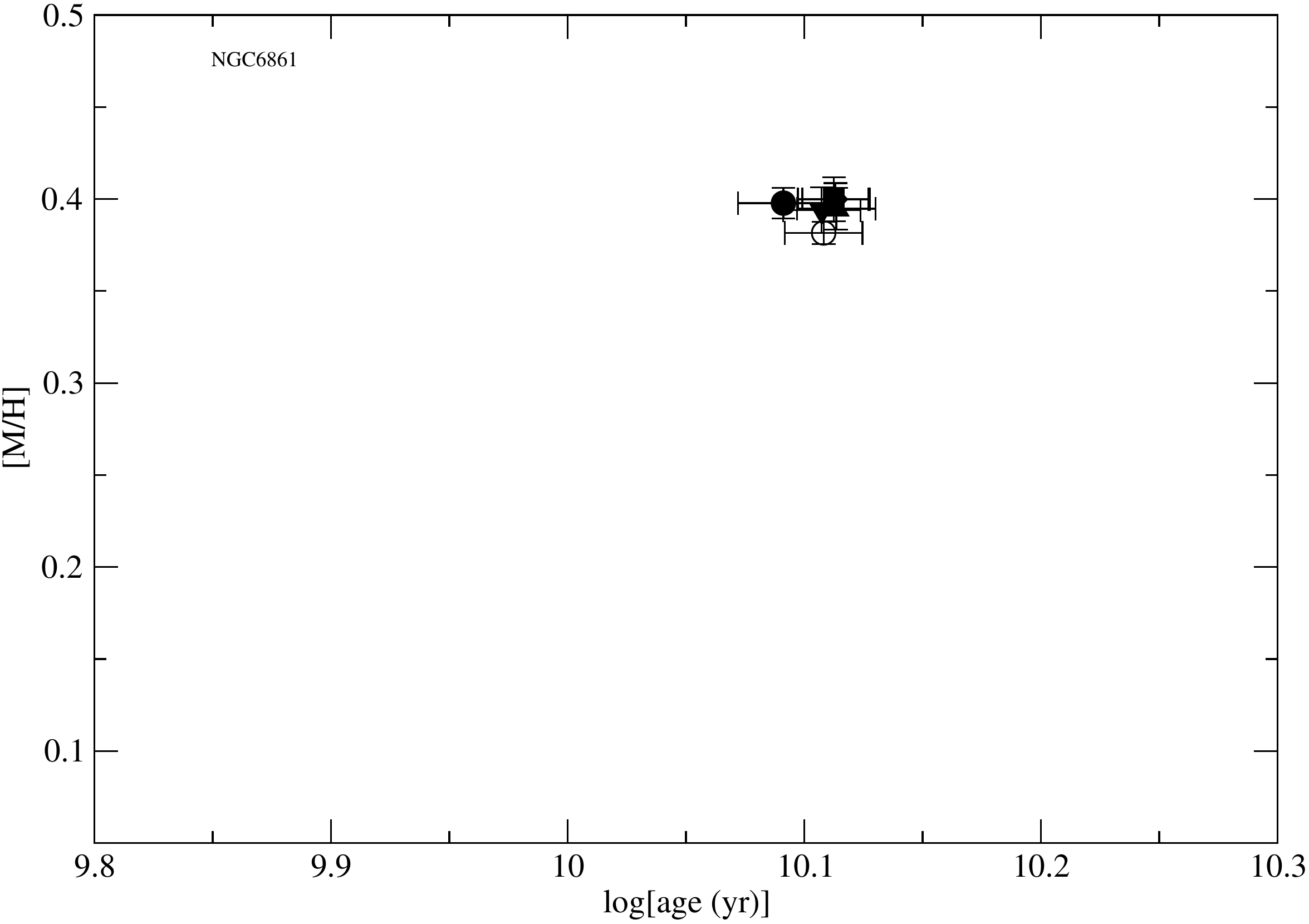}
\includegraphics*[angle=0,width=0.43\textwidth]{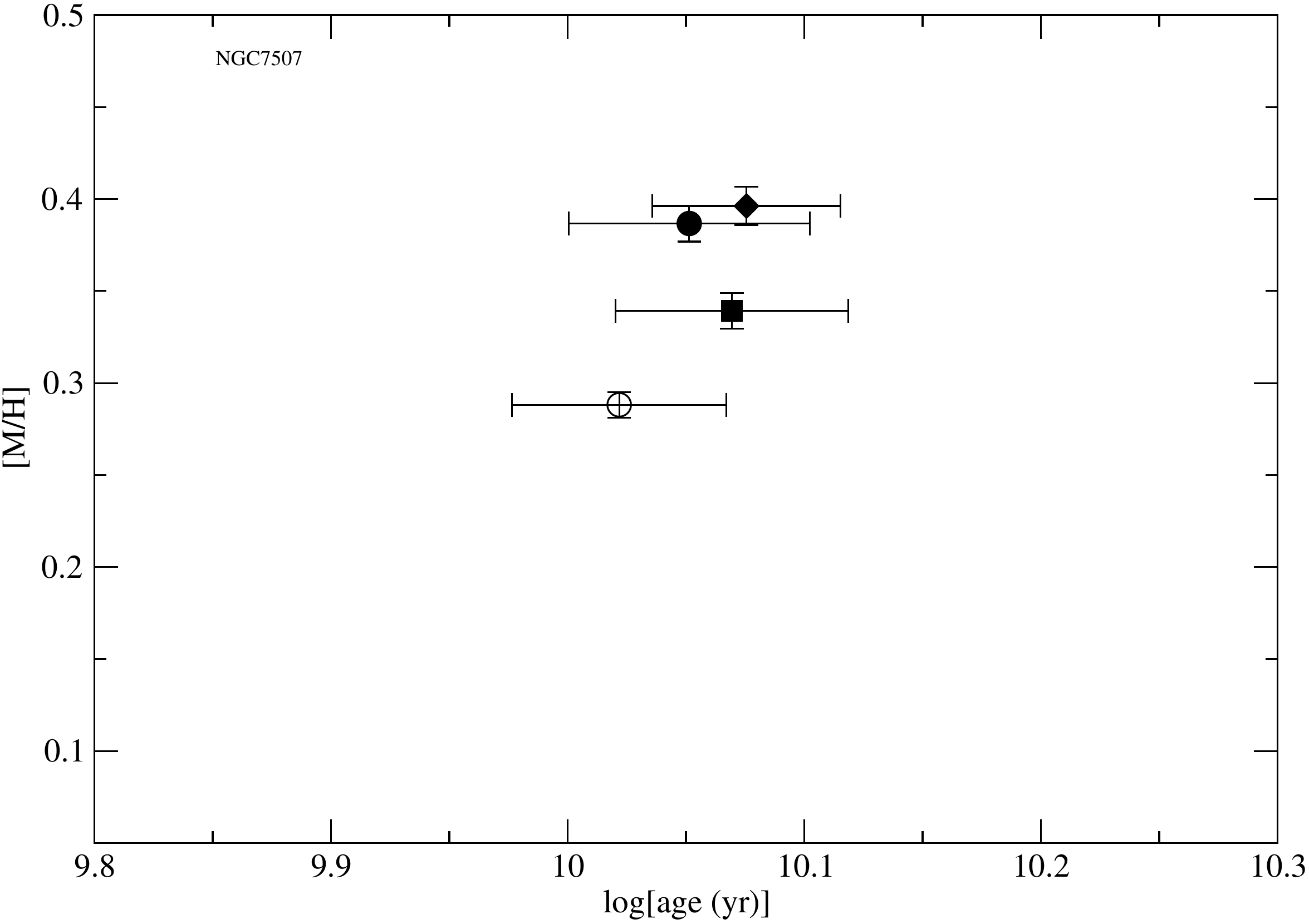}
\includegraphics*[angle=0,width=0.43\textwidth]{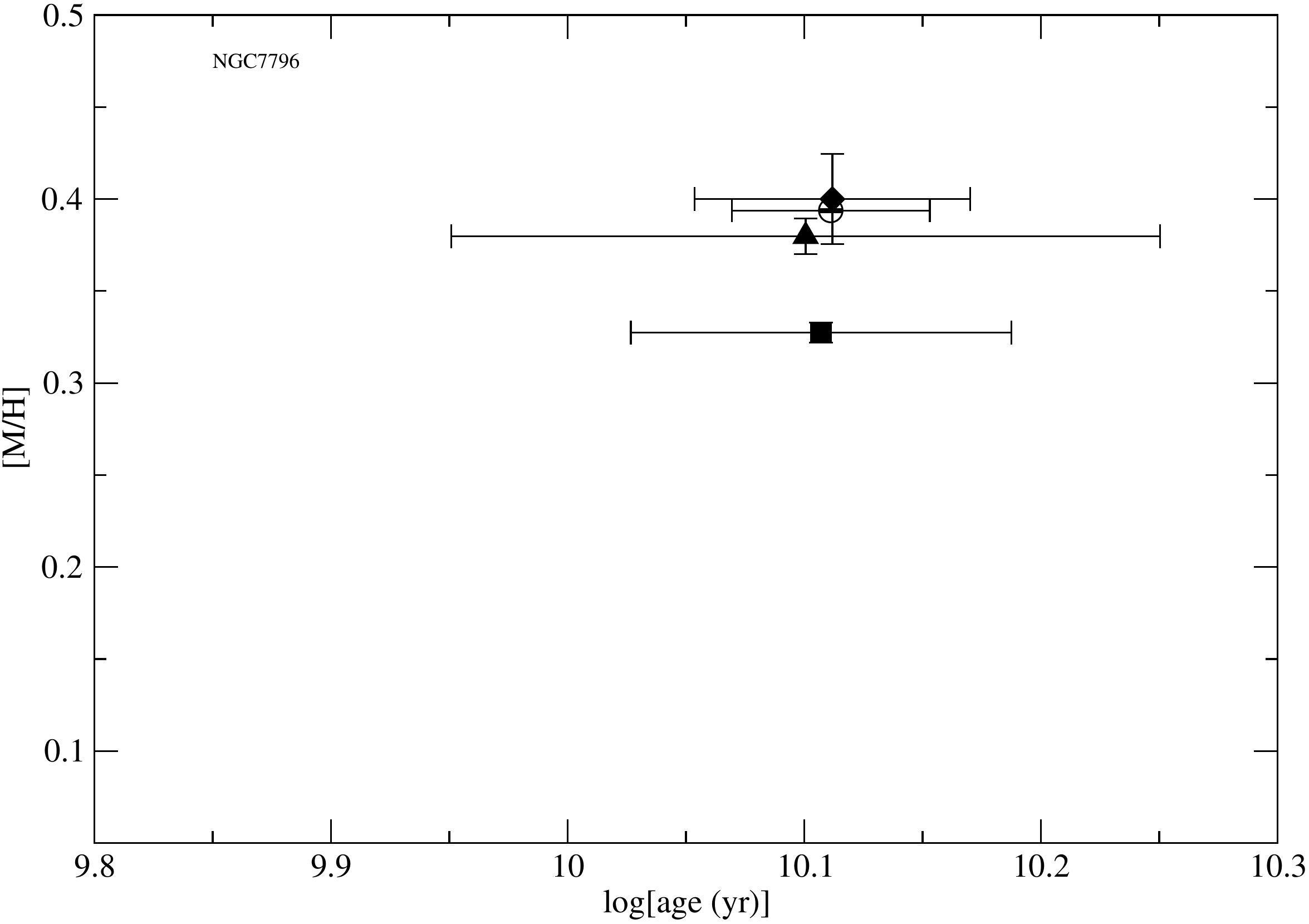}
\caption{Same as Fig.~\ref{zsolar}, but for six sample galaxies separately shown as indicated 
(stellar population synthesis results across a single slit direction: major axis for IC\,5328, NGC\,6758,
NGC\,6861, and NGC\,7796, and EW for NGC\,5812 and NGC\,7507).}
\label{che2}
\end{figure*}

\label{lastpage}

\end{document}